\documentclass[reprint,onecolumn,showpacs,footinbib,amsmath,amssymb,aps,prf,longbibliography]{revtex4-1}

\usepackage{graphicx}
\usepackage{bm}

\usepackage[normalem]{ulem}
\usepackage{soul}
\usepackage[abs]{overpic}
\usepackage{pinlabel}
\usepackage{xcolor}
\usepackage[export]{adjustbox}
\usepackage{multirow}
\usepackage{booktabs}
\usepackage{fixmath}

\newcommand\Rey{\mbox{\textit{Re}}}
\newcommand\etal{\mbox{\textit{et al.\ }}}

\begin{document}

\title{A DNS Study of entrainment in an axisymmetric turbulent jet as an episodic process}

\author{Prasanth Prabhakaran}
\email[]{prasantp@mtu.edu - presently at Michigan Technological University}
\affiliation{Max Planck
Institute for Dynamics and Self-Organization, Goettingen, 37077, Germany\\
Engineering Mechanics Unit, Jawaharlal Nehru Center for Advanced Scientific Research,
Bangalore 560064, India}
\author{Sachin Y. Shinde}
\email[]{sachin@iitk.ac.in}
\affiliation{Department of Mechanical Engineering, Indian Institute of Technology, Kanpur, 208016,, India\\
Engineering Mechanics Unit, Jawaharlal Nehru Center for Advanced Scientific Research,
Bangalore 560064, India}
\author{Roddam Narasimha}
\email[]{roddam@jncasr.ac.in}
\affiliation{Engineering Mechanics Unit, Jawaharlal Nehru Center for Advanced Scientific Research,
Bangalore 560064, India}

\date{\today}

\begin{abstract}
This investigation is based on a Direct Numerical Simulation (DNS) of a steady self-preserving incompressible axisymmetric turbulent jet at a Reynolds number of $2400$ (based on orifice diameter $d_0$ and mean exit velocity $\overline{w}_0$ at floor level $z~=~0$). The DNS data enable accurate maps of the outer irrotational flow field, and also the vorticity field $\mathbold{\omega({x}},t)$ in the turbulent core of the jet.  It is found necessary to define two separate boundaries of the jet. The first is an inner boundary (turbulent/nonturbulent, T/NT) at $|\mathbold{\omega(x,}t)|=0.5$ (quoted in units of local center-line velocity and half-width), from where $|\mathbold{\omega(x},t)|$ rises steeply towards the turbulent core whose boundary is located around $|\mathbold{\omega(x},t)|=1.0$. The second is an outer rotational/irrotational (R/IR) boundary at $|\mathbold{\omega(x},t)|=0.1$, beyond which the flow may be considered irrotational. In the diametral and axial sections, the separation between the inner and outer boundaries can vary from $2\eta$ to $5\lambda$ ($\eta$ and $\lambda$ are the Kolmogorov and Taylor length scales respectively). In the latter case, $|\mathbold{\omega(x},t)|>0.1$ between the two boundaries but is not stochastic, so the inter-boundary region is in the nature of a viscous laminar sheath.  This region can often be traced to a fossil of a vortical tongue that has  a long life of the order of a few hundred flow units ($d_0/\overline{w}_0$),   because of the very low advective velocity towards the edge of the jet.  The velocity field beyond the outer boundary often has ordered, nearly irrotational circulatory motions. These can be shown, in  simpler cases, to be the velocity field induced by one or more vorticity elements in a coherent structure in the turbulent core.  A detailed examination of axial and diametral sections indicates that there are periods when there is a large inrush of ambient fluid into parts of the T/NT interface, which gets distorted into a ‘gulf’ or `well' that can be both twisted and deep.  Sections of these wells often appear as what may be called as `lakes' of irrotational fluid in diametral sections of the jet flow.  Part of the inrushing fluid crosses the T/NT interface within the well and is entrained into the turbulent core, by a process that can legitimately be called nibbling.  The duration of such an inrush process can be of the order up to $20$ flow units  and suggests that entrainment can be an episodic process in which an inrush event accelerates ambient fluid even as it is pushed into a narrowing  gulf, where it penetrates the T/NT interface of the gulf by nibbling.  This supports the view that engulfment and nibbling are successive stages in the life of an entrainment episode. 
In view of the well-known difficulties associated with the analysis of fluid flow past a highly convoluted 3D fractal interface, the flow is analyzed here by the device of enclosing the jet in $z_1\le z\le z_2$ by a minimal cylindrical disk of height $z_2-z_1$ and radius nearly touching the radially farthest point on the inner edge of the jet anywhere within the disk.  By analyzing the radial flow at the edge of the disk with flow at the T/NT interface in the same angular sector, it is found that the two flows are correlated in about $90$\% of the time and anti-correlated the rest of the time. A mass flux budget across the surfaces of the disk and the turbulent core of the jet enables determination of the inrush events around the polar angle in the diametral section, and suggest an average of 12 to 16 events in $360^{\circ}$. This episodic view permits the introduction of such flow parameters as burstiness, which is a measure of the compactness and intensity of the events under study, as used in the analysis of momentum flux events in the atmospheric boundary layer \cite{narasimha2007turbulent}.  In the present case, similar episodes are responsible for entrainment.  In the turbulent round jet, the entrainment burstiness is found to be of order $0.75$, comparable to the momentum flux burstiness found in a turbulent boundary layer.

\end{abstract}


\maketitle

\section{Introduction}
\label{sec:intro}

The process by which ambient fluid is entrained into a turbulent shear
flow has been a question of great importance for a long time. The
first significant study of the problem, due to Corrsin (1955)
\cite{Corrsin_NACA_1955}, was inspired by a striking instantaneous
shadowgraph of the turbulent wake of a bullet traveling at supersonic
speeds. This picture showed a sharp boundary between the turbulent
core of the wake and the ambient fluid. Corrsin proposed that
entrainment occurred due to what has come to be called the
\textquoteleft{nibbling}' of the ambient fluid at the turbulent /
non-turbulent (T/NT) interface of the flow across a viscous super-layer whose thickness scales with the Kolmogorov length. On the other
hand, the work of Brown \& Roshko (1972, 1974)
\cite{Brown_1972_structure, Brown_JFM_1974} presented striking optical
visualizations of flow in turbulent free-shear (or mixing) layers
showing highly organized structures, and concluded that entrainment
occurred by \textquoteleft{engulfment}' of ambient fluid through the
velocity induced by the coherent structures present in the turbulent
core flow. Studies by Dimotakis (1983) \cite{Dimotakis_PoF_1983} on an
axisymmetric jet showed that the instantaneous boundary of the
turbulent core (as marked by passive scalar dye concentration) was
highly convoluted, and supported the idea for a dominant role of
coherent structures in entrainment, by engulfment of ambient fluid.

The main issue in much recent work on entrainment in turbulent shear
flows can be summarized as a debate on nibbling vs. engulfment as
primary candidates for the entrainment mechanism. While it is
realized that it is not easy to make a precise distinction between the
two (Da Silva (2014) \cite{Silva_ARFM_2014}), recent work, conceding
that engulfment may be responsible for most of the entrainment in
mixing layers (Westerweel \etal (2002)
\cite{Westerweel_ExpFluids_2002}), has generally concluded that it
accounts for little in other turbulent flows like jets, wakes (Bisset et al (2002)
\cite{Bisset_JFM_2002}) and grid
turbulence (Holzner \& Luethi (2011) \cite{Holzner_PRL_2011}). A related part of the
debate has been concerned with determining the length and velocity
scales relevant to the entrainment process, and in particular the
roles played by the Kolmogorov and Taylor length scales therein \cite{Westerweel_JFM_2009, Silva_PoF_2010, Silva_ARFM_2014}.

On these issues the jet has been the most widely studied
turbulent shear flow (experiments on round jets reported by Westerweel
\etal (2002) \cite{Westerweel_ExpFluids_2002}, Wolf \etal (2012)
\cite{Wolf_PoF_2012}; DNS results on a temporal round
jet by Mathew \& Basu (2002) \cite{Mathew_PoF_2002}, on a temporally evolving plane
jet by Reeuwijk and
Holzner (2014) \cite{Reeuwijk_JFM_2014} and on a spatially
developing plane jet by Da Silva (2010) \cite{Silva_PoF_2010}). The analysis of Mathew \& Basu
\cite{Mathew_PoF_2002} concludes that most of
the entrainment occurs due to small-scale action at the T/NT interface. This idea was further supported by Westerweel \etal (2005, 2009)
\cite{Westerweel_PRL_2005, Westerweel_JFM_2009} and Wolf \etal (2012)
\cite{Wolf_PoF_2012} based on PIV measurements on
a round jet. All these studies report that small-scale activity at the T/NT interface plays a dominant role in the
entrainment process. Westerweel \etal (2009) \cite{Westerweel_JFM_2009} concluded that large scale engulfment contributed about $8$\% of the net mass flux. Wolf \etal (2012)
\cite{Wolf_PoF_2012} used particle tracking velocimetry to generate the 3D velocity field and compute the interface velocity using this data. Their studies suggested that the entrainment velocity is dominated by viscous action due to the small scale eddies at the T/NT interface. They also show that the entrainment velocity is strongly influenced by the local curvature of the interface with higher entrainment at the concave region of the T/NT interface. The analysis on the DNS of temporal jets by Reeuwijk \& Holzner (2014)~\cite{Reeuwijk_JFM_2014} shows the presence of a viscous super layer (VSL) enveloping the vortical core. They also define a buffer region similar to those in wall bounded flows, which connects the VSL with the turbulent core. 

Philip and Marusic (2012) \cite{Philip_PoF_2012}
proposed a large eddy model to reproduce some of the features of the
large scale motion in jets and wakes. Recent work by Mistry \etal
(2016) \cite{mistry2016entrainment}, using simultaneous PIV and PLIF measurements, stresses the
multi-scale nature of the entrainment velocity and the
scale-independence of mass flux in a turbulent jet. These studies, 
along with the work of Turner (1986) \cite{turner1986turbulent}, Dahm \& Dimotakis (1987) \cite{dahm1987measurements}, Mathew
\& Basu (2002), Wolf \etal (2012)
\cite{Wolf_PoF_2012} and Reeuwijk \& Holzner (2014)~\cite{Reeuwijk_JFM_2014} propose that entrainment dynamics is
dictated or imposed by large-scale coherent motions, to which small scale
viscous action adjusts itself to maintain a scale-independent mass flux rate.

In a recent work Burridge \etal (2017)~\cite{burridge_parker_kruger_partridge_linden_2017} have
reported simultaneous PIV/PLIF measurements on the entrainment dynamics in a turbulent
axisymmetric plume. They present conditional statistics highlighting the role of large scale
motions in the entrainment process. Their studies suggest that significant momentum is imparted to
the fluid outside the T/NT interface in the zones between the coherent structures. Moroever, they conclude
that turbulent entrainment in a plume is dominated by engulfment in the first stage followed by
viscous nibbling. Similar conclusions on the role of the large scale motions were reported earlier by
Plourde \etal (2008) \cite{Plourde_JFM_2008}. They associate the presence of large scale organized
motions with higher instantaneous entrainment levels.

To the best of our knowledge there has been no
numerical study of the round jet from the entrainment point of view,
and the present work was undertaken in part to fill this gap. The big advantages of having DNS data are that the vorticity vector field can be accurately determined, enabling a vorticity view of the T/NT interface, and the flow field in the ambient fluid beyond the edge of the jet can be accurately determined. The
present report starts by presenting, as a first step, a physical picture of entrainment largely
derived from DNS imagery. This is followed by an analysis of the entrainment budget across a
cylindrical disc of finite thickness circumscribing the jet in the self-preserving zone of the jet. The
analysis strongly suggests an episodic
(spatially discrete and  temporally intermittent) nature of the entrainment
process. One reason for this approach is that while much attention
has been given in earlier studies to flow within and very close to the
fully turbulent core, surprisingly little attention has been given to
the structure of the flow in the ambient fluid outside the turbulent core in
a slightly more extended neighborhood (over an area with a length
scale of order $b_w$, the radius of the jet at half the mean
center-line velocity). This may partly be due to
difficulties in making accurate measurements of the very low
velocities in the ambient fluid. Furthermore, early models for what
may be called the \textquoteleft{ambient near-field}' visualized it as
resulting from a distribution of sinks along the center-line of the
jet (Phillips (1955) \cite{Phillips_1955}, Bradshaw (1976)
\cite{Bradshaw_1976}), but this approach does not throw any light on the spatio-temporal structure
of the flow in the ambient fluid. A very useful recent review of these
developments has been presented by Da Silva \etal (2014)
\cite{Silva_ARFM_2014}.

Several definitions of entrainment have been proposed in the literature (Turner (1986) \cite{turner1986turbulent}). All the experiments and simulations on jets and wakes discussed so far define entrainment as the velocity at which the turbulent flow spreads into the irrotational ambient. Here we define entrainment as the rate at which the irrotational ambient fluid enters into the vortical turbulent core of the jet. The two definitions have significantly different implications which are best understood by looking at a planar wake. According to the former definition of entrainment, a planar wake has non-zero entrainment as the wake spreads in the streamwise direction. On the contrary, the wake has a net zero entrainment using the latter definition as the average mass flux in the streamwise direction is a constant. It is important to distinguish between encroachment (as in the spreading wake with no change in mass flux) and entrainment (which involves an increase in turbulent mass flux). In a turbulent round jet, the net mass flux in the streamwise direction as well as the diameter increases downstream. We investigate here the entrainment dynamics close to the T/NT interface in a turbulent round jet.

Against this background we present data on the instantaneous flow
field outside the turbulent core in a round jet, as obtained by a
direct numerical simulation of the jet flow.  This makes it possible
to visualize the general structure of the ambient flow field, and
examine its possible relation to the vorticity field (or other flow
characteristics) within the jet. A general schematic view of the
velocity field, as sketched for example by Philip \& Marusic (2012)
\cite{Philip_PoF_2012}, implies that the ambient fluid flows into the
turbulent jet along more or less parallel streamlines almost right up
to the T/NT interface, where it is entrained into the jet core across
its edge. A somewhat more structured velocity field is indicated by
the measurements reported by Westerweel \etal (2009)
\cite{Westerweel_JFM_2009} (Figs.~19 and 20).  Here we present
evidence, chiefly through flow imagery, indicating that the
instantaneous flow in the ambient neighborhood of the turbulent edge
of the jet is much more organized than thought previously, and the entrainment across the T/NT interface is episodic in nature.  A more
detailed quantitative study will appear separately.

The paper is organized into ten sections. We start with the description of the numerical procedure used for the simulation and the details on the accuracy of the data, followed by a discussion on the identification of the boundaries separating the turbulent core from the (nearly) irrotational ambient. Later, the axial and diametral sections are analysed to understand the dynamics near the T/NT interface. Based on these details we conduct a quantitative analysis of the mass flux near the T/NT interface followed by concluding remarks.

\section{Simulation Details}
\label{sec:simu}

\begin{figure}[!h]
\begin{overpic}
[width = 8.0 cm, height = 9.30 cm, unit=1mm]
{Fig_box.eps}
\put(0,0){\includegraphics[width = 8.0 cm]{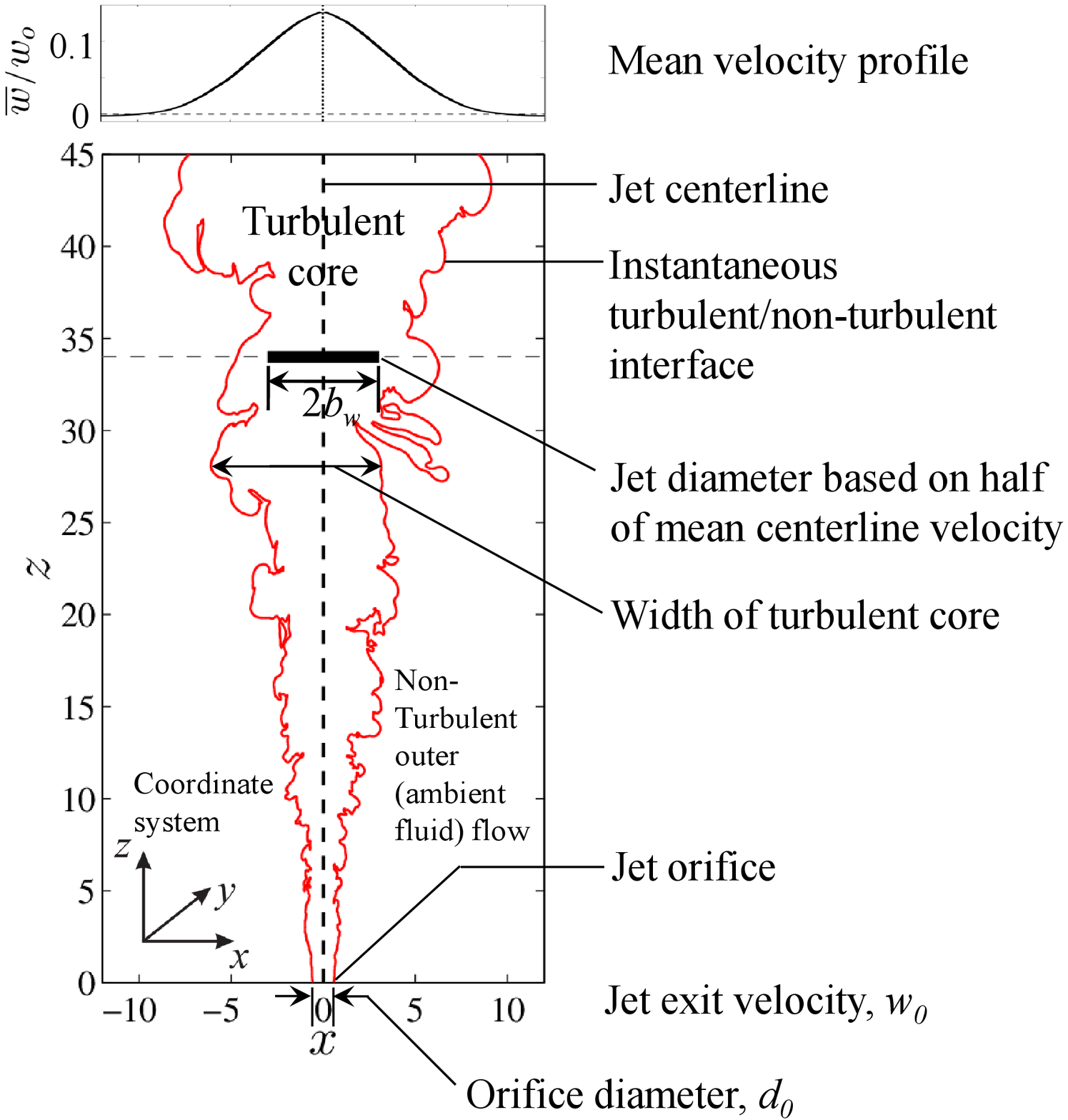}}
\end{overpic}
\caption{Diagram of an axial section to illustrate main terminology and notation for jet flow. (See additional explanation in text.)}
\label{fig:schmt}
\end{figure}

Fig.~\ref{fig:schmt} shows a schematic of an axial section of the jet,
where $d_0$ represents the orifice diameter from which the jet issues
with a top hat velocity profile characterized by a uniform jet exit mean velocity $\overline w_0$.  The flow is
governed by the incompressible Navier-Stokes equations (in
non-dimensional form with jet exit velocity scale $\overline w_0$ and nozzle exit
diameter $d_0$ as velocity and length scales respectively),
\begin{equation}
\label{eq:cont}
\nabla \cdot \bm{u} = 0,
\end{equation}
\begin{equation}
\label{eq:NS}
(\partial{\bm{u}}/\partial{t}) + (\bm{u} \cdot \nabla)\bm{u} = - \nabla{p} + (1/\Rey)\nabla^2\bm{u},
\end{equation}

\noindent
where $\bm{u}$, $p$ represent the flow velocity vector and pressure
respectively, and $\Rey = \overline w_0 d_0/\nu$ is the Reynolds number (= 2400
in the present simulations). Unless otherwise specified, all
distances, velocities and time presented here are non-dimensionalized
with $d_0$ and $w_0$ as scales, and vorticity with the time-mean local scales at height $z$, \textit{viz.}, centerline
velocity $\overline{w}_{c}(z)$ and half-velocity jet width $b_{w}(z)$.
Also note
that, unless otherwise stated, all flow variables are non-dimensional.

The equations are solved using the extension of Harlow \& Welsh's
scheme \cite{Harlow_PoF_1965} to non-uniform grids as proposed by
Verstappen \& Veldman (2003) \cite{Verstappen_JCP_2003}, using a
second-order finite volume framework. The governing equations are
integrated in time using the second-order Adams-Bashforth scheme with a time step $\Delta t = 0.005$. A
non-uniform Cartesian mesh is used to resolve all the relevant scales
in the turbulent part of the flow-field. The velocity and pressure variables are stored in a staggered arrangement to prevent pressure-velocity decoupling. The pressure variable is stored in the cell center and the velocity components are stored on the cell faces \cite{Harlow_PoF_1965}. More details on the scheme and the code are discussed in Prasanth (2014) \cite{Prasanth_MS_Thesis}. The computational domain
extends up to $z = 55$ in the streamwise ($z$) direction and $-50$ to
$+50$ in the cross-stream directions ($x, y$). The total grid size is
$480 \times 480 \times 600$ in the $x$, $y$ and $z$ directions
respectively (yielding 138 million grid points in all). In the
streamwise direction the grid size varies from 0.05 at the orifice to
0.18 at the outflow. In the cross-stream plane the grid is uniform
(of size 0.05) from $x$, $y$ = $-9.25$ to $+9.25$, thus ensuring adequate
resolution in the turbulent core; beyond this the grid is gradually stretched towards the lateral side walls. The mean viscous dissipation $\overline{\epsilon}$ is calculated
from the fluctuating strain rate $s_{ij}^\prime$ as
\begin{equation}
\label{eq:dissip}
\overline{\epsilon} = 2 \nu \overline{s_{ij}^\prime s_{ij}^\prime},
\end{equation}
\begin{equation}
\label{eq:strain}
s_{ij}^\prime = \frac{\partial {u_{i}^\prime}} {\partial x_{j}} +
\frac{\partial {u_{j}^\prime}} {\partial x_{i}},
\end{equation}
where the primes denote fluctuating quantities ($u_{i}^{\prime} = u_{i} - \overline{u}_{i}$, etc.)
The velocity derivatives are calculated using a second order central differencing
scheme. The Kolmogorov and Taylor microscales are
respectively given by
\begin{equation}
\label{eq:kolmo}
\eta = {\nu}^{3/4} \, {\overline{\epsilon}}^{-1/4},
\end{equation}
\begin{equation}
\label{eq:taylor}
\lambda = (15 \nu \overline{{w^\prime}^2} / {\overline{\epsilon}})^{1/2},
\end{equation}
where $\overline{{w^\prime}^2}$ is the turbulence intensity in the streamwise direction. The dissipation (normalized using $w_{c}^3/(z-z_0)$, where
$z_0$ is a virtual origin) near the jet centerline at $z\approx33$ in the present
simulation is $0.18$, to be compared with $0.18$ reported by
Panchapakeshan \& Lumley \cite{Panchapakesan_JFM_1993} at $Re = 11,000$, $0.22$ by Bogey
\& Bailly (2009) \cite{Bogey_JFM_2009} at $Re = 11,000$ and $0.25$ by Taub \etal (2013) at $Re = 2000$
\cite{Taub_PoF_2013}. At $z/d \approx 33$ near the jet center-line, $\lambda / b_w \simeq 0.2$, to be
compared with 0.18 (\cite{Westerweel_JFM_2009}, round jet, $\Rey =
2000$, center-line). The Kolmogorov length scale ($\eta$) at $z\approx33$ is about $0.04~d_0$, and the corresponding grid size in the $x$, $y$, $z$ directions is approximately $\eta$, $\eta$ and $3\eta$ respectively. 

The bottom plane at $z = 0$ outside the orifice is treated as a no-slip wall in the plane
of the orifice exit. A fixed, uniformly distributed
disturbance of 5\% amplitude with zero mean is superposed
on the top hat velocity profile at the orifice to trip the flow to a
turbulent state. The disturbance is superposed only on the streamwise
velocity component, however, and at a random set of points also chosen
from a uniform distribution over the orifice area. At the outflow
boundary we use the zero normal derivative condition for all the
variables, with a layer of viscous padding from $z = 52$ to $z = 55$
to ensure smooth exit of the turbulent flow from the computational
domain. The lateral boundaries of the computational domain are also
treated as no slip walls, but (at $x, y = \pm 50$) they are
sufficiently far away from the jet axis to have any significant effect
on the momentum balance. The net momentum flux $ \int (\overline{w}^2
+ \overline{{w^\prime}^2} -0.5[\overline{{u^\prime}^2}+\overline{{v^\prime}^2}] )dS$, where the overbar
denotes time average, prime denotes fluctuation and the domain of
integration is $-50\le x, y\le50$, is conserved to better than 99\% over the whole domain in the
$z$ direction (Shinde \etal (2019) \cite{shinde2018}). Other details about the solver and its validation are
discussed in Prasanth (2014) \cite{Prasanth_MS_Thesis}. The simulation took up to $t =
900$ to attain a stationary state. Data for computing flow statistics
was acquired over the time interval $900<t<2650$. The simulation was
run on 108 nodes of the 360 TF supercomputer at CSIR-4PI, Bangalore.

\section{Data analysis}
\label{sec:data-analysis}

The mean
streamwise component of the velocity at $r, z$ is computed as
\begin{equation}
\label{eq:time-avg}
\lim_{t \rightarrow \infty} \frac{1}{t} \int_{t_1}^{t_2}  \frac{1}{2\pi}\int_{0}^{2\pi}\bm{w}(r, \bm{z}, \phi, t^{\prime}) d\phi dt^\prime = \overline{\bm{w}}(\bm{r, z}),
\end{equation}
\noindent
where $t_1 = 900$, $t_2 = 2600$, $t = t_2 - t_1$, and $\phi$ is the azimuthal angle. It is found that the jet is in a self-similar state over the range $27
\leqslant z \leqslant 40$ to within $0.6\%$ in the mean velocity
($\overline{w}/\overline{w}_c$) and in an equilibrium or self-preserving state over
the range $32 \leqslant z \leqslant 36$, the maximum in $(\overline{{w^\prime}^2}/\overline{w}^2_c)^{0.5}$ and Reynolds shear stress $- \overline{w^\prime u^\prime}/\overline{w}^2_c$ vary less than $1\%$ and
$1.5\%$ respectively. Additional details on the statistics are available in Shinde \etal (2019) \cite{shinde2018}.

We present here analyses of the solutions for the vorticity field
within the turbulent jet and the velocity field outside a suitably
determined \textquoteleft{edge}' of the turbulent core of the jet, defined as the contour surface of a specified threshold
value of the vorticity modulus, 
\begin{equation}
\label{eq:vort-mag}
\left|\bm{\omega} (\bm{x},t) \right| = \sqrt{\omega_{x}^2 + \omega_{y}^2 + \omega_{z}^2}
\end{equation}
\noindent
in terms of its Cartesian components.  Vorticity-based criteria are
particularly appropriate as the essential characteristic of any
turbulent flow is a time-dependent stochastically varying vorticity
field, which is logically a more fundamental parameter than the scalar
concentration field of a flow-visualizing dye, often used in
experiments (e.g. \cite{Westerweel_PRL_2005}, \cite{Westerweel_JFM_2009},
\cite{Dimotakis_PoF_1983}).  The definition of the edge is determined as
a characteristic value of $|\bm{\omega}|$ connected with the interface
layer of relatively rapid change that is encountered as a test point
moves inward toward the jet core from the ambient (nearly)
irrotational flow (see Fig.~\ref{fig:schmt}).  Based on extensive
studies of these vorticity profiles across the edge (Shinde et al (2018)), it is found
convenient to introduce two distinct edges or boundaries of the jet.
The first is located at the \textquoteleft{middle}' of the T/NT
interface at $|\bm{\omega}| = 0.5$, beyond which $|\bm{\omega}|$
varies rapidly to 1.0 or above in the turbulent core. This defines the
\textquoteleft{inner}' boundary of the turbulent jet.

For reasons that will be clear shortly, we find it necessary to define
an \textquoteleft{outer}' boundary at $|\bm{\omega}| = 0.1$, which
similarly locates the rotational / irrotational (R/IR) boundary of the
jet. The basis for this choice will not be presented here in detail,
chiefly because the images shown in the rest of the paper provide
ample supporting evidence for the choices made (see
Sec. \ref{Results1}).  The contour surfaces $|\bm{\omega}| = 0.1$
and $0.5$ will thus be referred to in the sequel as the outer and
inner boundaries of the jet respectively. There have been other
proposals for defining two boundaries, using slightly different
criteria. Reeuwijk \& Holzner (2014)~\cite{Reeuwijk_JFM_2014} define an inner and outer boundary that separate the turbulent core from the buffer region and the VSL from the buffer region respectively. The inner boundary in their analysis represents the enstrophy threshold for which the interface propagation velocity $v_n$ is zero and the outer boundary respresents the threshold below which the enstrophy production is negligible. 

In most of the images of jet cross-sections presented here,
it has been found illuminating to include a background vorticity field
within the core of the jet.  Depending on the purpose of the analysis being
carried out, different choices may be made for the particular measure
of vorticity that is most appropriate for the analysis. Thus
$\omega_{z}$ may be more relevant in a diametral section. Similarly,
in an axial section in the $x-z$ plane at $y = 0$, either the modulus $|\omega_{y}|$ or
the signed value of $\omega_{y}$ or the azimuthal component
$\omega_{\phi}$ may be more appropriate.

The axial section is taken over the range $23 \leqslant z \leqslant
45$. The flow is also studied in two diametral planes, one at
$z = 34.05$ which lies exactly in the middle of the self-preserving
zone, and the other at $z = 42.69$.

\section{Jet Boundaries}
\label{Results1}

We shall now present the results derived largely from an analysis of axial and
diametral sections. Particular attention is given to possible connections between the flow-field
outside the jet boundary and the vorticity field within the jet. Only a limited
number of 2D sectional data are analyzed, as this is adequate for the
conclusions drawn here. A more detailed analysis, including that of
selected 3D velocity and vorticity fields will be reported
separately.

\begin{figure*}[h!]
\begin{overpic}
[width = 16.0 cm, height = 12.5 cm, unit=1mm]
{Fig_box.eps}
\put(15,0){\includegraphics[width = 13 cm]{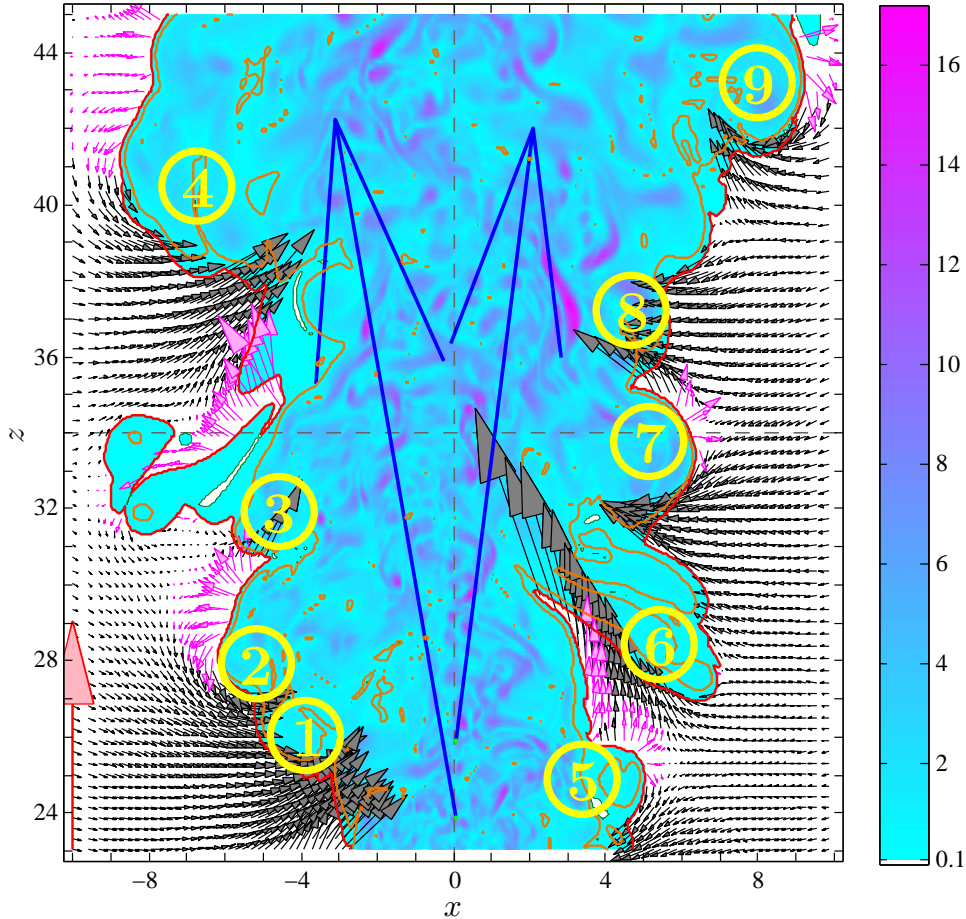}}
\end{overpic}
\caption{Axial section in the $xz$ plane at $t = 925$, showing the
instantaneous velocity and total vorticity modulus
$|\bm{\omega}(\bm{x},t)|$ fields. The black and pink velocity vectors
(respectively with radially inward and outward components) are shown
only in the ambient beyond the outer boundary of the jet, represented
by the red contour at $|\bm{\omega}| = 0.1$; the orange contour at a
threshold of 0.5 represents the inner boundary). For comparison two
typical flow velocity vectors (blue) inside the core of the jet
near the jet centerline are plotted. To avoid clutter, only every
fourth velocity vector in the $x$-direction and every third in the
$z$-direction are plotted. The red vertical arrow at the bottom-left
corner is a reference velocity vector of magnitude $0.1 w_0$.}
\label{fig:vel-vort} 
\end{figure*}

\begin{figure*}[h!]
\begin{overpic}
[width = 16.50 cm, height = 6.4 cm, unit=1mm]
{Fig_box.eps}
\put(0,0){\includegraphics[width = 4.10 cm]{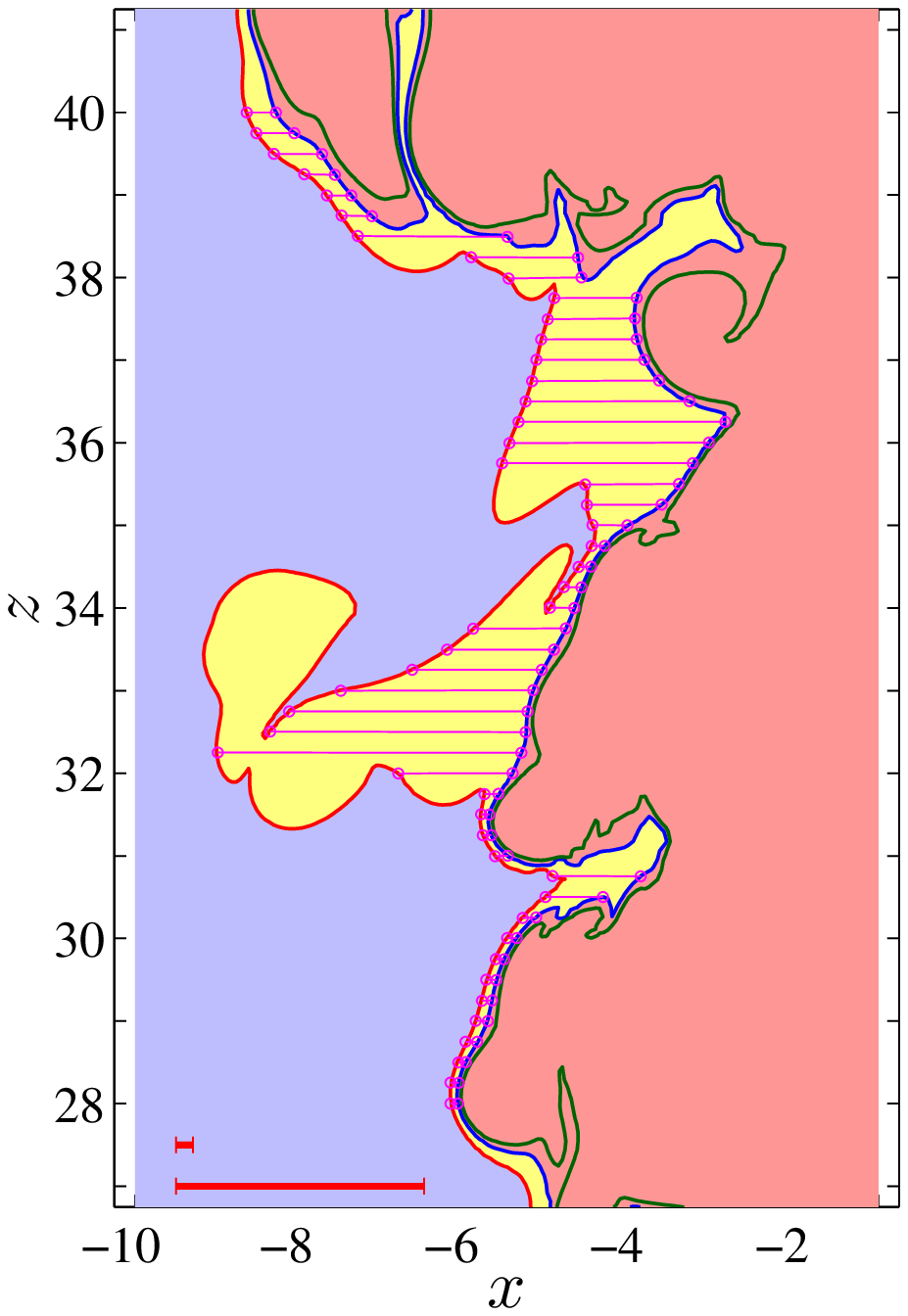}}
\put(41,0){\includegraphics[width = 3.75 cm]{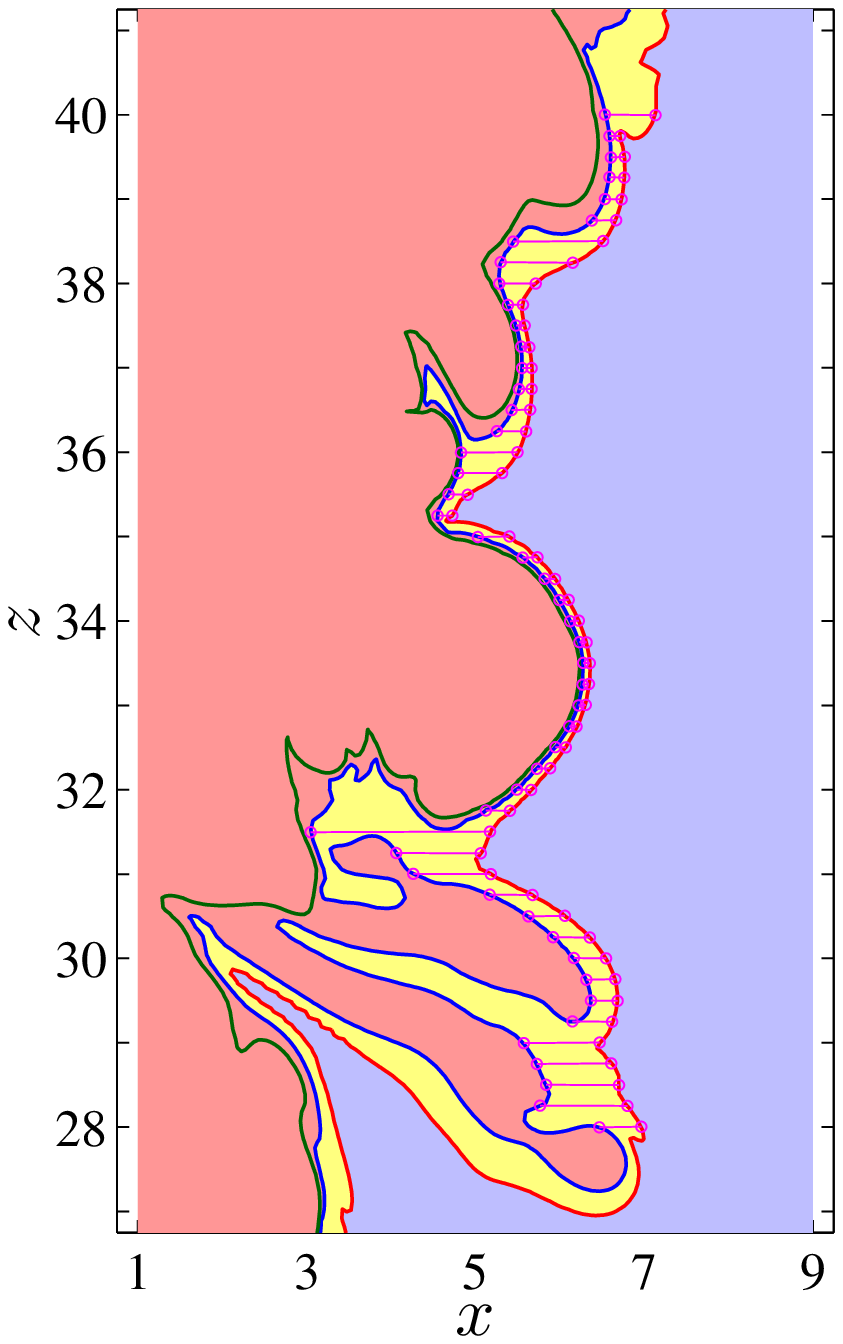}}
\put(80,0){\includegraphics[width = 8.40 cm]{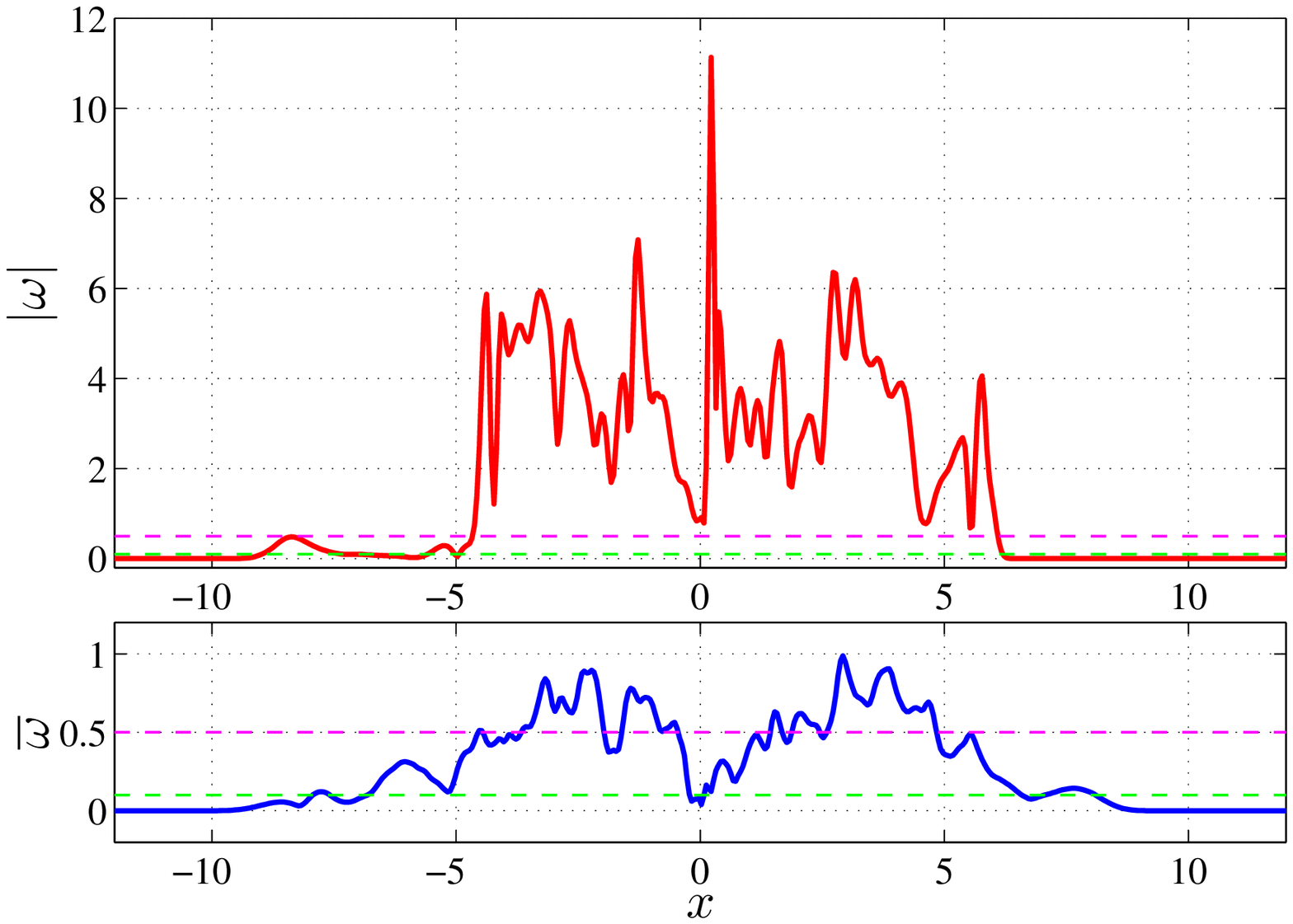}}
\put(1,61){(a)}
\put(81,61){(b)}
\end{overpic}
\caption{(a) Axial section in the $xz$ plane at $t = 925$. The
$|\bm{\omega}| = 0.1$ and $0.5$ contours in the axial cross section of
Fig.~\ref{fig:vel-vort} showing the wide variability in the separation
distance between the two contours that bound the yellow region; the
lighter reddish region is the turbulent core, the lighter lavender
region is the ambient. The red, blue and dark green boundaries are
marked respectively at thresholds $|\bm{\omega}| = 0.1, 0.5, 1$. The
thick red horizontal lines at the bottom in the left panel show $5\eta$ (short line) and $5\lambda$ (long line). (b) A typical
variation of the normalized instantaneous vorticity modulus
$|\bm{\omega}|$ (in red), compared with the time-averaged vorticity modulus,
$|\overline{\bm{\omega}}|$ (in blue). The bumps in $-10<x<-5$
correspond to the two intersections of the line $z = 34.05$ with the
outer viscous buffer zone. The broken horizontal lines are marked at
0.1 (green) and 0.5 (pink).}
\label{fig:vort-prof} 
\end{figure*}

The axial sections of the
jet flow used in the analysis below are within the self-similar region of the jet ($ 27
\lesssim z \lesssim 40$).  We begin with Figure
\ref{fig:vel-vort}, which shows an axial cross section of the jet
with contours of $|\bm{\omega}|$ in the turbulent core. The yellow circles circumscribing the
numbers indicate
the sub-regions (henceforth referred to as \textquoteleft{SR}'
followed by the number) where significant entrainment activity appears
to be taking place. Following the
inner boundary, we can discern in Fig.~\ref{fig:vel-vort} a
complete coherent structure with a base along a 2-6 line (i.e.,
joining SR2 and SR6), narrowing to a small nose around $z \approx 38$ penetrating the larger downstream
structure (of which we see only
the bottom third) at its base along the line 4-9. These have a strong resemblance to structures
observed in experiments (e.g., Mungal \& Hollingsworth \cite{Mungal_PoF_1989}) and other
numerical simulations (e.g., Basu \& Narasimha \cite{basu1999direct}). There is a similar
penetration by the upper third of the upstream structure along the base
2-6 of the central structure. The core of the central
structure $(30 \lesssim z \lesssim 38)$ slopes slightly rightwards whereas the upper structure has a tilted base 4-9
sloping up towards the left. On the whole, the core of 
the jet appears to be in a helical mode (of the kind observed by
Dimotakis \etal (1983) \cite{Dimotakis_PoF_1983}, Mungal and
Hollingsworth (1989) \cite{Mungal_PoF_1989}). It is known from studies
like the wavelet analysis of Narasimha \etal (2002) \cite{narasimha2002coherent} that the base of the
structure is likely to be a fluted vortex ring with ambient fluid
 below the base being drawn into the structure through the Biot-Savart
relation.  

Figures~\ref{fig:vort-prof} (a) and (b) respectively present axial sections near the edges of the jet and the distribution of $|\bm{\omega}|$ across a diameter of the jet. From Fig.~\ref{fig:vort-prof} it is seen that there is a relatively rapid change in
$|\bm{\omega}|$ values as the T/NT interface at the inner boundary
$|\bm{\omega}| = 0.5$ is crossed by a test point moving towards the
turbulent core. 
However, it is also seen that the separation between the contours of $|\bm{\omega}| = 0.5$ and $|\bm{\omega}| = 0.1$ can vary
appreciably at certain locations along the boundary of the jet. For example, in Fig.~\ref{fig:vort-prof}(a) the inner and outer
boundaries are close to
each other in some areas (e.g., $-8<x<-6$, $38<z<45$) and well
separated in others (e.g., $-8<x<-3$, $31<z<38$). In the latter case
we shall call the region between the inner and outer boundaries an
\textquoteleft{outer buffer zone}', to be distinguished from the
buffer zone defined by Reeuwijk and Holzner \cite{Reeuwijk_JFM_2014}
in a plane jet, which possibly is inwards of the present inner boundary, and
will here be called the \textquoteleft{inner buffer zone}'. The threshold values used for defining the boundaries by Reeuwijk \& Holzner (2014)~\cite{Reeuwijk_JFM_2014} are not applicable here as these thresholds are a function of Reynolds number and also of the dimensionality and nature of the flow. Figure
\ref{fig:vort-prof}(a) shows how the distance between the two
boundaries varies along the inner boundary of the jet.  A measure of this distance
is the separation along the $x$ axis from the inner to the outer boundary on
either side of the jet. It is seen that this inter-boundary separation
varies from $0.090$ to $3.67$ in the left boundary and from $0.077$ to
$2.13$ in the right boundary. 


For comparison, estimates of the two
relevant scales in the problem, namely the Kolmogorov length and the
Taylor microscale, are shown in Fig.~\ref{fig:vort-prof}(a) at $z\approx33$, $x,y=0$. It is seen that
inter-boundary separation varies from the order of the
Kolmogorov scale to the Taylor microscale; however, the changes in
$|\bm{\omega}|$ from 0.5 toward 1.0 and above in the turbulent core is
almost always sharp, and has a length scale comparable to the Kolmogorov scale.  Figure~\ref{fig:vort-prof}(b)
makes this point by displaying the variation of the total vorticity
modulus $|\bm{\omega}|$ at $z = 34.05$. For comparison the magnitude of the mean flow
vorticity is also shown in the figure. It is first of all seen that
the fluctuating vorticity $|\omega|$ in the core can be an order of
magnitude larger than the mean.  Furthermore, while the T/NT interface
at $x = 6.5$ drops steeply from about 4.0 to nearly 0, the interface
near $x = -5$ drops steeply all the way to about 0.5, then becomes
less steep, and finally takes the form of two noticeable but smooth
bumps, the one farther away being the bigger.  From comparison with
Fig.~\ref{fig:vel-vort} and Fig.~\ref{fig:vort-prof} (a) the bumps correlate with the outer buffer
zone with its small but apparently non-turbulent vorticity (as we shall shortly demonstrate). In
the light of the
many discussions on this issue in the literature \cite{Silva_PoF_2010, da2011role, da2008invariants, Westerweel_JFM_2009, hunt2006mechanics} we conclude that both scales seem to 
have a role to play, from the sharp T/NT interface layer of thickness
$\mathcal{O}(\eta)$ to the larger lateral extent of the viscous buffer
zone with a dimension of order $\lambda$. Understanding the role of these different scales and whether this difference in scales persists at higher Reynolds
requires further investigation.

\ul{Nature of the vorticity field from the T/NT to the R/IR interface}

Beyond the outer boundary at $|\bm{\omega}| = 0.1$, the flow is nearly
irrotational. It is seen (Fig.~\ref{fig:vel-vort}) that the core of
the jet has $|\bm{\omega}|$ values ranging from 1.0 to more than 17
over much of the cross-section, except in the very short section of
length $\Delta x \simeq 2$ at $z \simeq 25$ to 28. This could mark the
region of separation between two successive coherent structures in the
jet, as may be inferred from jet flow visualizations by the PLIF
technique \cite{Dimotakis_PoF_1983}, or by examining the computed
vorticity field of Basu \& Narasimha \cite{basu1999direct} in
a temporal round jet. Another such separation between two successive
structures is seen in Fig.~\ref{fig:vel-vort} around $z \simeq 38$ to 40. Here, the base of the structure is on the 4-9 axis, and clearly
carries a vortex ring, as we can infer from the velocity field in the
ambient. Taken together, this lends further support to the view that the flow exhibits a helical mode.

\begin{figure}[!h]
\begin{overpic}
[width = 16.0 cm, height = 8.7 cm, unit=1mm]
{Fig_box.eps}
\put(5,44){\includegraphics[width = 7.0 cm]{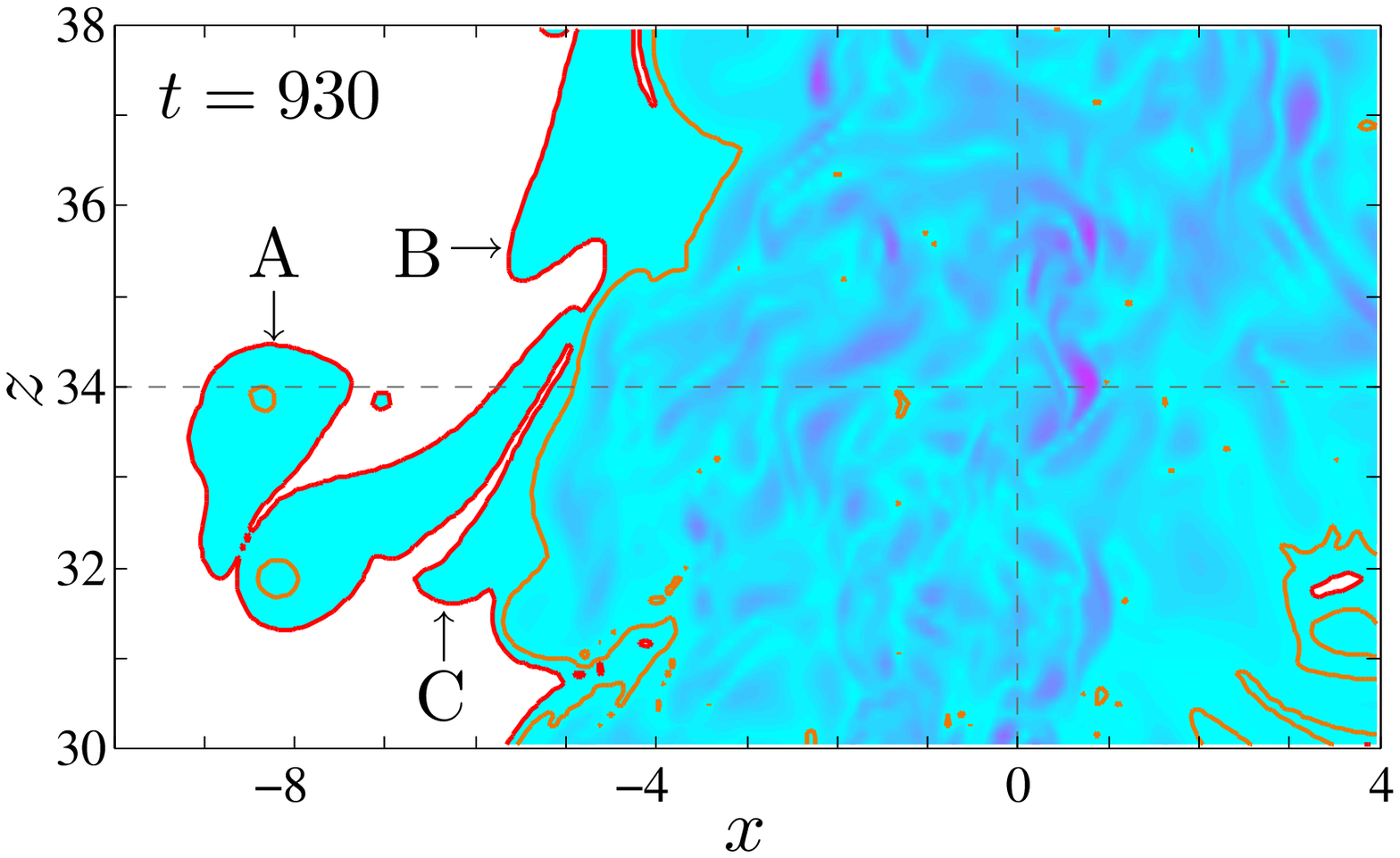}}
\put(76,44){\includegraphics[width = 7.0 cm]{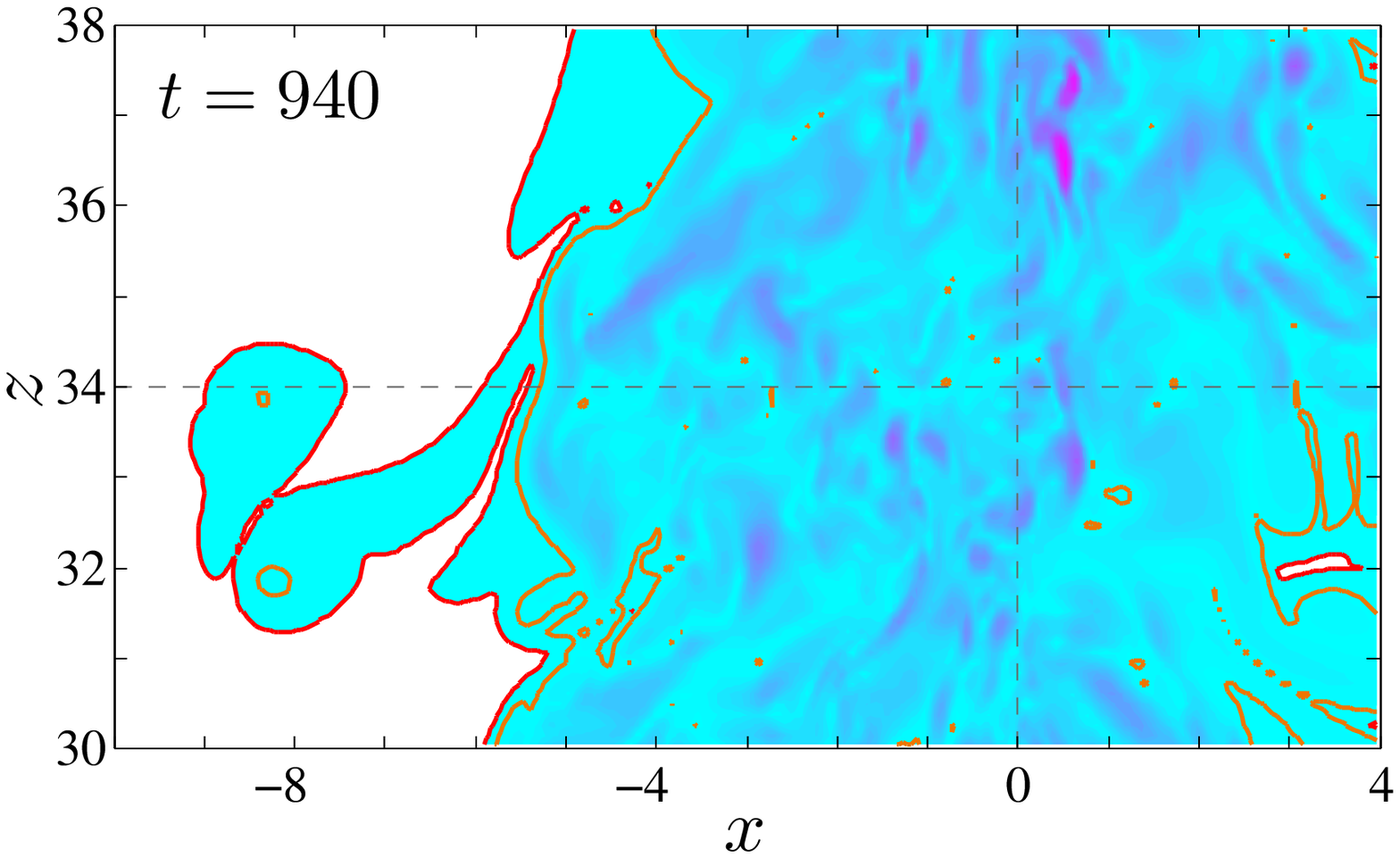}}
\put(5,0){\includegraphics[width = 7.0 cm]{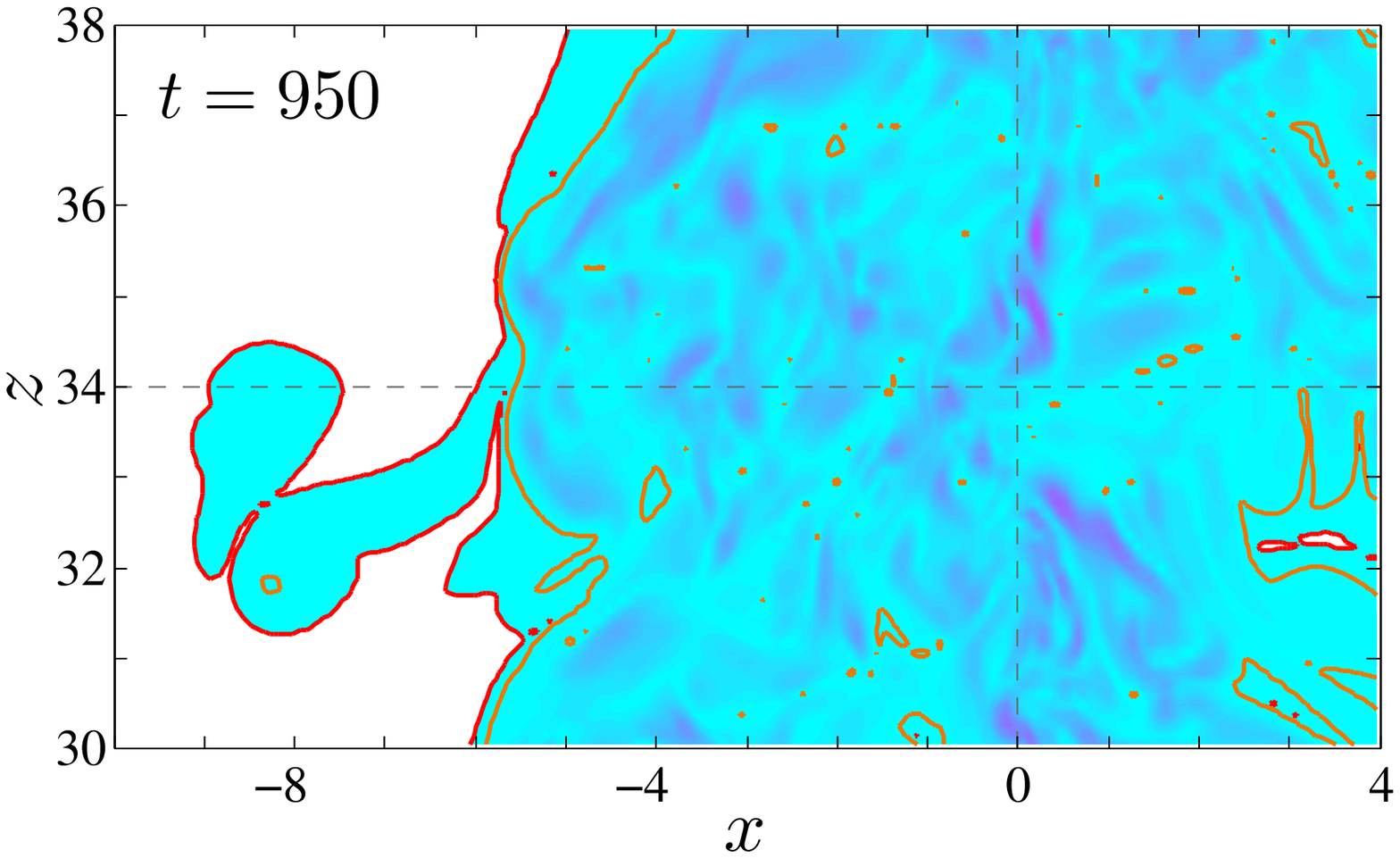}}
\put(76,0){\includegraphics[width = 7.0 cm]{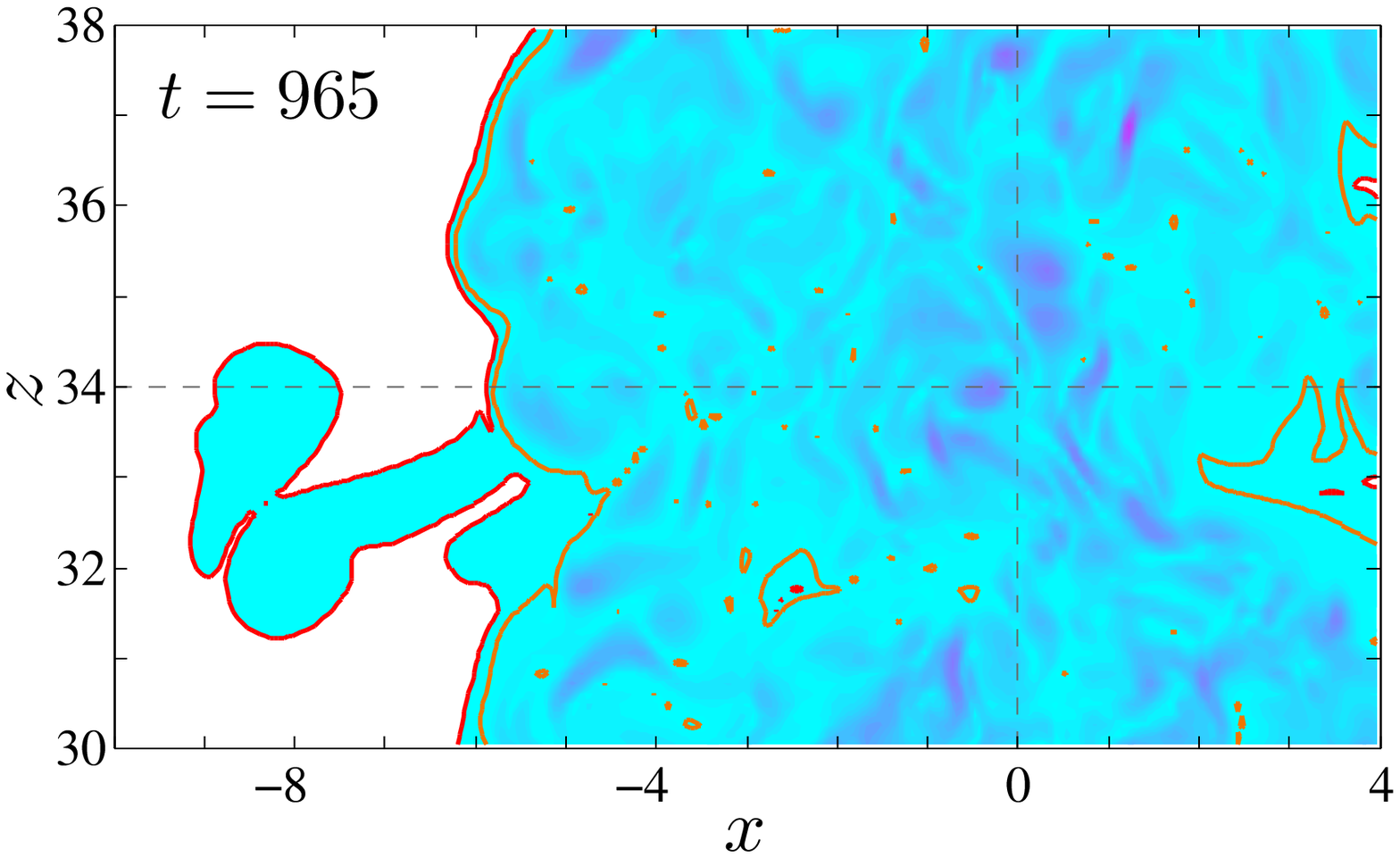}}

\put(148,17){\includegraphics[width = 1.0 cm]{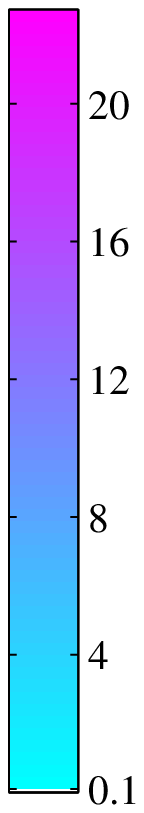}}
\put(12.5,51.5){\fontsize{10}{12}\selectfont {(a)}}
\put(83.5,51.5){\fontsize{10}{12}\selectfont {(b)}}
\put(12.5,7.5){\fontsize{10}{12}\selectfont {(c)}}
\put(83.5,7.5){\fontsize{10}{12}\selectfont {(d)}}
\end{overpic}
\caption{Temporal evolution of a part of the jet from
Fig.~\ref{fig:vel-vort}. Three areas in the figure labeled as A, B and
C illustrate important features of the evolution (see text for
discussion).}
\label{fig:vort-mag-axial}
\end{figure}

As expected, the edges of the jet are highly convoluted (especially
the inner boundary, which is a fractal curve (Sreenivasan \&
Meneveau (1986) \cite{Sreenivasan_JFM_1986} and Sreenivasan \etal 1989 \cite{sreenivasan1989mixing}). Figure
\ref{fig:vel-vort} includes the velocity field in the $xz$ plane
outside the inner edge of the jet.  The arrows in dark gray indicate a
radially inward component, and those in pink a radially outward
component.  A few long arrows within the core (with large heads) are
also shown, to give an idea of the relative values of the velocity
magnitudes within and outside the boundary of the jet. It is seen that in most of the selected
sub-regions the
velocity vectors crowd into a part of the jet boundary (dark gray
arrows), and represent entraining motions. In some cases (above SR1,
near SR4, between SR5 and 6) there are some pink arrows as well,
which could be detraining motions, but these are generally locally
diverging radially outward motions compared to the radially inward
converging and entraining motions. Whether the motions entrain or detrain also depends upon the
propagation velocity of the interface. However, Wolf \etal \cite{Wolf_PoF_2012} and Reeuwijk \& Holzner
\cite{Reeuwijk_JFM_2014} have shown that the interface velocity is of the order of the Kolmogorov velocity, so
generally negligible (at $z\approx33$, Kolmogorov velocity is $\approx0.01\overline{w}_0$). It will be seen that there is
little evidence for a parallel inward flow in the ambient impinging on
the jet boundary; instead, entrainment often seems to occur in
relatively intense events concentrated in a few narrower areas near
the edge (for example, below SR1, SR4, SR7 and SR9; this is further discussed in section~\ref{sec:Axial-Res}). In these areas
the flow velocities can go up to $0.4$ times the mean centre-line velocity $\overline{w}_c$ (e.g., in
SR7). Thus the fluid is pushing the turbulent edge inward and rushing
into the turbulent core across the T/NT interface through
\textquoteleft{gulfs}' or \textquoteleft{wells}' created by the inrush
of ambient fluid. (This is best seen in movies of the flow; see the
short sequence of stills in Fig.~\ref{fig:vort-mag-axial}.)
In all these cases the streamlines in the outer flow
exhibit, in part, one or two roughly circular patterns near the
entraining areas, suggesting that they represent a velocity field
induced by relatively strong vorticity in the near core (see analysis of SR9 in sec.~\ref{sec:Axial-Res}).
Occasionally, however (see e.g. the region between SR2 and SR3), the
flow appears to be radially organized, indicating a source-like flow,
which we shall further discuss below.

Figure \ref{fig:vort-mag-axial} shows the time evolution of a part of the
jet $(32<z<36)$ that is accurately self-preserving. Region A hardly changes in shape or in
vorticity range
over the whole time interval $930 \leqslant t \leqslant 965$ covered
by the images.  On the other hand, region B keeps continuously shrinking, and has
disappeared at $t = 965$.  This disappearance is largely due to the
outward motion of the inner boundary till, in (d), the two boundaries
are very close to each other in the interface above $z = 34$.  Region C
undergoes minor changes in shape, in part again because of the outward
movement of the inner boundary. Between (a) and (d) it is seen that
the turbulent core has expanded, with a general reduction in the
larger vorticity-sparse areas seen near the boundaries in the earlier
images.

\begin{figure}[!h]
\begin{overpic}
[width = 8.0 cm, height = 5.2 cm, unit=1mm]
{Fig_box.eps}
\put(5,0){\includegraphics[width = 6.5 cm]{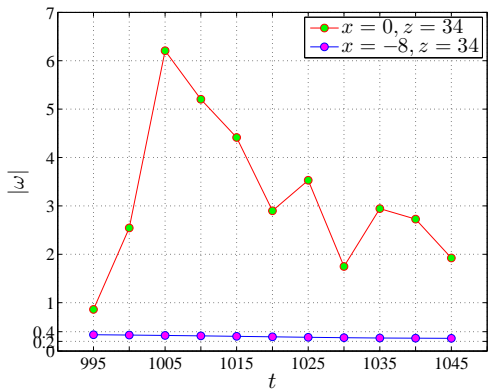}}
\end{overpic}
\caption{Temporal variation of the total vorticity modulus at two
points in the plane $z = 34$, one on the centerline $(0, 0, 34)$
and the other at $(-8, 0, 34)$ in the outer buffer region between
the inner and outer boundaries.}
\label{fig:vort-mag-prof}
\end{figure}

Figure \ref{fig:vort-mag-prof} compares the variation of
$|\bm{\omega}|$ with time at two points. The vorticity at the
center line shows substantial variation in time, whereas that in area A in the
outer buffer zone is more than an order of magnitude lower and exhibits a
slow and gentle decay with hardly any fluctuation that can be
associated with turbulence. This justifies why such a buffer zone can be
characterized as \textquoteleft{viscous}', and the structure at zone A in Fig.~\ref{fig:vort-mag-axial} may
be called a \textquoteleft{viscous tongue}', most probably a relic or
fossil from an earlier excursion of in- or
out-of-plane vorticity from the core or buffer zone into an ambient nearly at rest.

\begin{figure*}[h!]
\begin{overpic}
[width = 16.0 cm, height = 10.8 cm, unit=1mm]
{Fig_box.eps}
\put(0,49){\includegraphics[width = 4.32 cm]{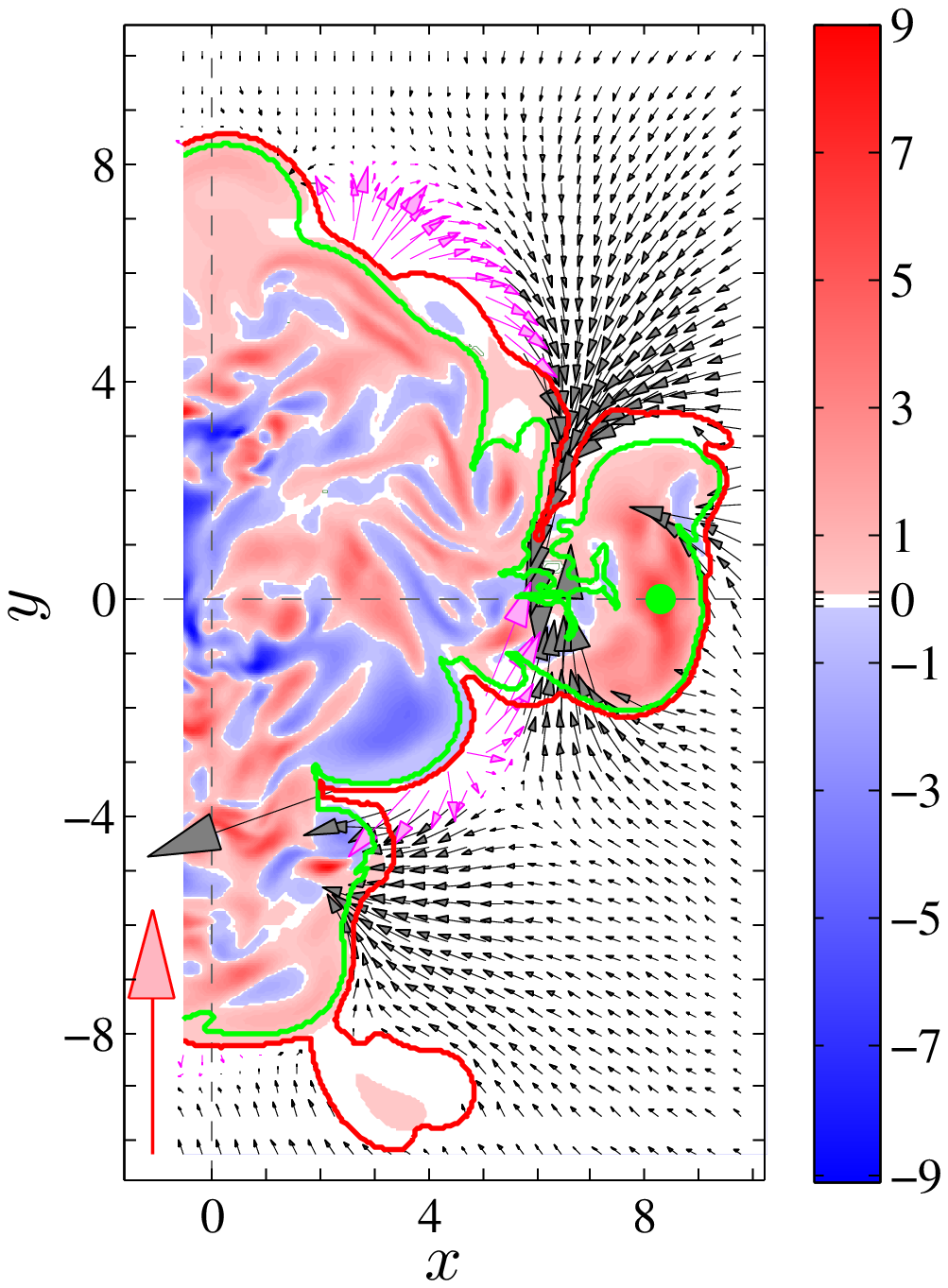}}
\put(45,49){\includegraphics[width = 6.45 cm]{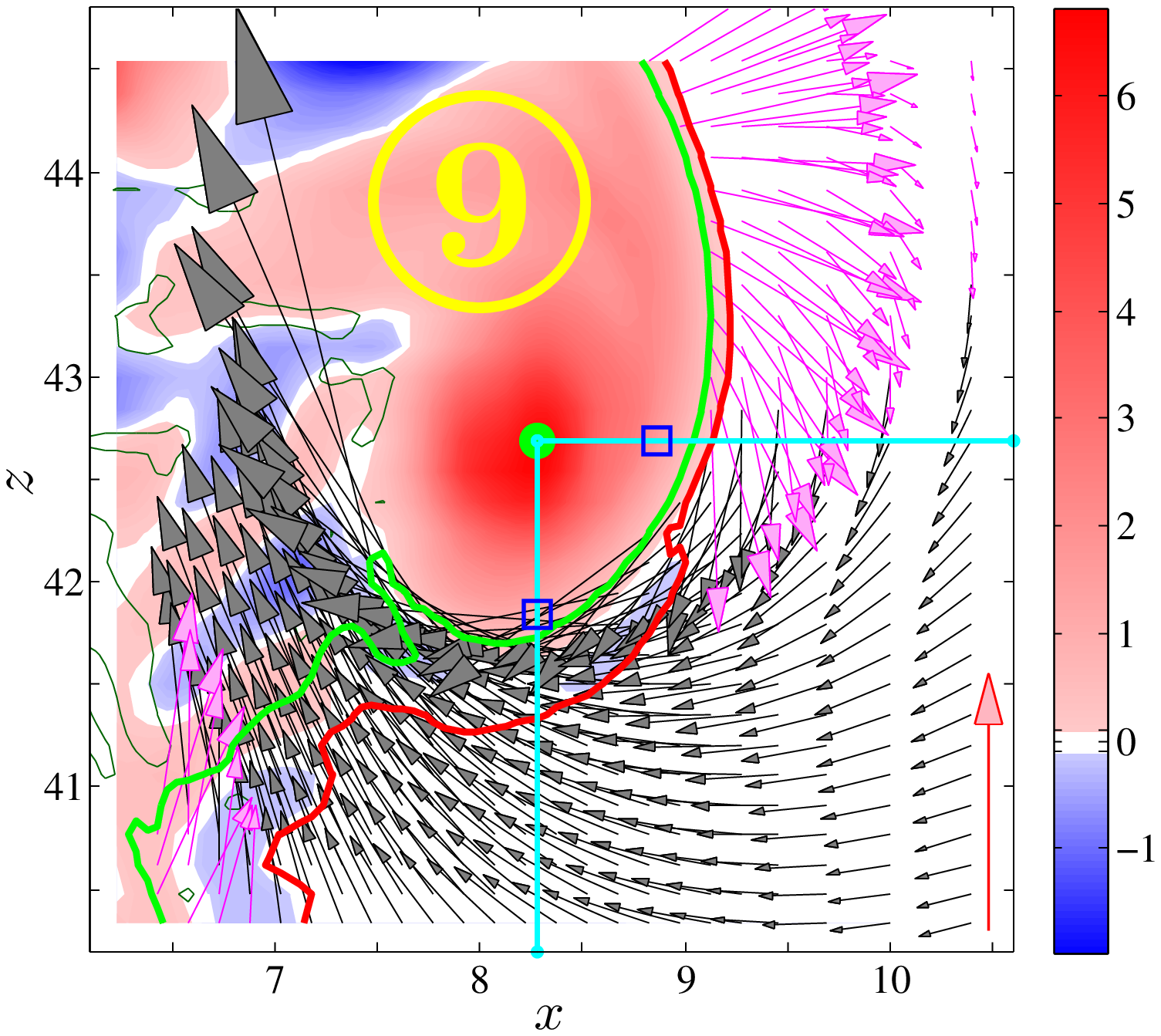}}
\put(111,49){\includegraphics[width = 5.01 cm]{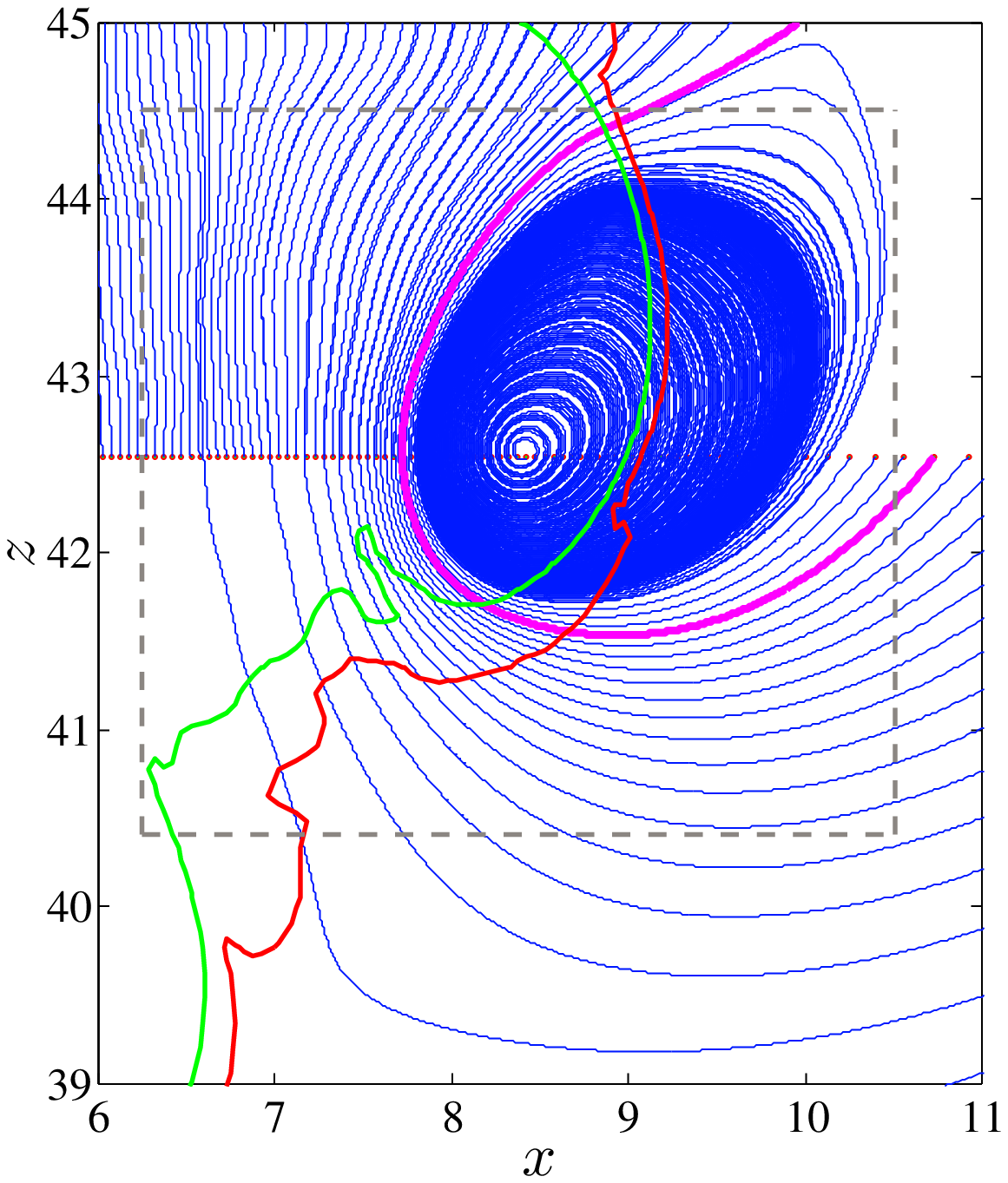}}
\put(19,0){\includegraphics[width = 6.0 cm]{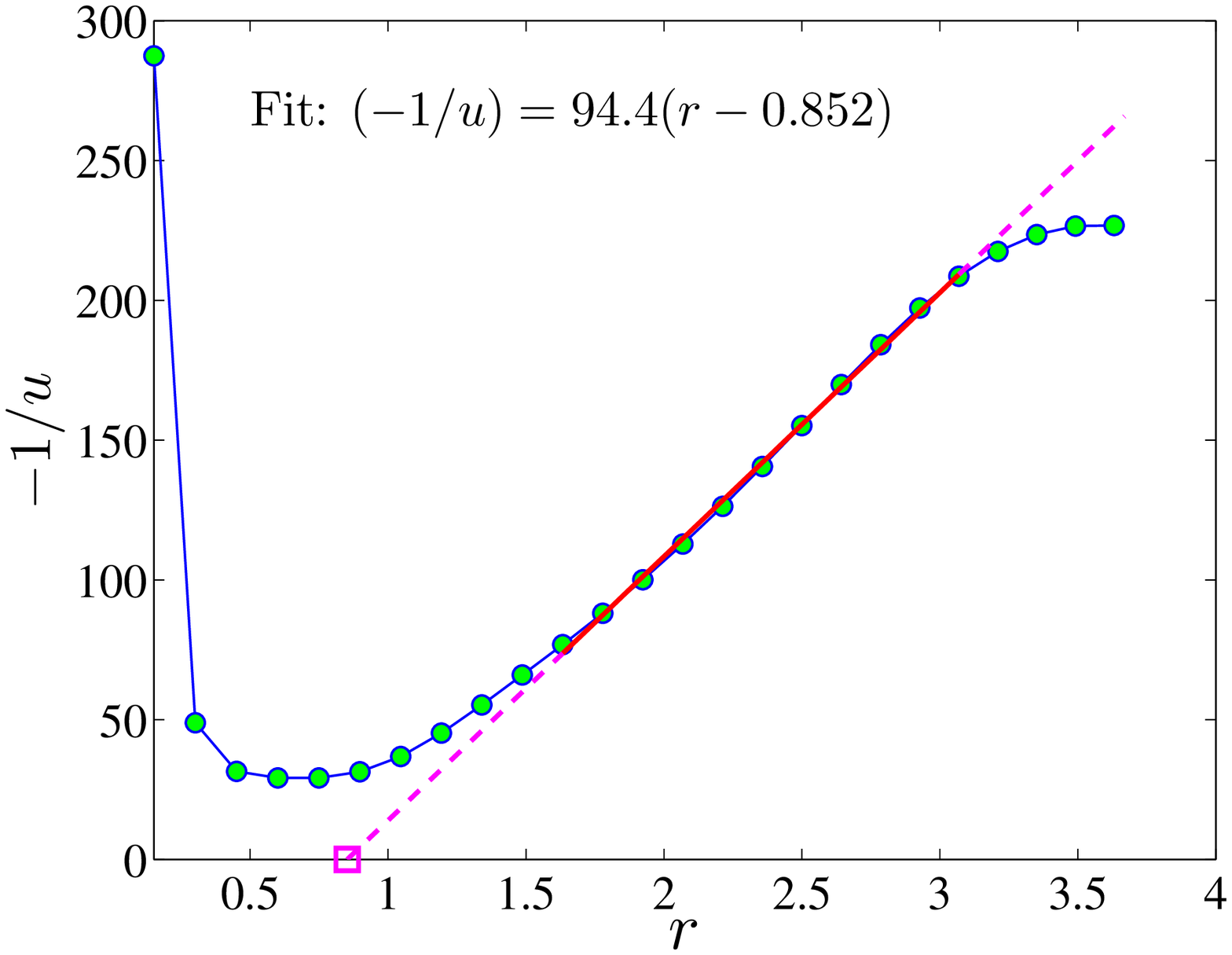}}
\put(81,0){\includegraphics[width = 6.0 cm]{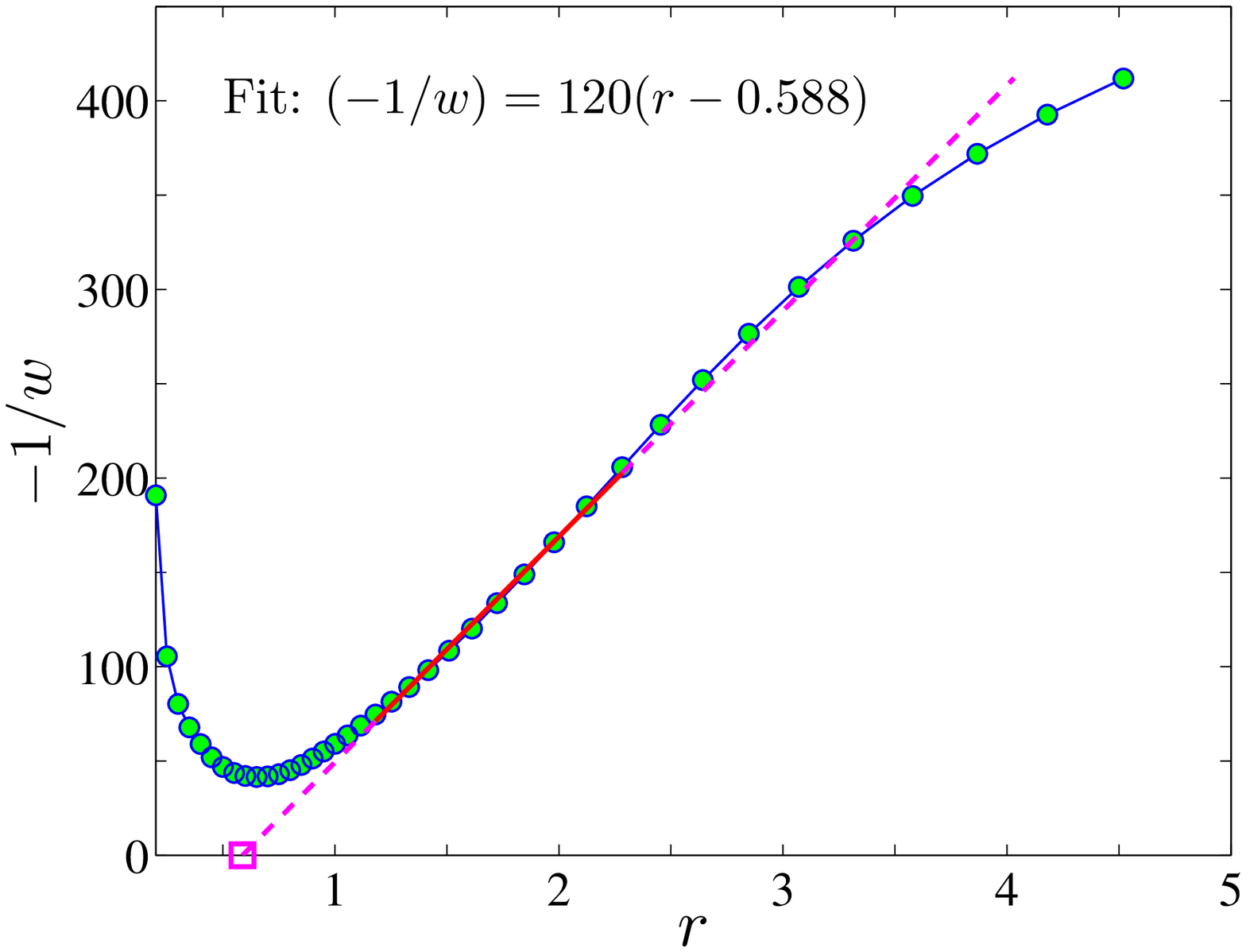}}
\put(0,103){(a)}
\put(44,103){(b)}
\put(108.5,103){(c)}
\put(14,43){(d)}
\end{overpic}
\caption{(a \& b) Instantaneous velocity in the ambient fluid and
azimuthal vorticity $(\omega_{\phi})$ field at $t = 925$.  Red and
blue colors indicate clockwise and counterclockwise vorticity
respectively.  The red and green lines show the outer and inner jet
boundaries respectively at thresholds 0.1 and 0.5 based on
$|\bm{\omega}|$. The green dot is nominal center of the red vorticity
patch (see main text). Two ticks on either side of zero on the color-bar correspond to
$+0.1$ and $-0.1$.  Black vectors show inward velocity, while the pink
vectors indicate outward flow velocities.  The reference velocity
vector shown in red color at the bottom-left corner of (a) is $0.05
w_0$ and at the bottom-right corner of (b) is $0.025 w_0$.  To avoid
clutter, only every seventh velocity vector is plotted in (a) and
every third in the $x$-direction in (b).  (a) A portion of the
diametral section at $z = 42.69$ showing a vortical object at the
right around $x = 8$.  (b) Zoomed-in view for SR9. (c) Instantaneous
streamlines for SR9 starting from $z = 42.5$. The pink curve marks the
streamline separating the fluid spiraling into the vortex from that
which drifts away from the vortex. The dashed grey rectangle marks the
zoomed portion shown in (b). (d) The inverse velocity relation with
distance along the horizontal (left panel) and vertical (right panel)
lines drawn in (b) showing a region of $1/r$ decay like that due to a
line vortex normal to the $xz$ plane at the location of the green dot (nominal centre).}
\label{fig:SR9}
\end{figure*}

\section{Structure of the axial sections at selected subregions}
\label{sec:Axial-Res}
To analyze the relation between the vorticity field in the turbulent core and the velocity field beyond the R/IR boundary, we consider selected subregions in axial and diametral planes. We start with the axial sections, namely SR 9, 1, 4
and 7.

\begin{figure*}[h!]
\begin{overpic}
[width = 16.5 cm, height = 14.0 cm, unit=1mm]
{Fig_box.eps}
\put(0,68){\includegraphics[width = 8.6 cm]{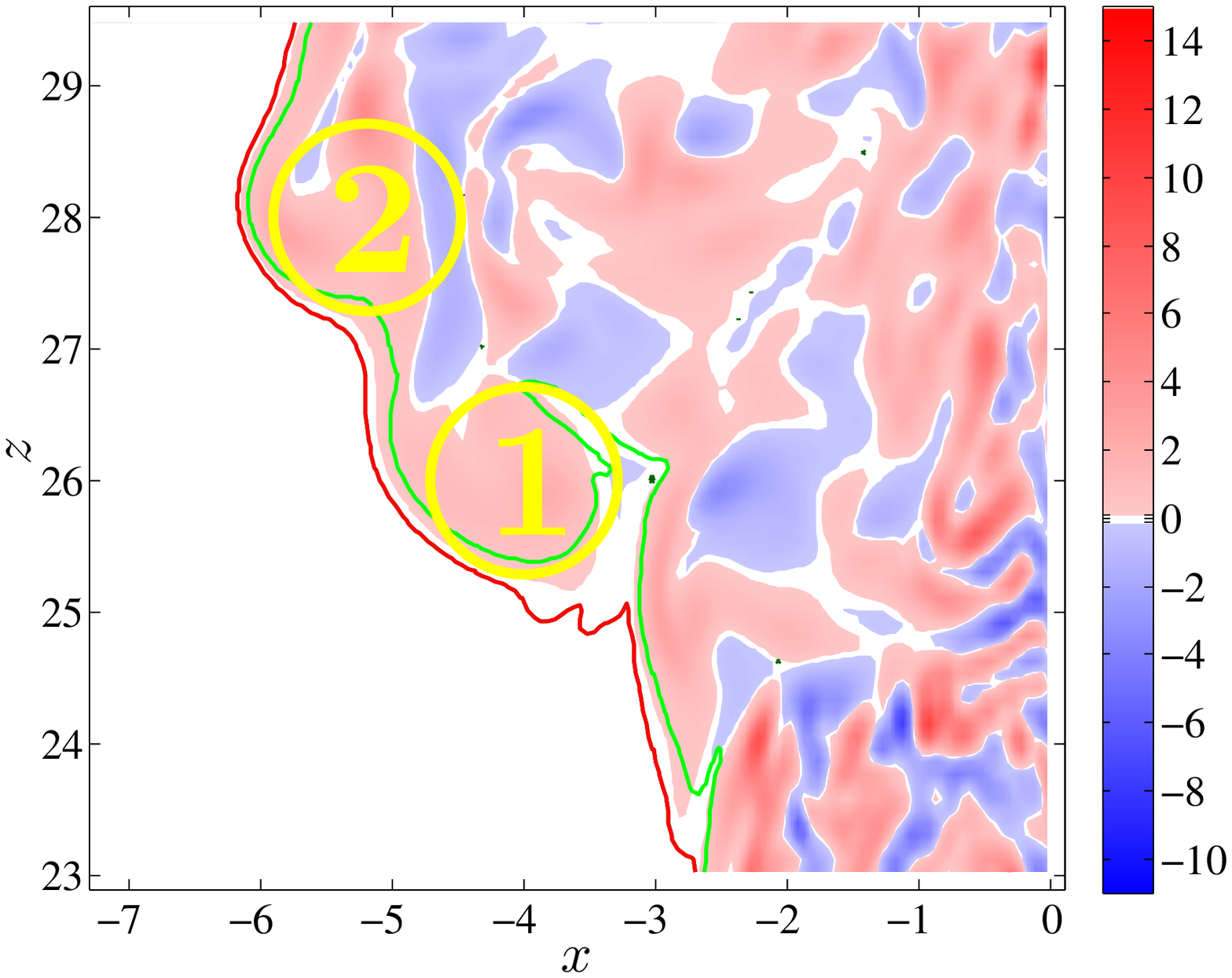}}
\put(84,68){\includegraphics[width = 7.8 cm]{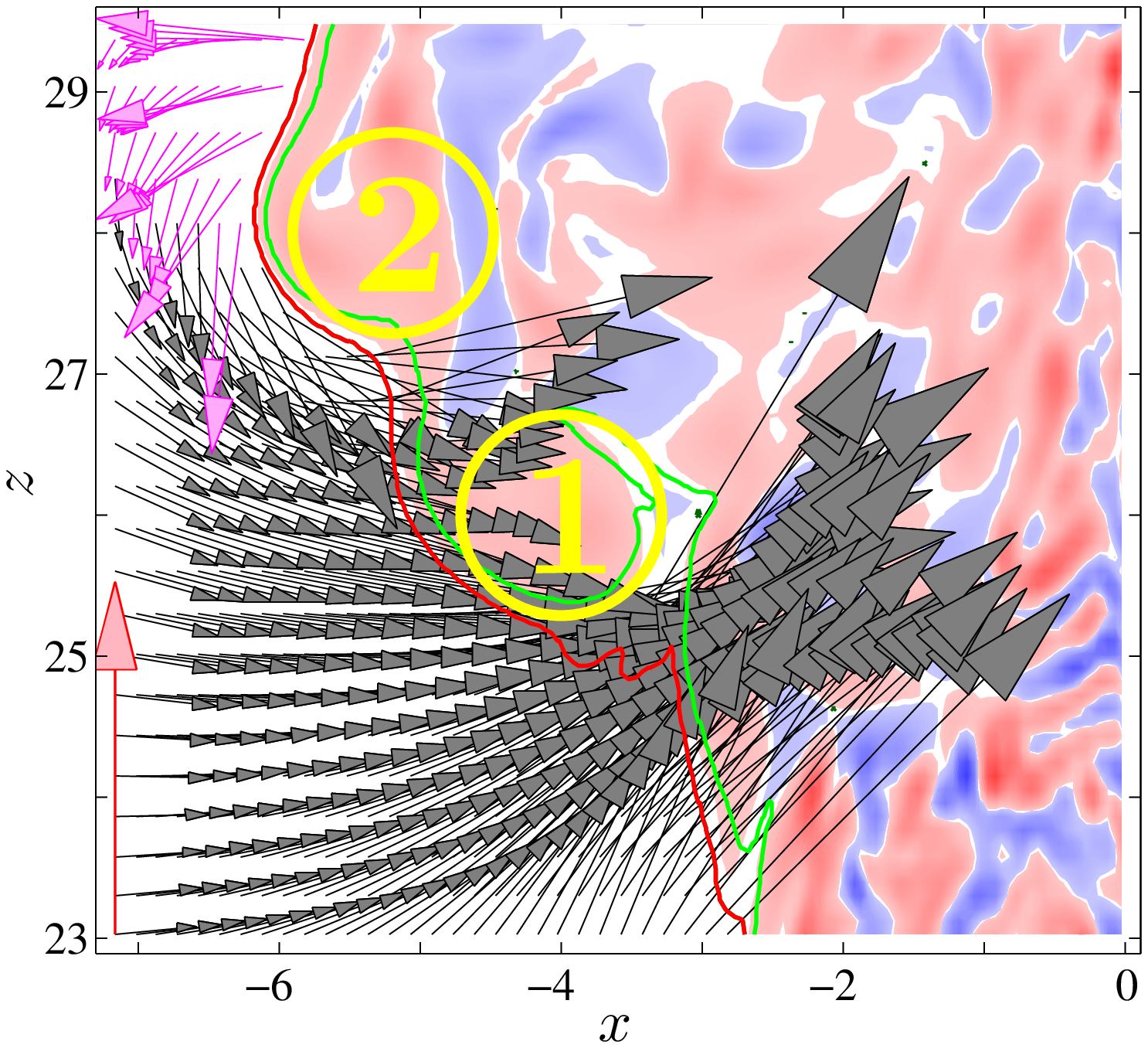}}
\put(0,0){\includegraphics[width = 7.8 cm]{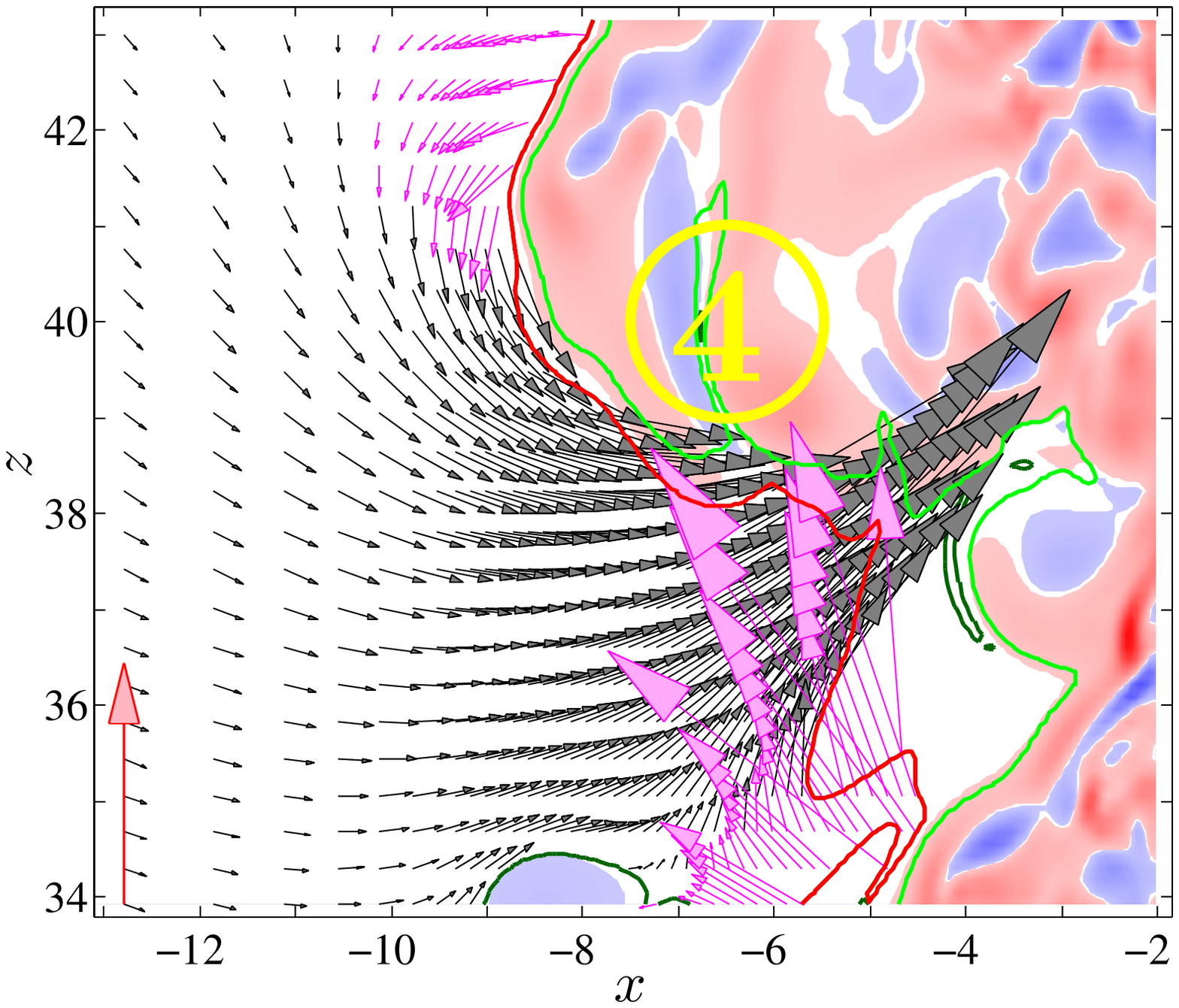}}
\put(80,0){\includegraphics[width = 8.4 cm]{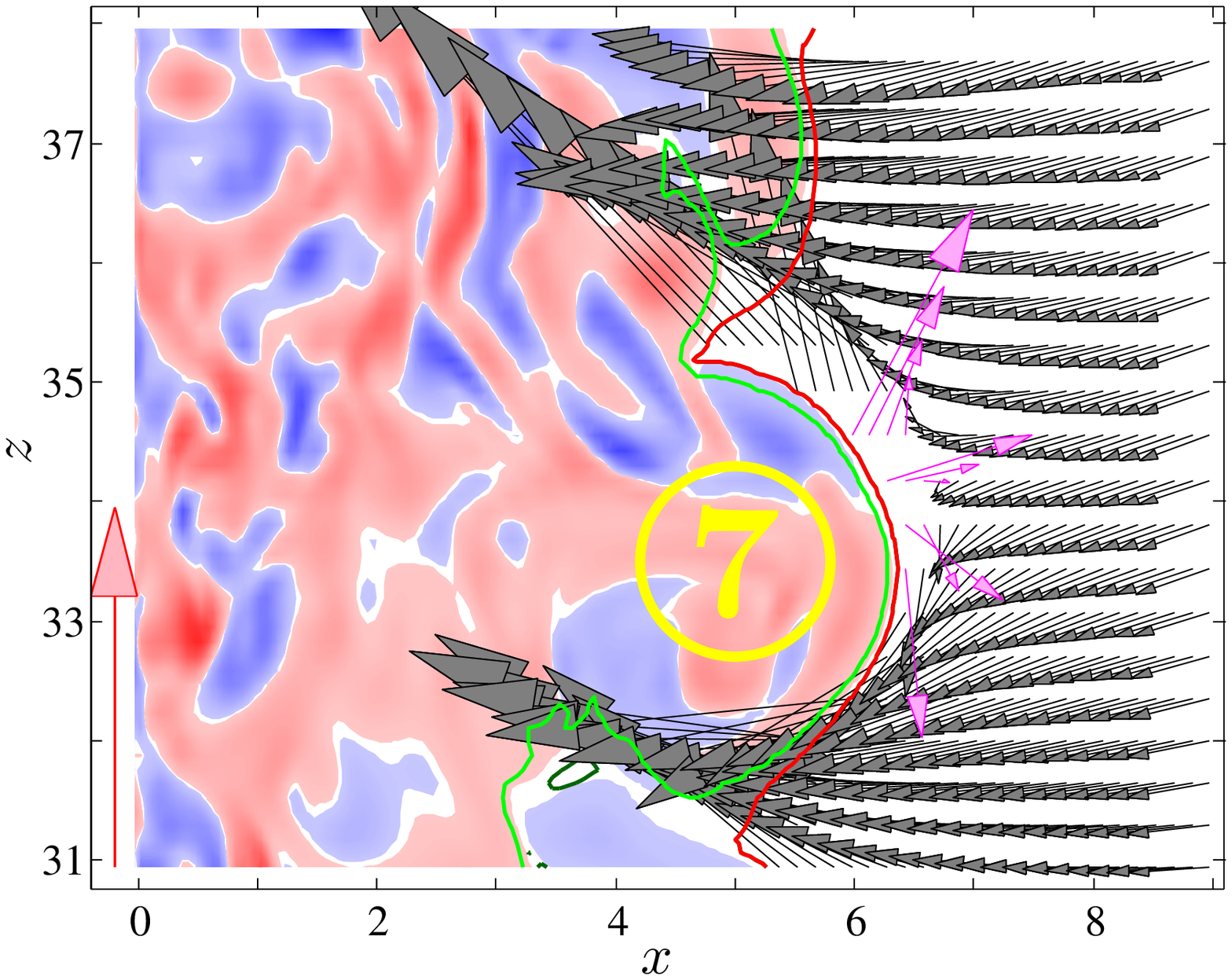}}
\put(0,136){(a)}
\put(80,136){(b)}
\put(0,63){(c)}
\put(80,63){(d)}
\end{overpic}
\caption{Ambient velocity and core azimuthal vorticity
$(\omega_{\phi})$ fields, showing the inrush of flow into the core of
the jet from the ambient fluid. All other conventions are the same as
in Fig.~\ref{fig:SR9}(b). (a) SR1, SR2 $(\omega_{\phi})$ field (b) SR1, SR2, with inrush velocity field in the ambient with the $(\omega_{\phi})$ field;
the tick marks shown above and below zero on the color-bar indicate
respectively $\omega_{\phi} = +0.1$ and $-0.1$ (c) SR4, (d) SR7. Magnitude of the reference vector shown in red at the bottom-left
corner of (a, c, d) is $0.025 w_0$. Color-bar shown in (b) also
applies to the vorticity contours shown in (a, b, c, d).}
\label{fig:velvort-SR-1467}
\end{figure*}

\noindent
\textbf{SR9:} This offers the simplest flow situation among those seen
in Fig.~\ref{fig:vel-vort} for the following reasons illustrated in
Fig.~\ref{fig:SR9}. (Note that the back-ground vorticity field
displayed in Figs.~\ref{fig:SR9} - \ref{fig:velvort-diam} is the
signed azimuthal component $\omega_{\phi}$, which is helpful in
correlating the velocity field in the ambient with the vorticity field in the
core.) The dominant feature of the flow within the jet core in SR9 in
Fig.~\ref{fig:vel-vort} is a large lump of clockwise (negative, blue)
vorticity. A diametral section of the jet around SR 9 at $z = 42.69$ is shown in
Fig.~\ref{fig:SR9}(a). By comparing with Fig.~\ref{fig:vel-vort}, it is seen that this
diametral section of the core flow in SR 9 contains an ovoid around $x\approx8,~-1<y\le1.5$, with $x,y,z$ dimensions of
about $2 \times 5 \times 2$. Thus the variation of the vorticity with $y$ in the central
region of the ovoid is not very strong. From Fig. \ref{fig:SR9}(b), which is a zoom on
SR 9 in Fig. \ref{fig:vel-vort}, it is seen that the highest azimuthal vorticity ($|\omega_\phi|=6.8$) is confined to
a nearly circular region of diameter less than 1.0. The maximum in the other components of vorticity are about a fifth of that of $\omega_\phi$. Although the structure is 3D, the azimuthal component dominates over the other components of vorticity, so to a first order approximation we can treat it like a 2D flow. The image clearly indicates that there is a
circulating motion around the vortex in a direction that is consistent
with the sign of the vortex. Very close to the region where vorticity is
intense and of one sign over the region (as often happens in coherent
structures) there appears to be relatively high correlation between
the vorticity and velocity fields. Figure \ref{fig:SR9}(b) thus depicts a
strong circulatory motion, first hugging the outer boundary and then
penetrating the inner boundary.

Figure \ref{fig:SR9}(c) shows instantaneous effectively $2D$ streamlines in the $xz$ plane for SR9 starting from
points located along $z = 42.5$, corresponding to the
flow field in Fig.~\ref{fig:SR9}(b). Note that all these streamlines
start from neighboring grid points (closer to each other in the core
than in the ambient, see Sec. \ref{sec:simu}), and therefore do not
represent equal differences in the stream function.  The red and black
contours are the same as in Fig.~\ref{fig:SR9}(b). A
\textquoteleft{dividing}' streamline, separating fluid which spirals
back into the vortex from fluid that moves away, is shown in pink.
Such an interpretation is encouraged by the weak variation of vorticity in the
$y$-direction, making the flow close to a 2D field as already noted.

From Fig.~\ref{fig:SR9}(c), with vorticity
in the object peaking at $x = 8.28$, $y = 0$, $z = 42.69$, the
domination of the azimuthal vorticity in determining the velocity
field in the neighborhood is understandable. This is further supported
by Fig.~\ref{fig:SR9}(d), which shows a plot of the reciprocal of the
magnitude of the $w$ and $u$ components of the velocity respectively
in the $xz$ and $yz$ planes, as a function of radial distance $r$ from
an approximate origin along the lines $z = 42.69$ and $x = 8.28$
respectively (shown by cyan colored lines in Fig.~\ref{fig:SR9}b).  A linear
fit is a good match to the data over the range of $r$ between 1.6 and
3.2, as marked by dashed red lines; the equations for the fits are
shown on the respective plots. This supports the conclusion that the outer velocity field approximates to that induced by a line vortex along the $y$-axis, by the Biot-Savart law. By extending the fitted line to the
zero of the inverse of the velocity, we can identify an ``effective
center'' for the vortex (marked by pink squares in
Fig.~\ref{fig:SR9}d). The fact that these effective centers are not
identical but not very far from each other (separation about 0.3) is consistent with the
finite size and the imperfect symmetry of the vortical structure seen
in Fig.~\ref{fig:SR9}(a).

\noindent
\textbf{SR1 \& 2:} Figure~\ref{fig:velvort-SR-1467}(b) (the azimuthal vorticity field alone is shown in Fig.~\ref{fig:velvort-SR-1467}(a)) illustrates how the ambient velocity field maybe understood in terms of 
two anticlockwise vortices next to each other near the
inner boundary.  In SR2 the vorticity is anticlockwise near the edge
but clockwise away from it further inward.  The net effect of this
somewhat more complex vorticity field is that in the region $28<z<29$,
$-7<x<-6$, the flow in the ambient is away from the jet (pink arrows),
with only a mild contribution to the inward circulatory motion. In the
zone $26.5<z<27.5$, however, the motion is more strongly circular in
the lower parts of SR2.

\noindent
\textbf{SR4:}
Here (Fig.~\ref{fig:velvort-SR-1467}(c)) the vortices in the upper reaches $38<z<42$
resemble the region near SR2 in Fig.~\ref{fig:velvort-SR-1467}(a). Because of the weaker
clockwise component in the vertical strip $38.2<z<41.8$,
$-7.5<x<-6.5$, the flow in $37<z<40$ is dominated by counterclockwise
vorticity.

\noindent
\textbf{SR7:} Fig.~\ref{fig:velvort-SR-1467}(d) is the most complex image among the ones selected
here for analysis.  There is first
of all the clockwise motion at the lower levels of SR7 ($31.5<z<32.2$,
$3.5<x<6.5$). Similarly there is a counter-clockwise circulation in the
upper levels of SR7 ($34.3<z<36$, $3.5<x<6.6$).  It is clear that the
former is strongly influenced by the large region of clockwise
vorticity in the lower half of SR7, and the latter is attributable to
the clockwise circulation due to the vorticity of the same sign in the
block $34<z <36$, $4.5<x<6$. The large values (up to $0.4\overline{w}_c$) of the inrush velocities here may
compensate for the
small area at the entrance to the gulf ($\sim d_o$), and hence can make a non-negligible contribution
to the entrainment rate.

The above four examples show some of the different types of organized
motion that occur due to interactions between vortex elements that
are part of the coherent structures in the core fluid, and the
Biot-Savart induced velocity due to these elements at the T/NT interface. The most striking
feature of the flow in the ambient is that it is organized largely in
terms of interacting circulatory motions.  Roughly speaking the
diameter of the region containing this ordered outer motion is about 20
$d_0$ at $z \simeq 40$ to 44, and about 12$d_0$ at $z \sim 28$. This is
approximately twice the local diameter of the turbulent core. These
estimates of the spatial extent of the velocity and vorticity
structures involved in the interaction are a strong indication that
they are related to \textit{elements} of the large-scale vorticity in coherent
structures in the core flow, and cannot be attributed to small scale vorticity.

This analysis may be compared
with two others in the literature. Phillip and Marusic (2012) \cite{Philip_PoF_2012} show, in a
\textquoteleft{caricature}' of a turbulent jet and the entrainment
(their Fig.~6), a flow of the ambient fluid towards the turbulent core in
parallel streamlines almost till they impinge on the edge of the
jet, in contrast to the ordered motion we see here in Figure
\ref{fig:vel-vort}. Westerweel \etal (2009)
\cite{Westerweel_JFM_2009} showed an instantaneous velocity field
(Fig.~19 in their paper) that appears (in one part of the domain covered in the
diagram) to transport turbulent fluid out of the turbulent core across
the interface into the ambient. However, in a frame of reference
fixed with respect to the vortex within the core, the velocity vectors
were almost parallel to the interface. Thus, they concluded that organized motions do not transfer
irrotational fluid into the turbulent core. In the immediate neighborhood
of the vortex, however, the relative velocity induced in the ambient fluid appears to be crossing into the core.
It is
thus seen that the picture presented by the present DNS analysis is
quite different from conclusions drawn by experimental results which may be affected by the limitations of PIV measurements in the low-velocity irrotational region.

\section{Analysis of diametral sections}
\label{sec:Res-diametral-section}

\begin{figure*}[tbp]
\begin{overpic}
[width = 16.50 cm, height = 17.40 cm, unit=1mm]
{Fig_box.eps}
\put(0,94){\includegraphics[width = 8.15 cm]{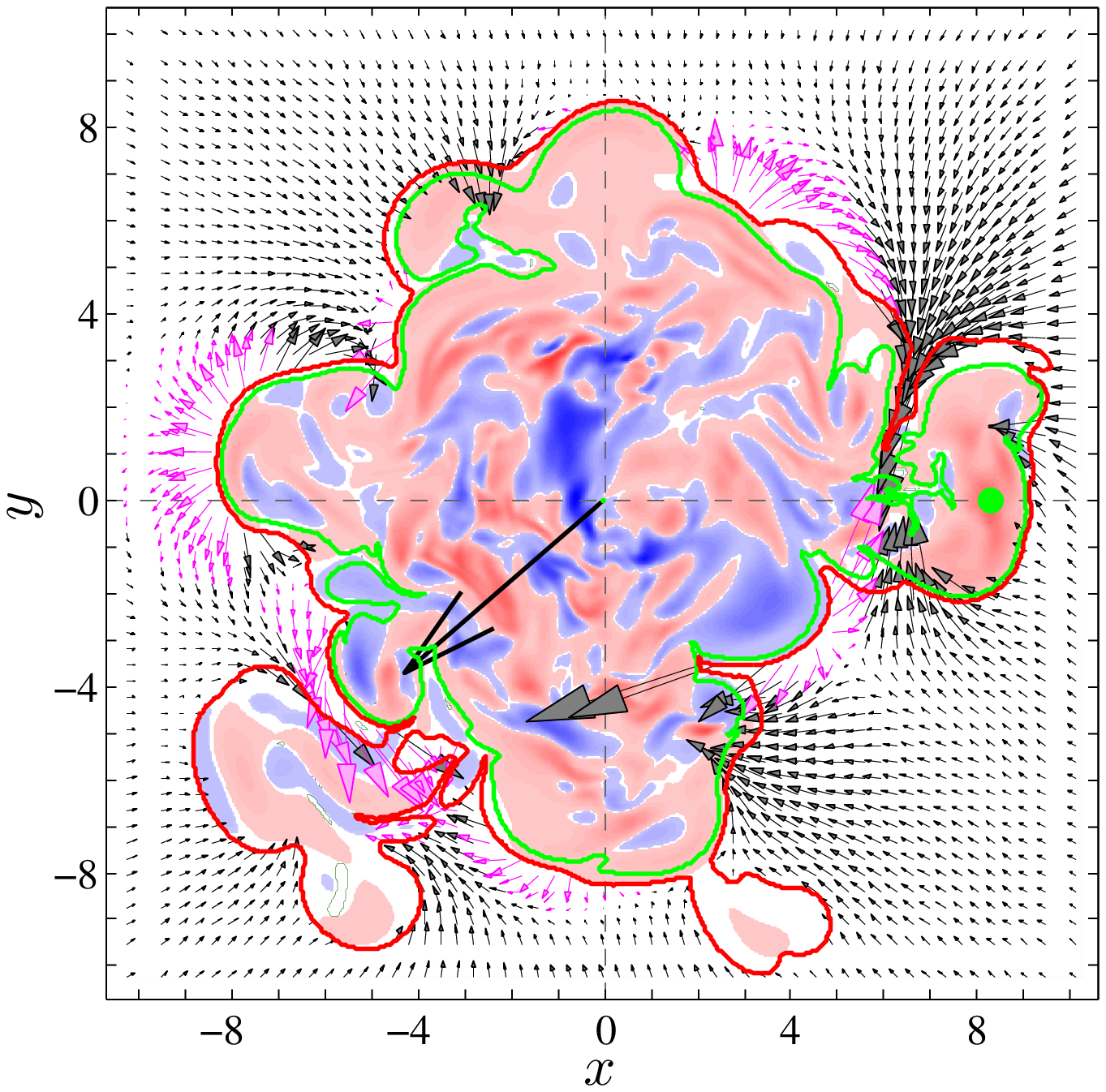}}
\put(82.5,94){\includegraphics[width = 8.15 cm]{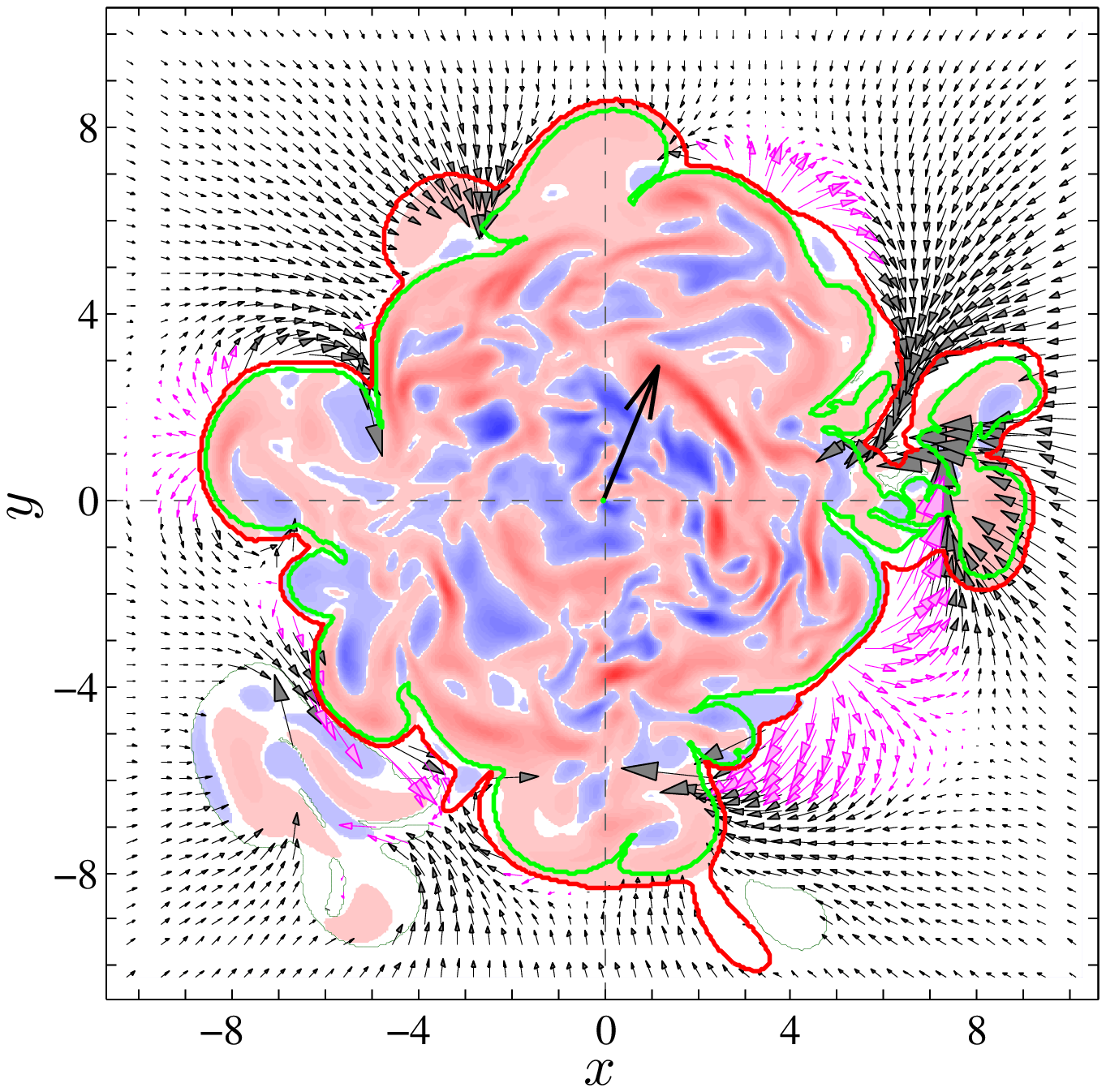}}
\put(0,13){\includegraphics[width = 8.15 cm]{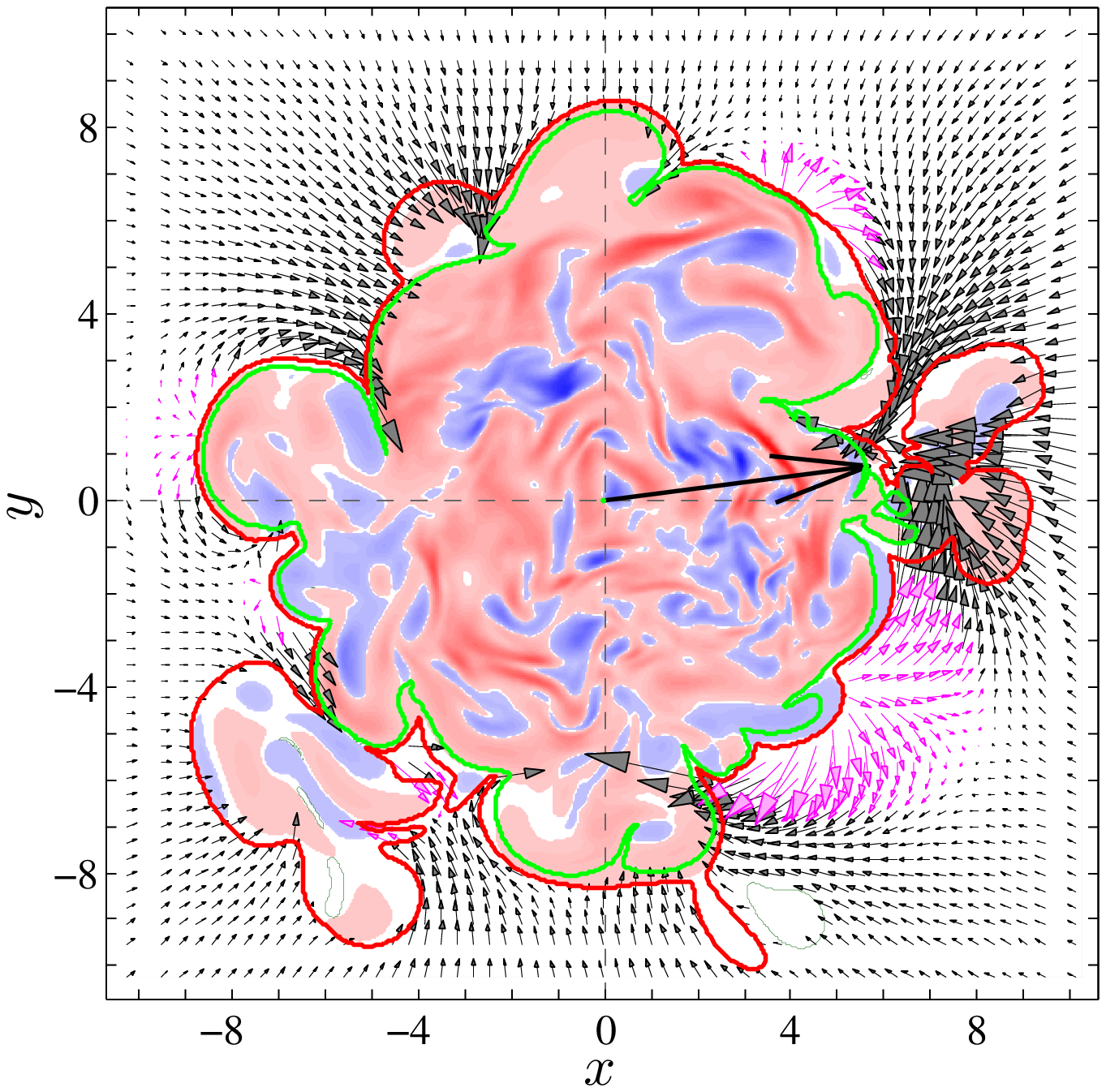}}
\put(82.5,13){\includegraphics[width = 8.15 cm]{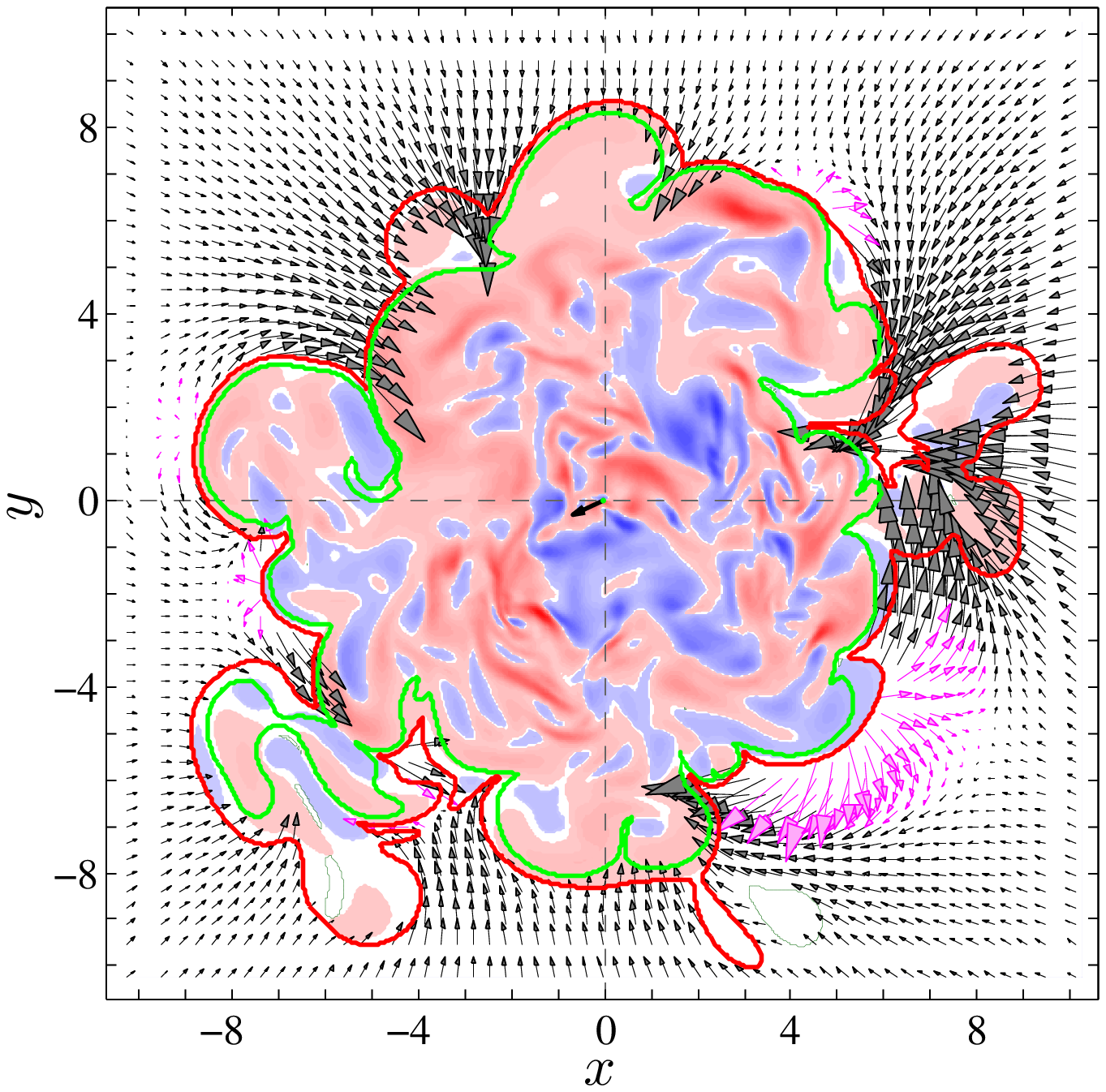}}
%
\put(26,0){\includegraphics[width = 1.85 cm]{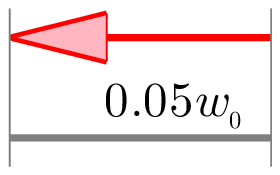}}
\put(74,0){\includegraphics[width = 8.15 cm]{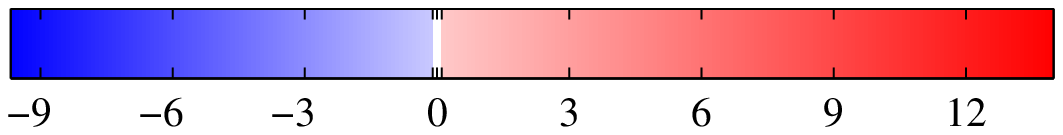}}
\put(10,6){\fontsize{10}{12}\color{gray}\selectfont \rotatebox{0}{Reference}}
\put(10,2){\fontsize{10}{12}\color{gray}\selectfont \rotatebox{0}{vector}}
\put(156,5){\fontsize{18}{20}\color{black}\selectfont \rotatebox{0}{$\omega_{\phi}$}}
\put(0,170){(a)}
\put(83,170){(b)}
\put(0,89){(c)}
\put(83,89){(d)}
\end{overpic}
\caption{Four instantaneous diametral sections of images at $z =
42.69$ at times $t = $ (a) 925, (b) 950, (c) 960, (d) 970. The
vorticity field in these images is the signed azimuthal component
$\omega_{\phi}$. Thick black vector indicates velocity inside the
jet at $~(0,0)$. The two ticks, one on each side of zero on the
color-bar, correspond to $+0.1$ and $-0.1$. To avoid clutter only every
seventh velocity vector is plotted.}
\label{fig:velvort-diam}
\end{figure*}

Figure \ref{fig:velvort-diam} presents a sequence of 
images of the diametral section at $z = 42.69$ at four different instants ($t = 925, 950, 960$ and
$970$). As before the
ambient velocity field is presented against a background of the
vorticity field $\omega_{\phi}$ within the turbulent core.

\begin{figure*}[tbp]
\begin{overpic}
[width = 16.50 cm, height = 17.40 cm, unit=1mm]
{Fig_box.eps}
\put(0,94){\includegraphics[width = 8.15 cm]{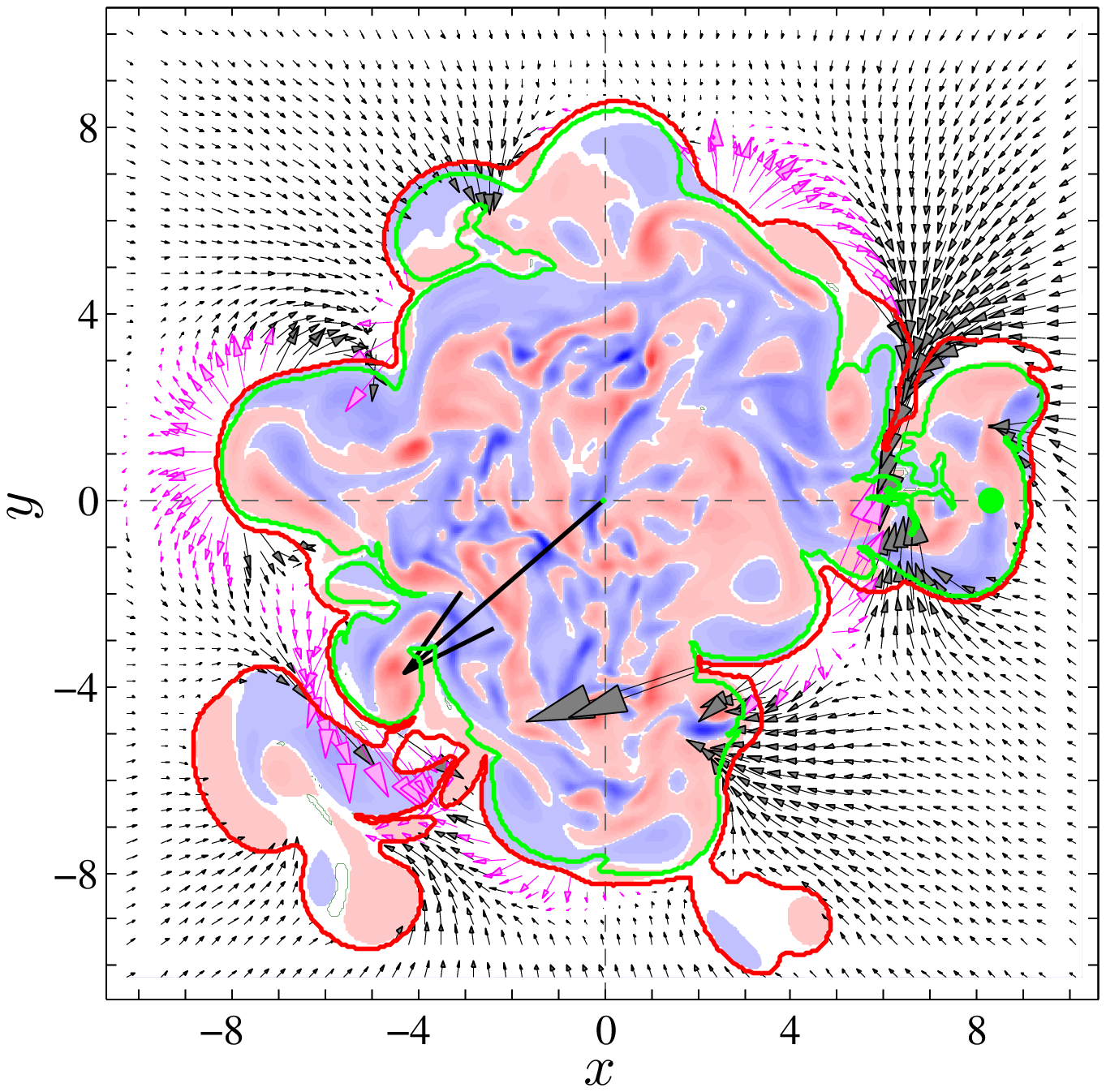}}
\put(82.5,94){\includegraphics[width = 8.15 cm]{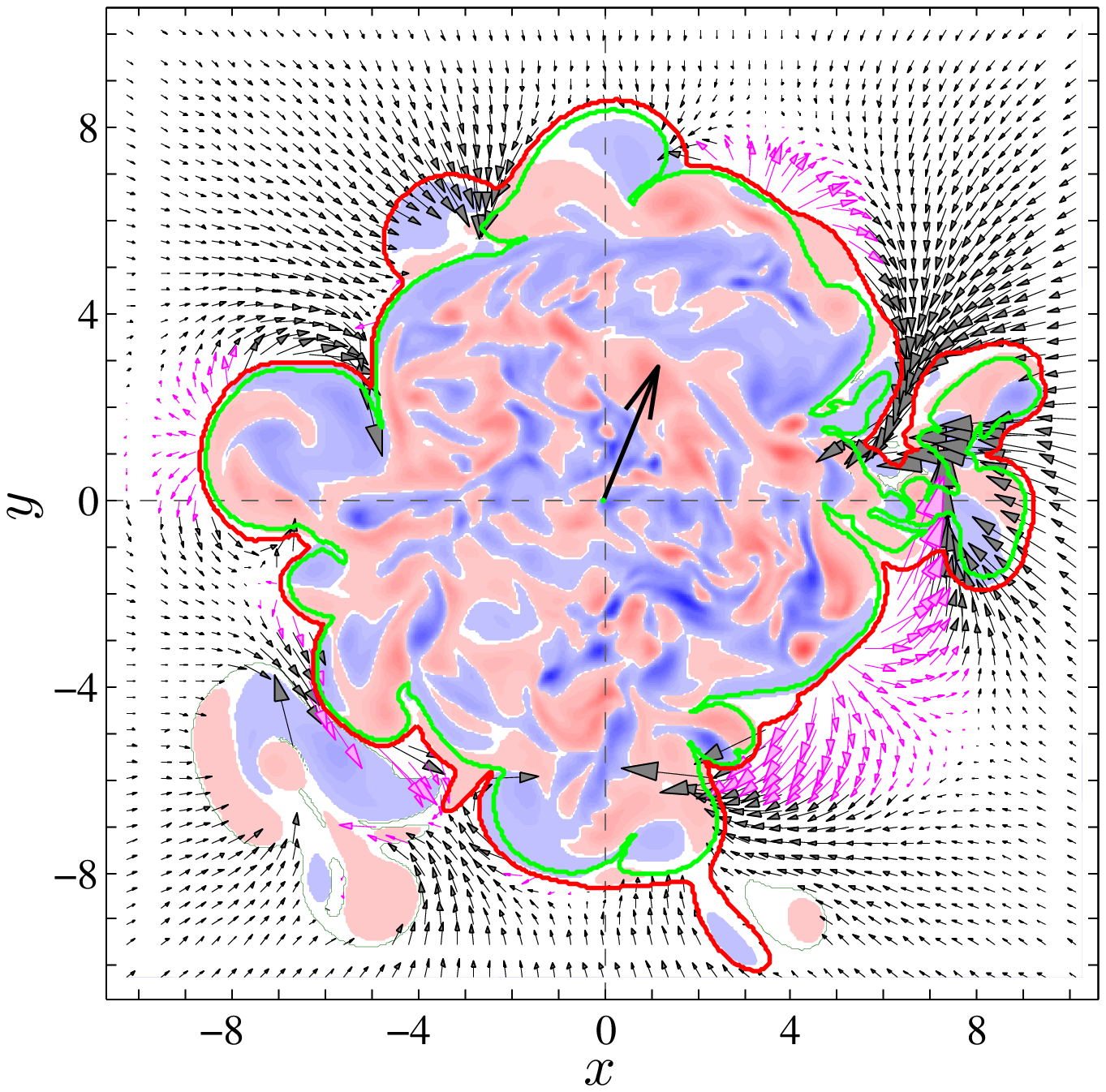}}
\put(0,13){\includegraphics[width = 8.15 cm]{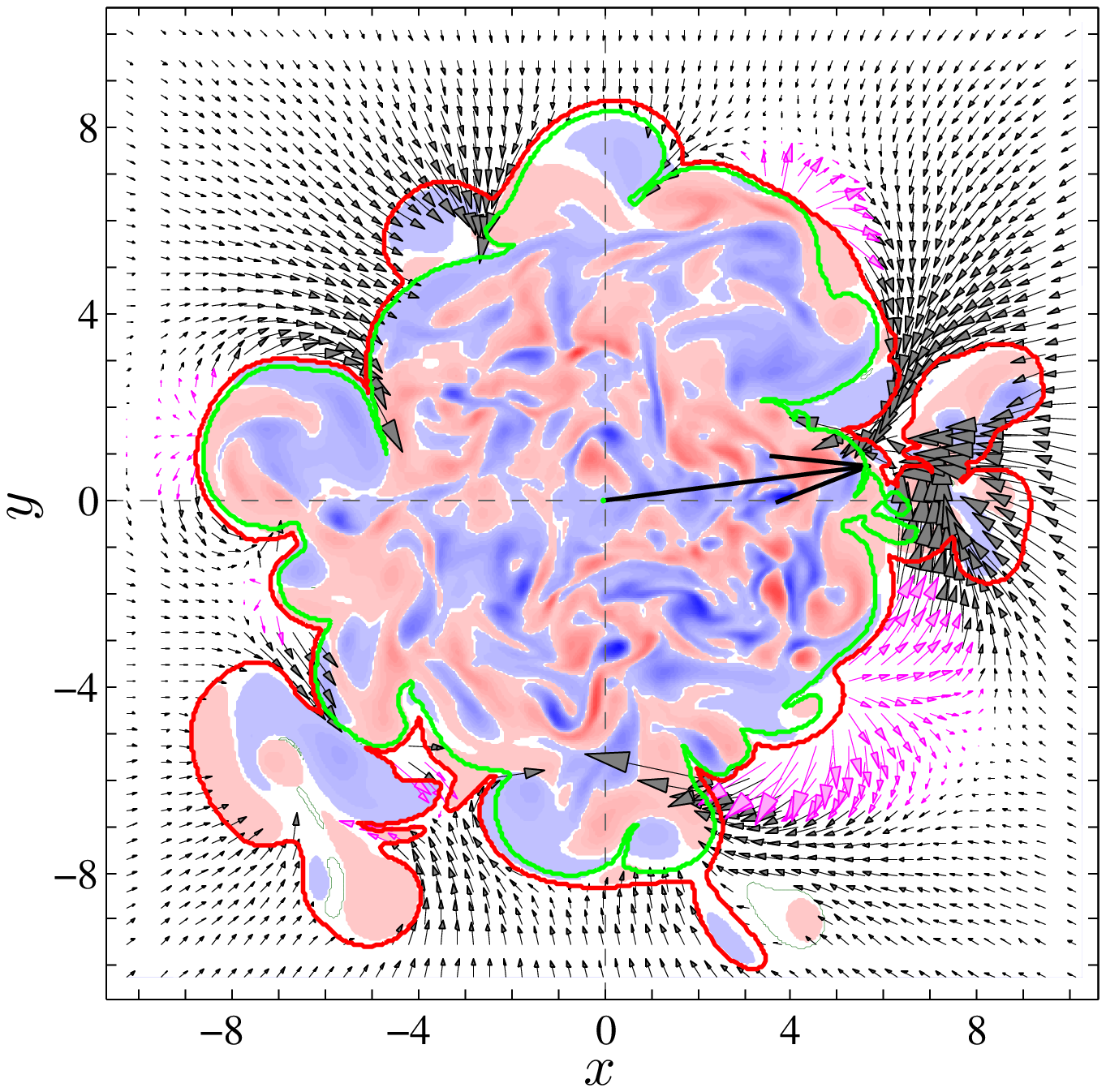}}
\put(82.5,13){\includegraphics[width = 8.15 cm]{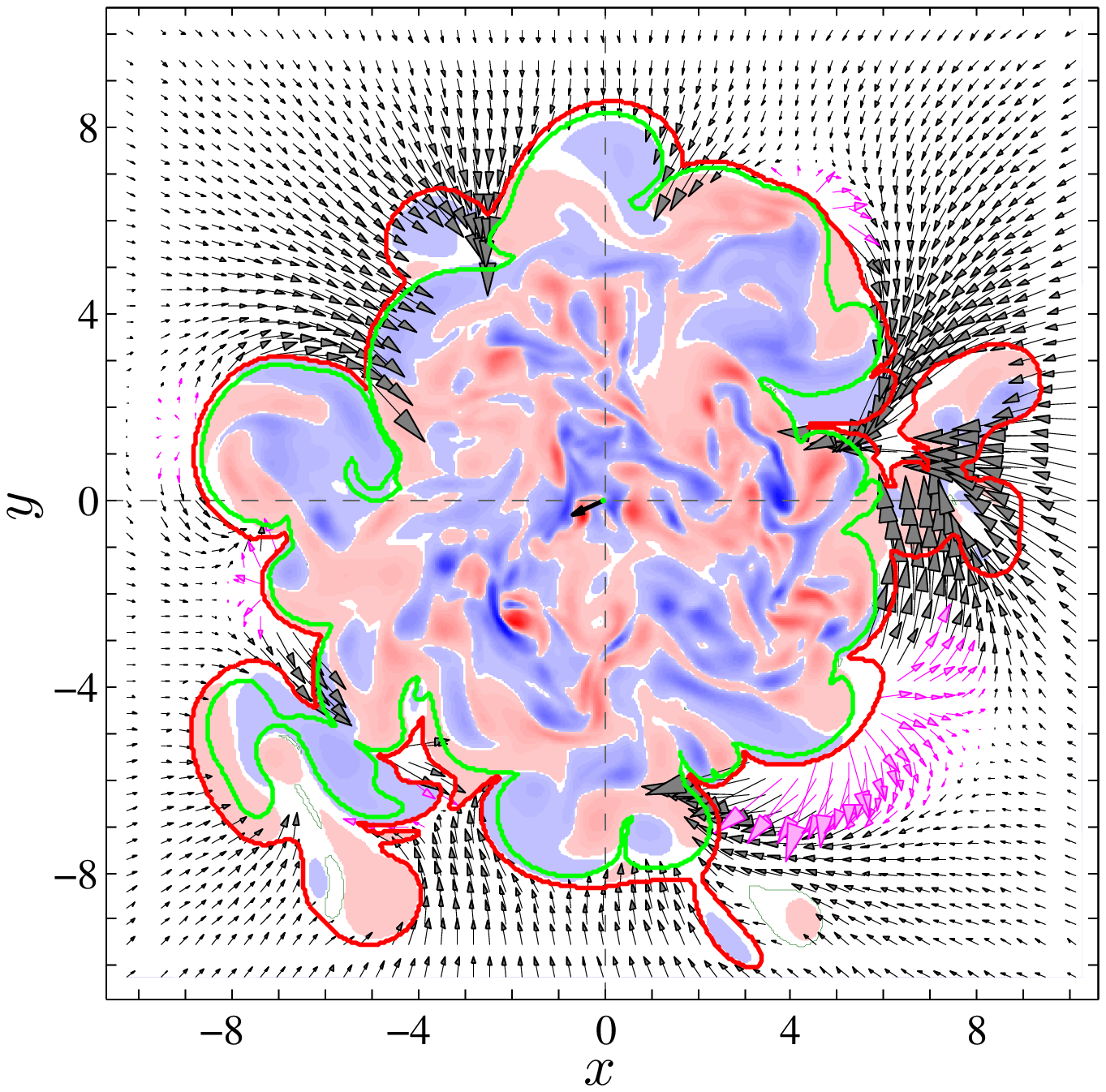}}
%
\put(26,0){\includegraphics[width = 1.85 cm]{Tst=185000_vel_vort_azim_Z=42p69_ref_vect.eps}}
\put(74,0){\includegraphics[width = 8.15 cm]{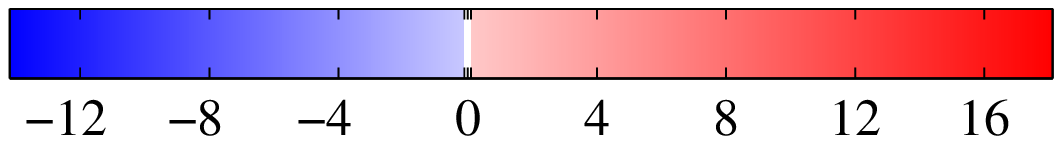}}
\put(10,6){\fontsize{10}{12}\color{gray}\selectfont \rotatebox{0}{Reference}}
\put(10,2){\fontsize{10}{12}\color{gray}\selectfont \rotatebox{0}{vector}}
\put(156,5){\fontsize{18}{20}\color{black}\selectfont \rotatebox{0}{$\omega_{z}$}}
\put(0,170){(a)}
\put(83,170){(b)}
\put(0,89){(c)}
\put(83,89){(d)}
\end{overpic}
\caption{Four instantaneous diametral sections at same $z$ and $t$ as in Fig.~\ref{fig:velvort-diam}. The background vorticity field is the signed axial component $\omega_{z}$. See Fig.~\ref{fig:velvort-diam} for additional details.}
\label{fig:velvort-diam-axial}
\end{figure*}

\begin{table}[tpb]
\begin{tabular}{|c|p{0.9\linewidth}|}
\hline
Zone & \bf{Flow events based on images in Fig.~\ref{fig:velvort-diam}(a-d) and an image at $\mathbf{t=975}$ not included in Fig.~\ref{fig:velvort-diam}}\\
\hline
\hline
    \multirow{1}*{A} & $\bullet$ Mild inrush at $t_0 = 925$, gets deeper but
                       bigger, begins to connect with H by $t = 950$ \\
\hline
    \multirow{2}*{B} & $\bullet$ Starts with outflow at $t_0$.\\
                     & $\bullet$ Boundary pushed outward by vorticity in the
                       turbulent core, gradually disappears by $t = 975$ \\
                       \hline
    \multirow{1}*{C} & $\bullet$ Starts at $t_0$ with mild inrush, grows slowly
                       and covers CD region in a broad inrush by $t = 970 - 975$ \\
\hline
    \multirow{2}*{D} & $\bullet$  Radial outflow prominent at $t_0$, decays slowly,
                       disappears at $t \simeq 970 / 975$.\\

                     & $\bullet$ Boundary is advancing throughout.\\
\hline
    \multirow{4}*{E} & $\bullet$ Modest outflow capped by inrush at bottom of D and lower part of E at $t_0$.\\
                     & $\bullet$ Upper part from D weakens, outflow disappears by $t = 950$.\\
                     & $\bullet$ Lower E squeezed by expansion of core.\\
                     & $\bullet$ Weak outflow reappears at E at $t = 975$.\\
\hline
    \multirow{1}*{F} & $\bullet$ Mild inrush at $t_0$, little change even at $t = 975$.\\
\hline
    \multirow{3}{*}{G} & $\bullet$ Modest outflow at $t = 925$, pushed out by both
                       inner and outer boundaries (which remain close to each other).\\
                     & $\bullet$ Spans almost whole quadrant [from $(-2,- 6)$ to $(7,0)$] by $t = 955$.\\
                     & $\bullet$ Shrinks by $t = 965$, making room for new
                       inrush at H, but still active at $t = 975$.\\
\hline
    \multirow{3}*{H} & $\bullet$ Inrush moves right, gets weaker by $t = 955$. \\
                     & $\bullet$ A and H inrushes appear to coalesce (from
                       opposite directions), and become a major feature
                       of the flow by \\
                     & \hspace{2 mm} $t = 970 / 975$, like C.\\
\hline
\end{tabular}
\caption{Flow events across a diametral plane}
\label{tab:time-evol-diam-velvort}
\end{table}


The inner and outer boundaries of the jet are indicated as before by contours of
$|\bm{\omega}|$ at 0.5 and 0.1 respectively. As with axial sections,
we see in Fig.~\ref{fig:velvort-diam} regions where the inner and
outer boundaries are very close to each other, as well as other
regions where they are much farther apart. The flow field in the
ambient is of two distinct types, the dominant one being radially
outward motions, with circulatory motion also present but not as
strong as in the axial section of Fig.~\ref{fig:vel-vort}. The
radial motions often issue from those regions where the inner and
outer boundaries are close, for the simple reason that core-flow vorticity elements are then close to the ambient fluid. For the same reason circulatory motions may also be
present in such an area (see for example Fig.~\ref{fig:SR9}, and
also the area straddling the $x$ axis at $y = 0$ across the first and
fourth quadrant in Fig.~\ref{fig:velvort-diam}).

The diametral section shown in Fig.~\ref{fig:velvort-diam}(a) is just
above the wide base S4-S9 of an existing coherent structure, the lower part of which is seen near
the
top around $z \simeq 38$ to 42 in Figure \ref{fig:vel-vort}. The
strong source-like radially outward motion seen in the ambient fluid across the
second and third quadrants, $-9<x<-5$ which we call zone A, has a striking degree of
circular symmetry, and which is not entirely consistent with the azimuthal component of vorticity shown in Figure \ref{fig:velvort-diam}(a). A
plausible explanation involves the effect of streamwise vortcity which has not been considered so far. Figure~\ref{fig:velvort-diam-axial}(a) shows data at the same stations as in Fig.~\ref{fig:velvort-diam}(a), except that at each of the four stations, the azimuthal component of vorticity is replaced by the streamwise component. Figure~\ref{fig:velvort-diam-axial}(a) shows the presence of a pair of streamwise vortices in zone A (red:counter-clockwise and blue:clockwise). The arrangement of these streamwise vortices along with the local azimuthal vorticity result in an outward movement of the fluid between the vortices from the turbulent core towards the T/NT interface, which in turn results in an instantaneous displacement effect (note that the inner and outer boundaries are
very close to each other and are pushed outward along with the fluid). This boundary expansion
is also evident from a comparison between Fig.~\ref{fig:velvort-diam}(a) and Fig.~\ref{fig:velvort-diam}(b).  Interestingly, a somewhat similar
radial flow is noticeable in the fourth quadrant of
Figs.~\ref{fig:velvort-diam}(a-d). Also, the vortical object of
SR9 (Fig.~\ref{fig:SR9}) is clearly seen in the diametral section in
Fig.~\ref{fig:velvort-diam}(a) where its length along the $y$ axis is
seen to be about 5.  However, the feature gradually changes its shape
with time, and is substantially different at $t = 970$
(Fig.~\ref{fig:velvort-diam}(d)). Between the data in
Figs.~\ref{fig:velvort-diam}(a) and Fig.~\ref{fig:SR9}(a), it can be
confirmed that the vorticity in the object around the point $x\approx0,~y\approx-7$ is confined to a $2 \times 5 \times 2$ ovoid.

It may be noted that the in-rush events, so conspicuous in the axial section
shown in Fig.~\ref{fig:vel-vort} and Fig.~\ref{fig:SR9}, are substantially
weaker in the diametral section. This suggests that the major contribution to
entrainment is made by the inward and upward velocity induced in the ambient fluid due
to elements of azimuthal vorticity in the coherent structure in the
core. 

To understand the evolution of the flow events in
the near ambient of the turbulent core, the flow region in the
diametral plane shown in Fig.~\ref{fig:velvort-diam} is divided into 8
approximately equal semi-quadrants A to H in the $xy$ plane, e.g. starting with A in the angular
sector $0^{\circ} < \phi < 45^{\circ}$ around the origin $(x,y)=(0,0)$ and going round anticlockwise to H in the sector
$335^{\circ}< \phi < 360^{\circ}$. The flow shows some very
interesting features as can be clearly seen in
Figs.~\ref{fig:velvort-diam}(a) to (d), and documented in Table~\ref{tab:time-evol-diam-velvort}. The turbulent core is largest at $t
= 950$ (Fig.~\ref{fig:velvort-diam}(b)), consistent with the fact that
the widest part of the coherent structure, with its base SR4-SR9 containing
a vortex ring, is passing through the plane $z = 42.69$ at the time.

 
\section{Relation between vorticity field at T/NT interface and its near-neighbourhood}
\label{sec:nature-outer-flow}

In most of the previous work, the boundary between the turbulent core
and the ambient fluid has been determined by thresholds on either
scalar concentrations \cite{Westerweel_JFM_2009, burridge_parker_kruger_partridge_linden_2017, Westerweel_PRL_2005} or an
out-of-plane vorticity component \cite{chauhan2014turbulent}. Boundaries defined by the former have often been the subject
of controversy mainly because of the vast difference between the diffusivity of the scalar and viscosity. Furthermore, the evolution equation of the passive scalar and vorticity are fundamentally different due to the absence of the stretching-tilting term in the former. One could argue that the gradient of a passive scalar and vorticity are orthogonal to each other in the limit of very high Peclet and Reynolds number as was shown in a recent work by Patwardhan and Ramesh (2018) \cite{pc}. They show that scalar gradient surface tracks the vorticity field if the two were aligned initially. However, they also showed that in the regions with sharp gradients the gradient of passive scalar and vorticity are not orthogonal. We know that at the T/NT interface the velocity and scalar gradients are sharp and as a consequence a scalar cannot faithfully represent the turbulent flow field. So vorticity is fundamentally the
more logical criterion to use, but to determine all three components
of the vorticity vector experimentally is difficult. The 3D
results from DNS studies can therefore provide valuable insights from
data on the full vector $\bm{\omega}$.

We have shown here that the use of $|\bm{\omega}|$ sheds much light on
the subject.  The contour $|\bm{\omega}| = 0.5$ is close to the
turbulent core of the jet, and is a good candidate for the threshold
defining the inner edge of an interface and is consistent with the
choice $\omega_{y} \simeq 0.5$ made by Westerweel \etal (2009)
\cite{Westerweel_JFM_2009} for the out-of-plane vorticity component
($\Omega_{z}$ in their notation). Thereafter $|\bm{\omega}|$ drops
steeply going outward; the $|\bm{\omega}| = 0.5$ contour is
approximately at the point where the outward gradient of
$|\bm{\omega}|$ is steepest in the T/NT interface(i.e., Corrsin's
viscous superlayer), and is taken here as defining the inner boundary
of the jet. 

The nature of the irrotational flow in the ambient near the T/NT interface has not received enough attention. With the present simulation more
precise pictures of the flow in the ambient have been constructed.
These pictures show a considerable degree of organization or order in
the flow, usually circulatory in nature but occasionally radial as
well.  It is often (but not always) possible to trace the circulatory
motions as those associated with the presence of elements of organized
large-scale vorticity in the turbulent core near the edge of the jet, making a highly plausible connection through the
Biot-Savart relation. Given this connection, it is not surprising that the inner boundary of the jet,
close to the region of transition from one structure to the next in
the core flow, often shows sharp and intense incursions at
velocities of order $0.4 \overline{w}_c$ as already pointed out. The highest velocity
in the outer field is $0.34 \, \overline{w}_c$ at $x \approx 0.3$ and
$z \approx 29.7$ in SR6 of Fig.~\ref{fig:vel-vort}.  Axial sections of
the flow show that along either inner boundary of the jet, about 4 or 5
\textquoteleft{inrush events}' of the above type and others (at an
average streamwise spacing of $~ 5$) can be inferred. Although the
length scale of a hypothetical `lid' for such an incursion into the
inner boundary covers only a small fraction of
the surface area of the jet, the higher inrush velocities compensate to some extent
for the lower area, and could contribute more significantly to the
total entrainment than has been estimated in recent work (less than
8\% according to Westerweel \etal (2009) \cite{Westerweel_JFM_2009}).
The DNS images leave no doubt however that a significant part of the
entrainment occurs through inrush events which have an episodic
character, reminiscent of the momentum flux events that contribute to
the Reynolds shear stress in a high Reynolds number atmospheric
boundary layer (Narasimha \etal 2007 \cite{narasimha2007turbulent}).


In summary, these observations suggest the following physical picture.

\begin{enumerate}
\item The propagation - outward or inward - of the inner boundary, as
well as the organized motion of the ambient fluid that rushes inward,
can result in penetration that pushes inwards \textquoteleft{vulnerable}' parts of the
outer boundary. This process is encouraged by the Biot-Savart induced
velocity in the ambient fluid due to elements of the vorticity field
associated with the coherent structures in the core, leading to what
has been called engulfment.

\item As long as there is a sharp boundary between turbulent and
non-turbulent flow, entrainment into the core flow must involve fluid
crossing a fractal inner T/NT boundary (which may itself be
propagating into non-turbulent fluid); this could legitimately be called 
\textquoteleft{nibbling}'.

\end{enumerate}



It would be interesting to find out how the radial and axial components
of velocity at the T/NT interface, which are primarily responsible for the fluid
entrainment from the lateral and cross-sectional sides of a turbulent
jet, are correlated to the relavant velocity components at the
surfaces of a short circular cylinder circumscribing the T/NT interface. There
is entrainment also in the axial or streamwise direction, \textit{e.g.} just
below S4-S9.

\begin{figure}[h!]
\begin{overpic}
[width = 16.0 cm, height = 6.8 cm, unit=1mm] {Fig_box.eps}
\put(40,0){\includegraphics[width = 8.0 cm]{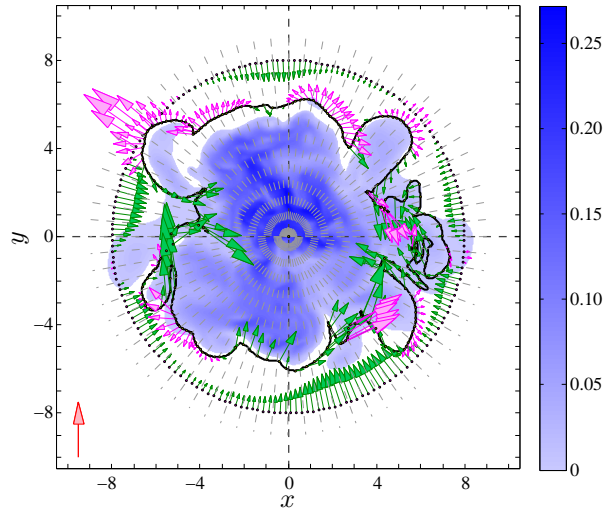}}
\end{overpic}
\caption{Conventions followed for overlap analysis; see text for
details. Diametral section at $z = 34.05$ and $t = 2050$ showing the
contours of axial velocity and the velocity vectors at the T/NT interface and
the cylinder circumference. The reference velocity vector shown in red
at the bottom-left corner is $0.05 w_{0}$. Velocity vectors at
the 	disc circumference are scaled up by a factor of four relative to the vectors at the T/NT interface. Note that radially outward motions are rare and radially inward
motions dominate.}
\label{fig:OA-conventions}
\end{figure}

We first choose the radial components for analysis. Consider the
diametral section at $z = 34.05$, in the middle of the
self-preserving zone. We now choose a cylinder of height $6d_0$ ($32\le z \le38$), and a radius of $8
d_{0}$ which is just adequate to lie close to but outside the T/NT interface almost
everywhere around the circumference. Radial velocity vectors are
plotted at both the T/NT interface and the cylinder circumference at the
selected instant $t = 2050$ as shown in
Fig.~\ref{fig:OA-conventions}; vectors in magenta and green indicate
respectively radially outward (generally detraining) and radially inward
(generally entraining) velocities (neglecting the interface propagation velocity, see section~\ref{Results1}). The contours within the T/NT interface show axial
velocity.

The procedure for finding the correlations mentioned above (along with
its limitations) is now outlined. The angular circumference of the
disc (360 degrees, also periphery of the T/NT interface) is divided into 72
sectors, each subtending $5^{\circ}$ at the center as shown in
Fig.~\ref{fig:OA-conventions}; this gives us enough sectors to make reliable estimates of the correlations.

\begin{figure}[h!]
\begin{overpic}
[width = 16.0 cm, height = 6.6 cm, unit=1mm]
{Fig_box.eps}
\put(0,0){\includegraphics[width = 8.0 cm]{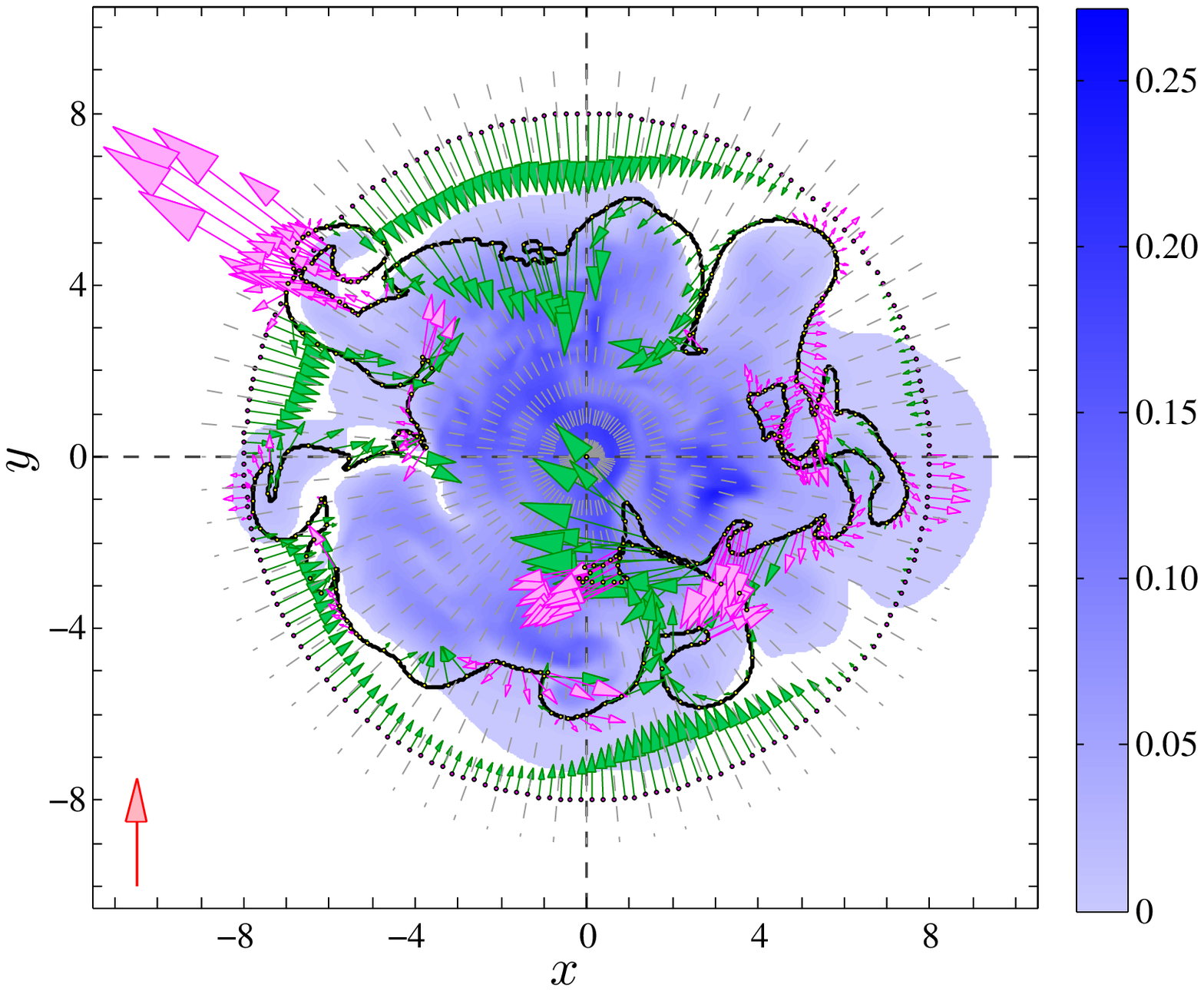}}
\put(81,0){\includegraphics[width = 8.0 cm]{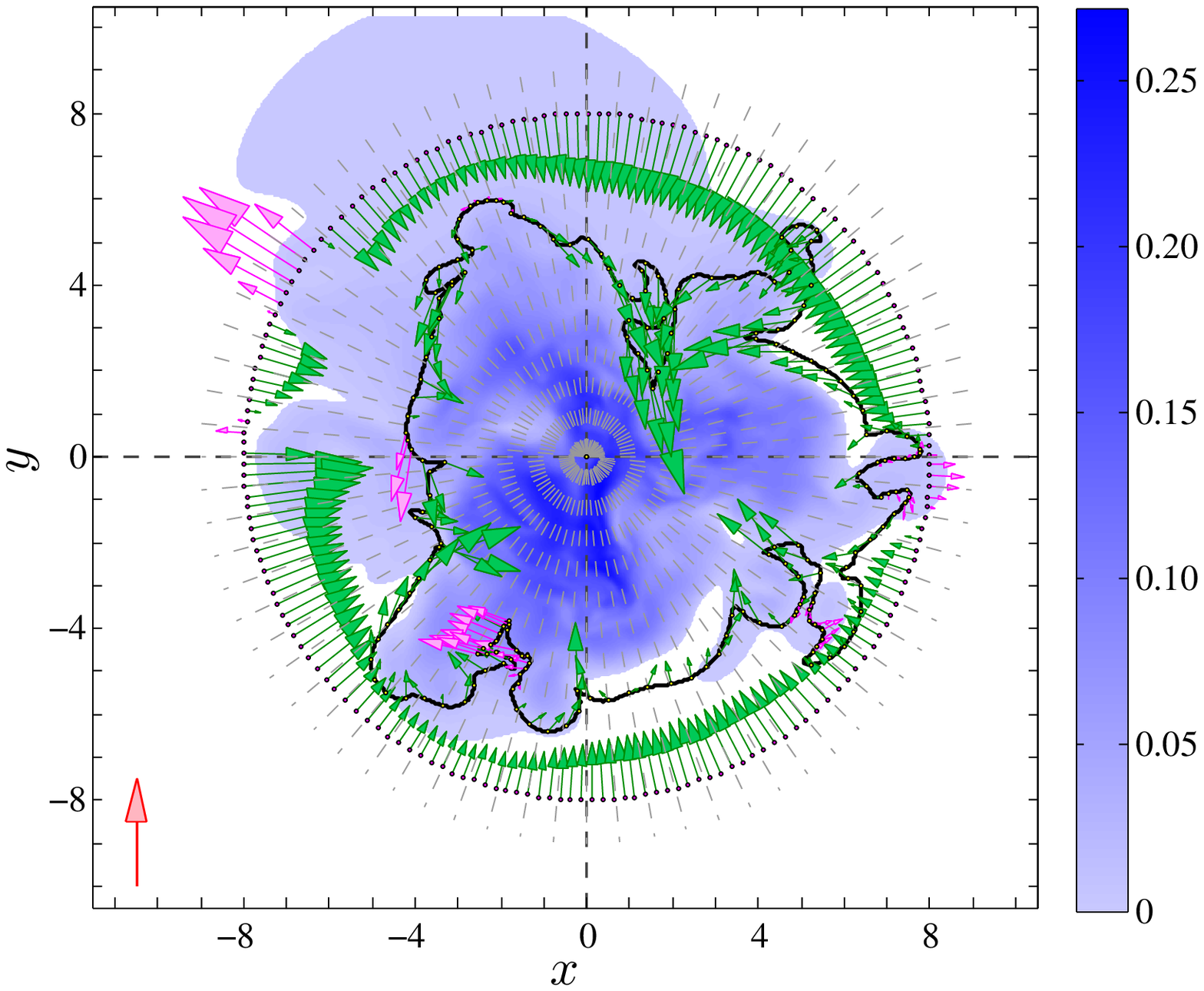}}
\put(8.5,60){(a)}
\put(89.5,60){(b)}
\end{overpic}
\caption{See Fig.~\ref{fig:OA-conventions} for conventions. (a)
Instantaneous image at $t = 2075$ indicating the difficulty in the
selection of highly convoluted T/NT interface for finding the correlations, especially in the fourth quadrant. (b) An instantaneous event at $t = 2175$ showing strong entraining correlations between the radial velocity vectors at T/NT interface and cylinder
circumference.}
\label{fig:OA-selection-difficulty}
\end{figure}

We use the following conventions while calculating the correlation:

\begin{enumerate}
\item Correlated: If velocity vectors at both the T/NT interface and the
cylinder are in the same direction, either inward or outward,
then that event is said to be ``correlated''.

\item Anti-correlated: If velocity vectors at the T/NT interface and the
cylinder are in the opposite directions, either inward at the T/NT interface and
outward at cylinder or vice versa, then that event is said to be 
``anti-correlated''.

\item Non-correlated / non-decisive: We classify the event as ``non-correlated /non-decisive''
\newline
(i) if the velocities are negligibly small ($<0.05 w_c$). By inspection and trial, the threshold was
determined to be about $0.001 w_{c}$; for example, look at the
sectors between $185^{\circ}$ to $195^{\circ}$ in Figure~\ref{fig:OA-conventions};
\newline 
(ii) when it is difficult to decide about the correlation because the
velocity at either T/NT interface or the outer circle is zero or very
small ($\sim 0.001 w_{c}$); for example, the velocity vectors at
cylinder circumference in sector between $330^{\circ}$ to
$335^{\circ}$ in Figure~\ref{fig:OA-selection-difficulty}(a);
 \newline 
(iii) when the T/NT interface goes beyond the outer circle, which has been found
to occur for three time instants viz. $t =
2100, 2125, 2150$ 
out of 8 time instants for less than about $40^{\circ}$ in the total
of $2880^{\circ}$ studied, i.e., 1.4\% of the circumference.  
\newline (iv)
When the T/NT interface is highly convoluted, correlations are estimated using the velocity vectors at T/NT interface closest
to the cylinder, and ignoring the inner T/NT interface. Figure~\ref{fig:OA-selection-difficulty}(a) presents one such convoluted T/NT interface.
\end{enumerate}

The eight time instants chosen for analysis are separated from their
neighbours by 25 flow units adding upto a total duration of 200 flow
units. We find that the velocity vectors are correlated for 73\% of
the circumference, anti-correlated for 24\%, and non-correlated for
3\%.

Figure \ref{fig:OA-selection-difficulty}(b) shows one instant when the
velocity vectors at T/NT interface and cylinder circumference both inward and strongly correlated. Note that 65 out of 72 sectors (more than 90\% of the circumference) are correlated and inward, while 7 out of 72 sectors are anti-correlated (i.e., less than 10\% of the circumference). This suggests that entraining motions at the T/NT interface are strongly correlated with the radially inward component of the velocity at the cylinder's surface.

\section{Mass Flux Budget} 
\label{sec:mass_budget}

\begin{figure}[h!]
\begin{overpic}
[width = 16.0 cm, height = 10.6 cm, unit=1mm]
{Fig_box.eps}
\put(30,0){\includegraphics[width = 10.0 cm]{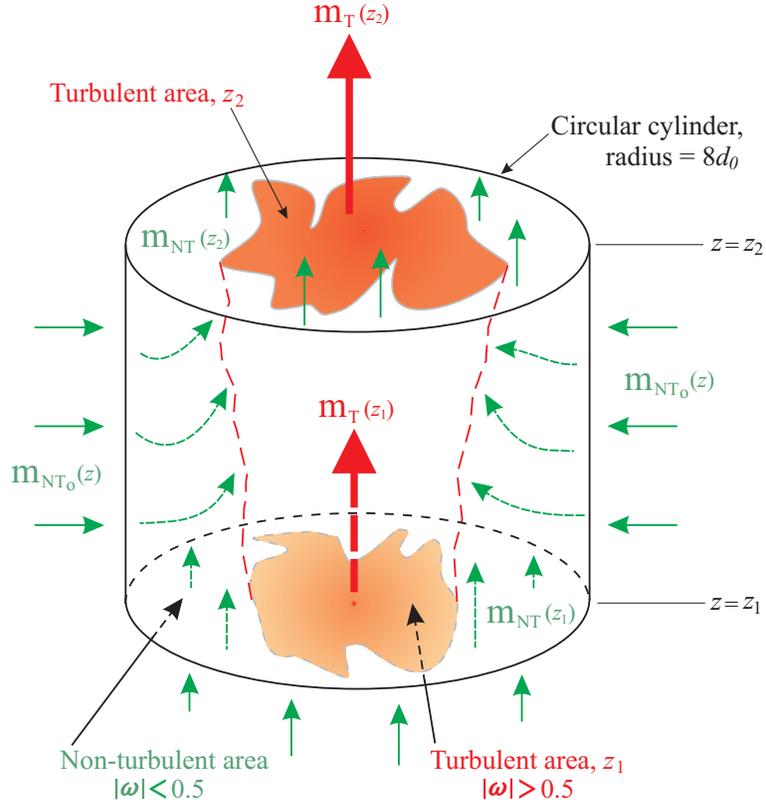}}
\end{overpic}
\caption{Schematic for the mass flux budget (see text for
details). Color code: red for turbulent motion (T), green for
non-turbulent motion (NT).}
\label{fig:mass_budget}
\end{figure}

This section presents the net mass flux budget within the turbulent core determined by $|\omega|=0.5$ and also within the cylindrical disc at $t=1995.25$ to demonstrate the accuracy of the mass flux entrainment calculations. This is done by calculating the streamwise variation of the mass flux in the turbulent core and correlating it with the mass flux entrained into a fixed cylindrical disc of radius $8d_0$ and height $\approx 6.5d_0$ circumscribing the T/NT interface. A schematic of the disc and the T/NT interface is shown in Fig.~\ref{fig:mass_budget}. The volume between the cylindrical disc surface and the T/NT interface will be called the sheath. The mass flux at the cylinder edge is computed as 
\begin{eqnarray}
\label{eq:massflux1}
\int\textbf{u}\cdot d\textbf{S}
\end{eqnarray}
\noindent 
where $d\textbf{S}$ represents the local surface area vector at the edge of the cylinder. The mass flux in the turbulent core is computed using Eq.~\ref{eq:massflux1} where $d\textbf{S}$ now represents the local area vector in the streamwise direction. The coordinates of the interface were detected with the iso-surface function in MATLAB using the criterion $|\omega| = 0.5$. The detected interface was mapped onto the nearest grid points in the computational domain, thus pixelising the interface. The advantage of this mapping is that the mass flux budget is accomplished to a very high accuracy as shown below and is not affected by any interpolation errors. The results are as follows: \\
\uline{Mass flux budget: cylindrical disc}\\
Mass flux into the disc from the bottom surface ($m_{z_1=32.35}$) $=~7.472$\\
Mass outflux from the top surface of the disc ($m_{z_2=38.64}$) $=~7.959$\\
Mass flux into the disc from the lateral boundary ($m_{NT_o}$) $=~0.486$\\
Net mass flux into the disc accounting for surface normal=$m_{z_1}-m_{z_2}+m_{NT_o}$ $\approx~10^{-11}$\\
\uline{Mass flux budget: turbulent core}\\
Mass flux into the turbulent core from the bottom surface ($m_T(z_1=32.47)$) $=~7.327$\\
Mass outflux from the top surface of the turbulent core ($m_T(z_2=38.51)$) $=~7.126$\\
Mass outflux from the top sheath surface ($m_{NT}(z_2)$) $=~0.798$\\
Mass flux into the bottom sheath surface ($m_{NT}(z_1)$) $=~0.155$\\
Mass flux into the lateral boundary of the cylinder ($m_{NT_o}$) $=~0.443$ \\
Total mass flux from the sides of the cylinder ($m_{NT_o^*}$)=$m_T(z_1)-m_T(z_2)+m_{NT}(z_1)-m_{NT}(z_2) =~0.443$\\
Mass flux balance=$m_{NT_o}$ - $m_{NT_o^*}$=$10^{-7}$\\

It must be noted that the $z$ location of the mass flux budget in the core and the cylinder is slightly different due to the staggered location of velocity and vorticity. This results in a slightly different mass flux from the lateral boundary of the cylinder. From the numbers given above its is seen that the entrainment mass flux calculations at the T/NT interface are accurate up to $10^{-7}$ and the continuity equation up to $10^{-11}$. At $t=1995.25$, for $32.5\le z \le38.5$, we note that the disc had a net mass flux entraining into the cylinder  and the turbulent core had a net mass flux detraining from the core. This would seem to suggest that the dynamics at the T/NT interface and the cylinder are not in sync as opposed to the conclusion derived in the previous section. This may well be due to the fact that conclusions in the previous sections did not account for the magnitude of these events. However, it must be noted that the present inference is based on the analysis of a single time instant. A more detailed analysis is required to understand the entrainment dynamics close to the T/NT interface. We also compute the Taylor entrainment coefficient $\alpha_E=\frac{1}{2\pi W_c b_{we}}\frac{d\overline{m}}{dz}$ \cite{taylor1945} using the data averaged over 1750 flow units where $\overline{m}$ represents the mean mass flux within the turbulent core, $W_c$ is the mean centreline velocity and $b_{we}$ is the mean velocity width, defined by $w(b_{we})=W_c/e$. The entrainment coefficient in the self-preserving region in the present data is 0.065 which is smaller than the $\alpha_E=0.074$ at $z\approx34$ reported in Carazzo \etal (2006) \cite{carazzo2006route}. This was expected, as the present calculation does not account for the mass flux outside the turbulent core, in contrast to the analysis in Carazzo \etal (2006). Additional details are discussed in Shinde \etal (2019).
\section{Entrainment as an episodic process}
\label{sec:Entrainment-dynamics}

\begin{figure}[h!]
\centering
\includegraphics[width = \textwidth]{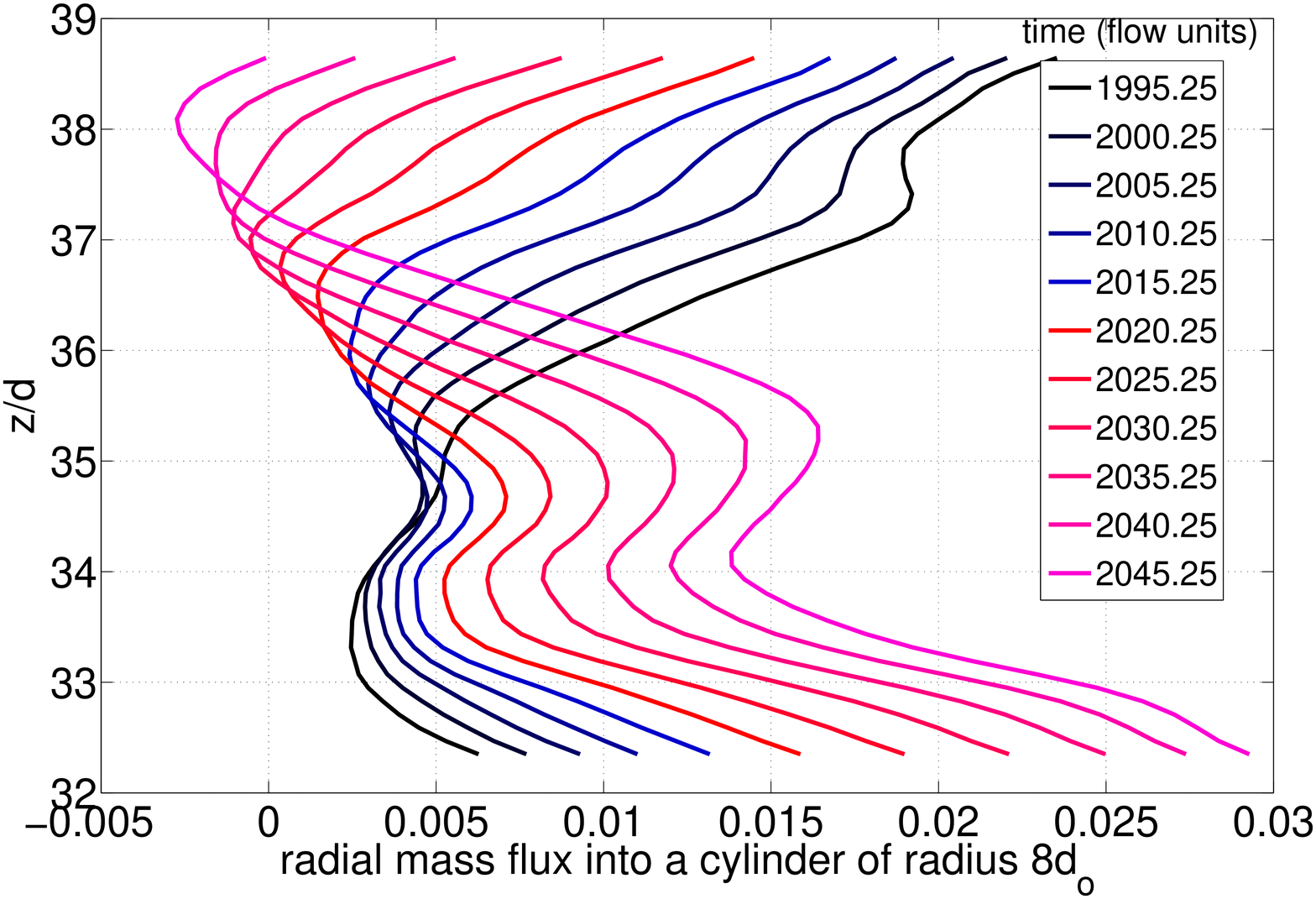}
\caption{Streamwise variation of radial mass flux into the $8d_0$ diameter cylindrical surface. Time indicated in flow units.}
\label{fig:event_fig1}
\end{figure}

\begin{figure}[!h]
\includegraphics[width = \textwidth]{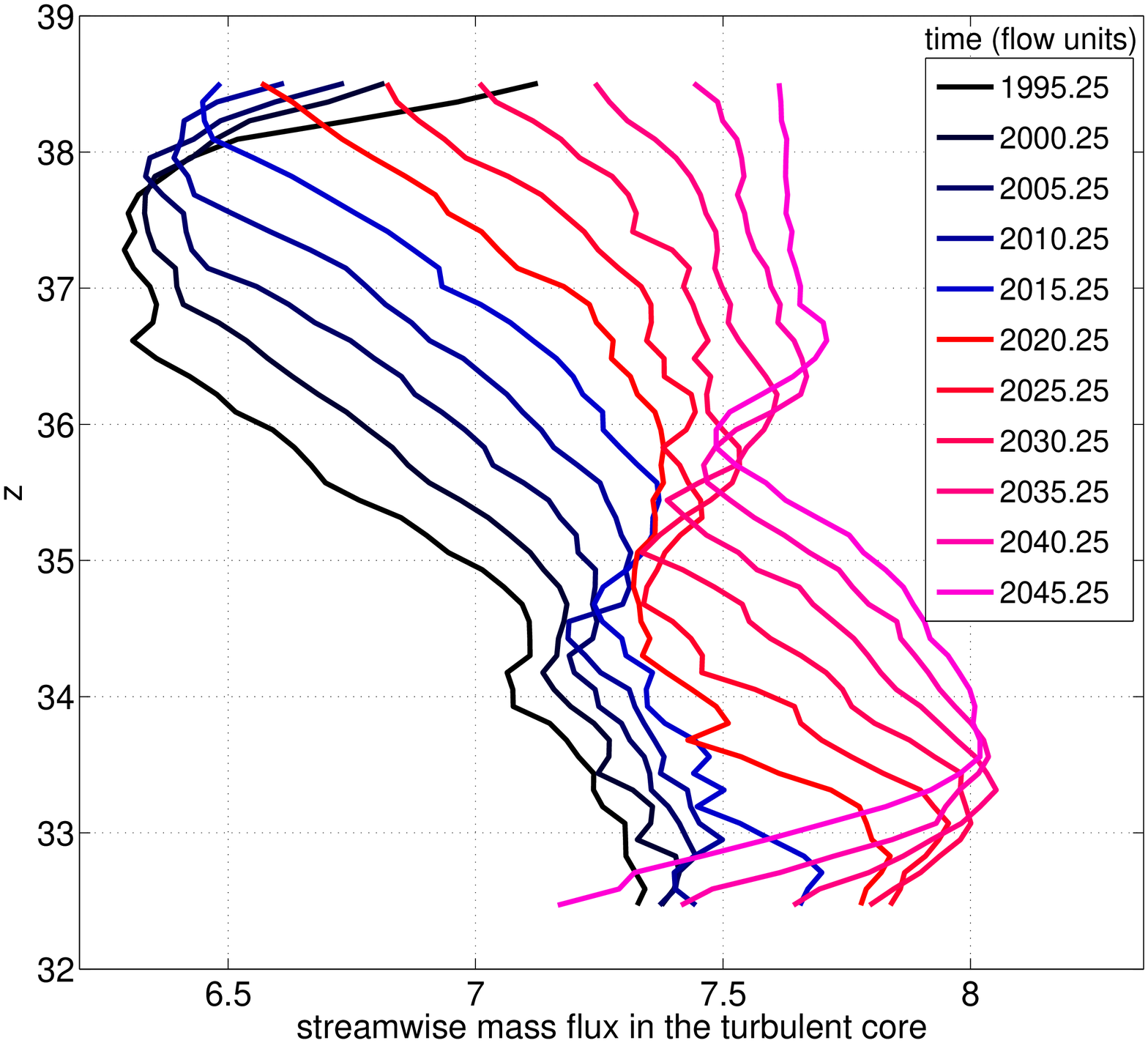}
\caption{Streamwise variation of the mass flux within the T/NT interface ($|\omega|=$0.5). Time indicated in flow units.}
\label{fig:event_fig2}
\end{figure}

To study the nature of entrainment in a turbulent jet it is
important to understand how the irrotational ambient fluid enters into
the turbulent jet. Figure~\ref{fig:event_fig1} shows the streamwise variation of the radial mass flux entering into the lateral surface of the $8d_{o}$ disc at various instants. We see that entrainment varies in both space and time, and it can be spatially anti-correlated in $z$. Thus, the net entrainment in the region $32 < z < 35$ at a given instant is anti-correlated with the net entrainment in $35 < z <
39$. Figure~\ref{fig:event_fig2} shows the variation of streamwise mass flux in
the turbulent core, obtained by integrating the streamwise velocity over the area within the T/NT interface at any $z$. The
anti-correlation observed at the edge of the cylinder is not as
strongly visible in the streamwise mass flux data. Some of the mass flux entrained into the cylinder may be retained in the non-turbulent region outside T/NT interface. We note that the mass flux in the turbulent core of the jet increases between $34\le z \le38.5$ over the time span shown in Fig.~\ref{fig:event_fig2}. This is not surprising as it is linked to the passage of large scale coherent structures. The presence of these structures locally enhances entrainment events that explains the increase in the mass flux in the turbulent core. And in the absence of these structures the mass flux in the core decreases. The decrease in the mass flux in the core is visible in the region between $32.5\le z \le34$ for the duration $2030\le t \le2045$. This trend would propagate downstream at later times, and the cycle repeats. Similar observations were reported in the recent studies on turbulent plumes \cite{burridge_parker_kruger_partridge_linden_2017, Plourde_JFM_2008}. To understand the
nature of these observations it is necessary to examine the entrainment dynamics close to the T/NT interface. 

\begin{figure}[!h]
\includegraphics[width = \textwidth]{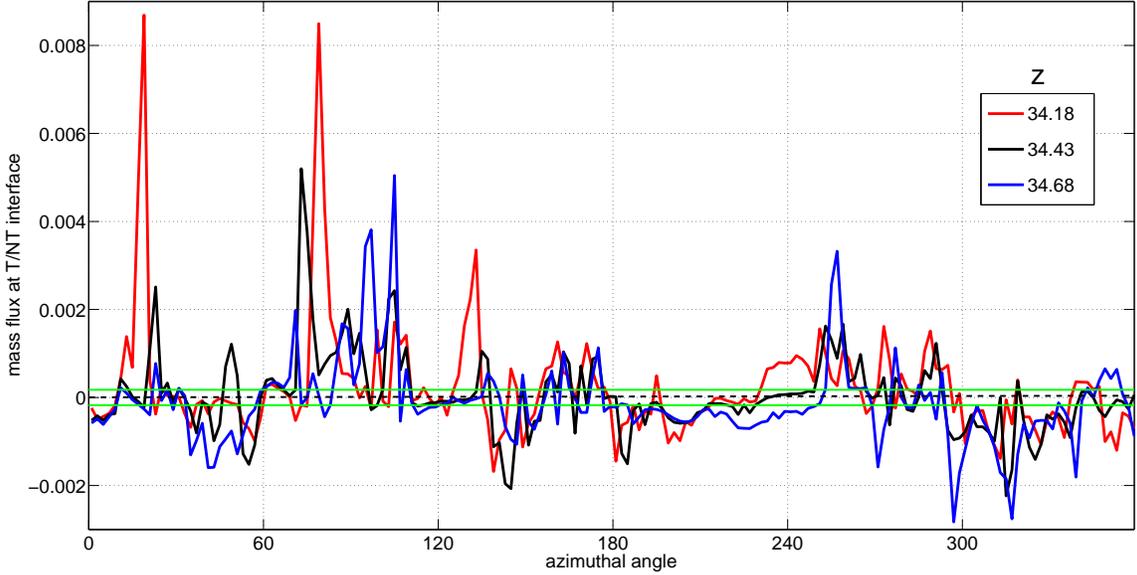}
\caption{Azimuthal variation of mass flux at the R/IR interface (threshold level on $|\omega|=$0.1). The green lines indicate $\pm$ 30\% of \textit{r.m.s} of the mass flux.}
\label{fig:event_fig3}
\end{figure}

Figure~\ref{fig:event_fig3} shows the azimuthal variation of mass flux at the R/IR interface at intervals of $2^{\circ}$. In certain angular regions there exist several interface co-ordinates due to the contortions in the interface. At these locations only the mass flux spatially averaged over $2^{\circ}$ is plotted. Figure~\ref{fig:event_fig3} shows that there are instances when localized regions of very high mass
flux entraining into or detraining from the turbulent core. At such times, the mass flux contributions from the positive events are significantly larger than from the negative events. From
Fig.~\ref{fig:event_fig3} and earlier images (e.g., see
Fig.~\ref{fig:vel-vort}) it appears that
entrainment is episodic. We identify about 13 to 16 such events at each $z$ location shown in Fig.~\ref{fig:event_fig3}. Some of these events or episodes may be spatially correlated. To better describe this episodic character we use the algorithm described by Narasimha \etal
(2007) \cite{narasimha2007turbulent} where they adopted a threshold method to identify the productive and
counter-productive periods in an atmospheric boundary layer momentum
flux time series. Following their analysis we adopt the
r.m.s. value $m^*$ as an appropriate unit for the threshold, and an event
is detected if the instantaneous mass flux $m(t,z,\phi)$

\begin{eqnarray}
\label{eq:eq1}
|m| > k m^*,
\end{eqnarray}
\noindent
where $k$ is a constant to be selected. This is done by examination of the data at
the T/NT interface at intervals of 2 degrees in the azimuthal angle ($\phi$). The angular
extent of an event in the azimuthal direction is defined as a
continuous interval in the azimuthal angle where Eq.~\ref{eq:eq1} holds true. This entire
interval is then considered as one spatial event at a given instant.

\begin{figure}[!h]
\includegraphics[width = \textwidth]{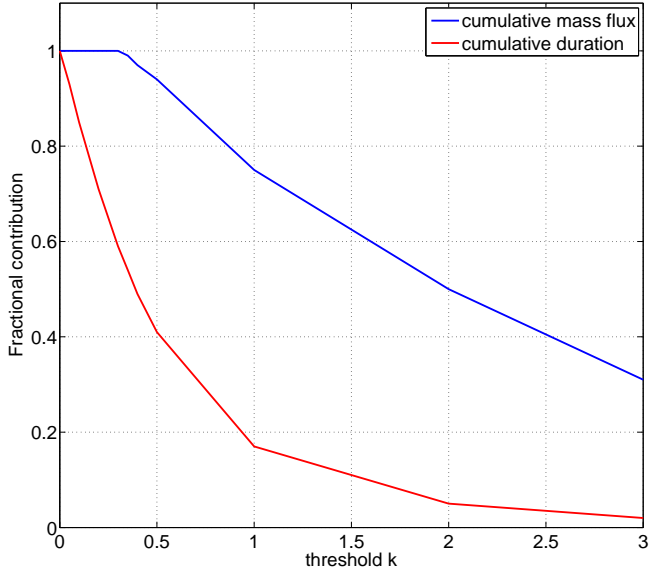}
\caption{Variation of fractional contribution to mass flux at the R/IR interface and the corresponding cumulative duration as a function of threshold $k$ for event detection based on Eq.~\ref{eq:eq1}.}
\label{fig:event_fig4}
\end{figure}

\begin{figure}[!h]
\includegraphics[width = \textwidth]{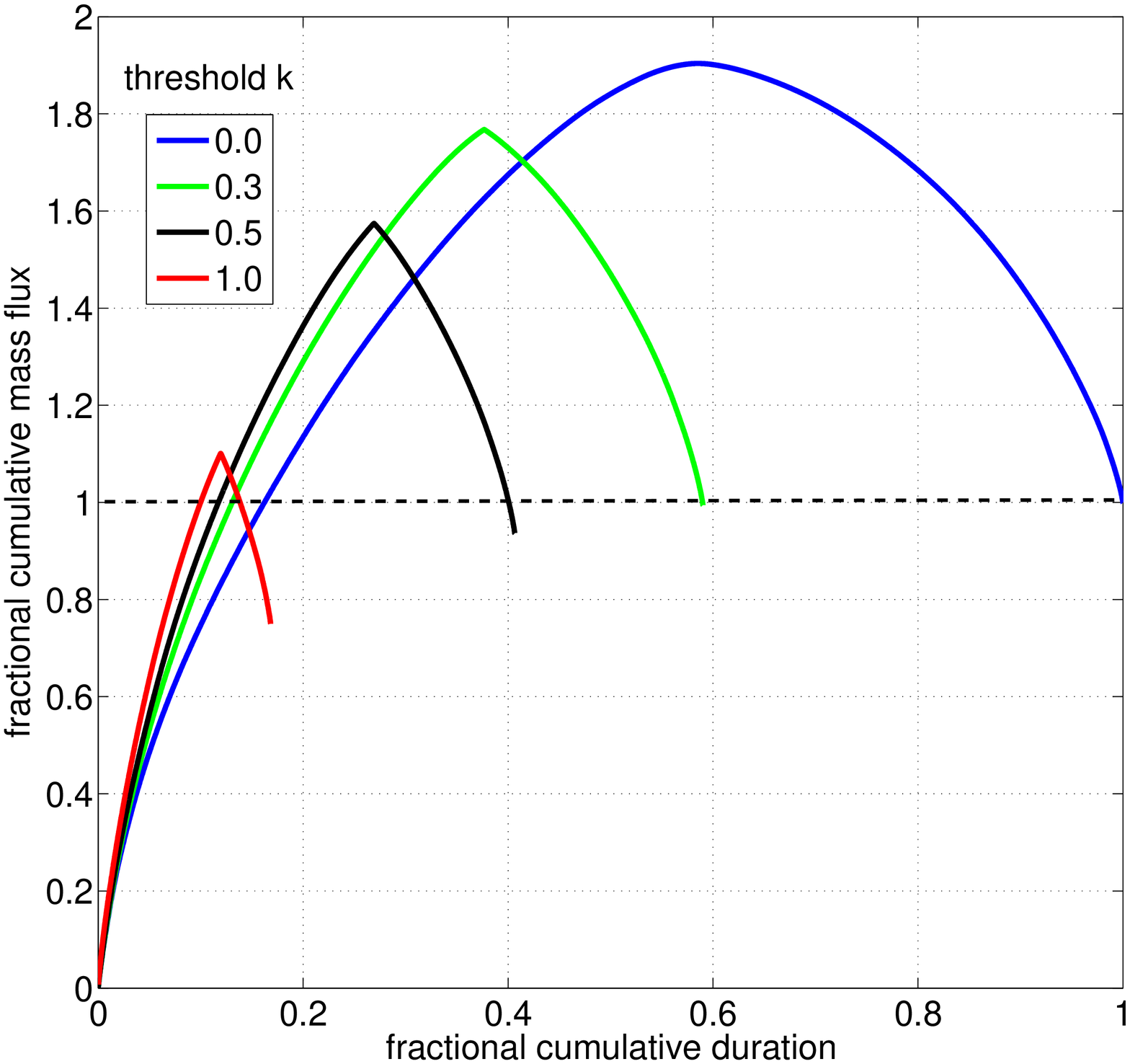}
\caption{Variation of cumulative mass flux as a function of cumulative duration for various event detection thresholds $k$. The events are arranged in descending order of the (signed) magnitude for calculating cumulative mass flux and cumulative duration.}
\label{fig:event_fig5}
\end{figure}

We define a few additional parameters to characterize the events. The
amplitude of an event $i$ is defined as

\begin{equation}
\label{eq2}
A_i(t) = \frac{1}{\overline{m}}\int_{\phi_{i\hat{1}}}^{\phi_{i\hat{2}}} \frac{m(t,z,\phi)}{\phi_{i\hat{2}} -
\phi_{i\hat{1}}} d\phi,
\end{equation}
where $\overline{m}$ is the mean mass flux (averaged over time and space) across the T/NT interface and $\phi_{i\hat{2}} -
\phi_{i\hat{1}}$ is the angular extent of event $i$. Thus the amplitude is the ratio of the contribution from an event
to the mean value of the total net mass flux. The fractional mass flux
contribution from an event $i$ is given by

\begin{equation}
\label{eq:contribution}
m_i = A_i*\frac{\phi_{i\hat{2}} -\phi_{i\hat{1}}}{T},
\end{equation}
where $T$ is the total length of the data set in degrees, over an appropriate number of realizations. So the net
contribution of an event is a product of both amplitude and
duration. Every event has a finite extent over time and space. Note that the events are not tracked over time, only the azimuthal extent of the event at a given time is accounted for in the present analysis. 

To determine the flux events the mass flux data at the R/IR interface were analyzed in $34 < z < 35$ for a total time of $185$ flow units. In Fig.~\ref{fig:event_fig4}, the ratio of the cumulative contribution of detected events to the total mass flux for various values of the threshold is presented. The objective of this plot is to determine
the appropriate threshold required to detect the minimum number of
events that will account for almost the whole flux to describe the mass flux balance. From
Fig.~\ref{fig:event_fig4} such a threshold is around $0.3$, as it
accounts for about $99.5\%$ of the total mass flux.

Narasimha \etal (2007) \cite{narasimha2007turbulent} also used a burstiness parameter to determine the
compactness and significance of the detected events. The contribution
from each event is calculated using eq.~\ref{eq:contribution}.  The events are arranged in
descending order of magnitude of their contributions from the largest positive through to the largest negative contribution. The
cumulative contribution from this re-ordered sequence of events is then plotted
against the cumulative duration of the events as shown in
Fig.~\ref{fig:event_fig5}. The burstiness curve reaches a maximum when the contributions of all the entraining (positive) events have been integrated, and then decreases thereon due to the detraining (negative) events. The curve is
plotted for various thresholds from 0.0 to 1.0. For a threshold
greater than about $0.3$, the detected events do not account for the total
mass flux. This would suggest that for higher thresholds the weaker
events are required for a complete description of the interfacial
dynamics. For thresholds less than $0.3$, contributions from
sub-threshold events are like noise with zero mean with little net contribution to the
net mean mass flux $\overline{m}$. For the threshold of $0.3$, the total contribution from positive
events reaches about $1.77\overline{m}$ over
37$\%$ of the total duration, and that from the negative
events is about $0.77\overline{m}$ over about $22\%$ of
the total duration. Following the terminology from Narasimha et al the
positive events are called productive and the negative events
counter-productive. The rest of the time,
over which the net contribution is negligible, is called the idle
phase.

\begin{figure}[!h]
\includegraphics[width = \textwidth]{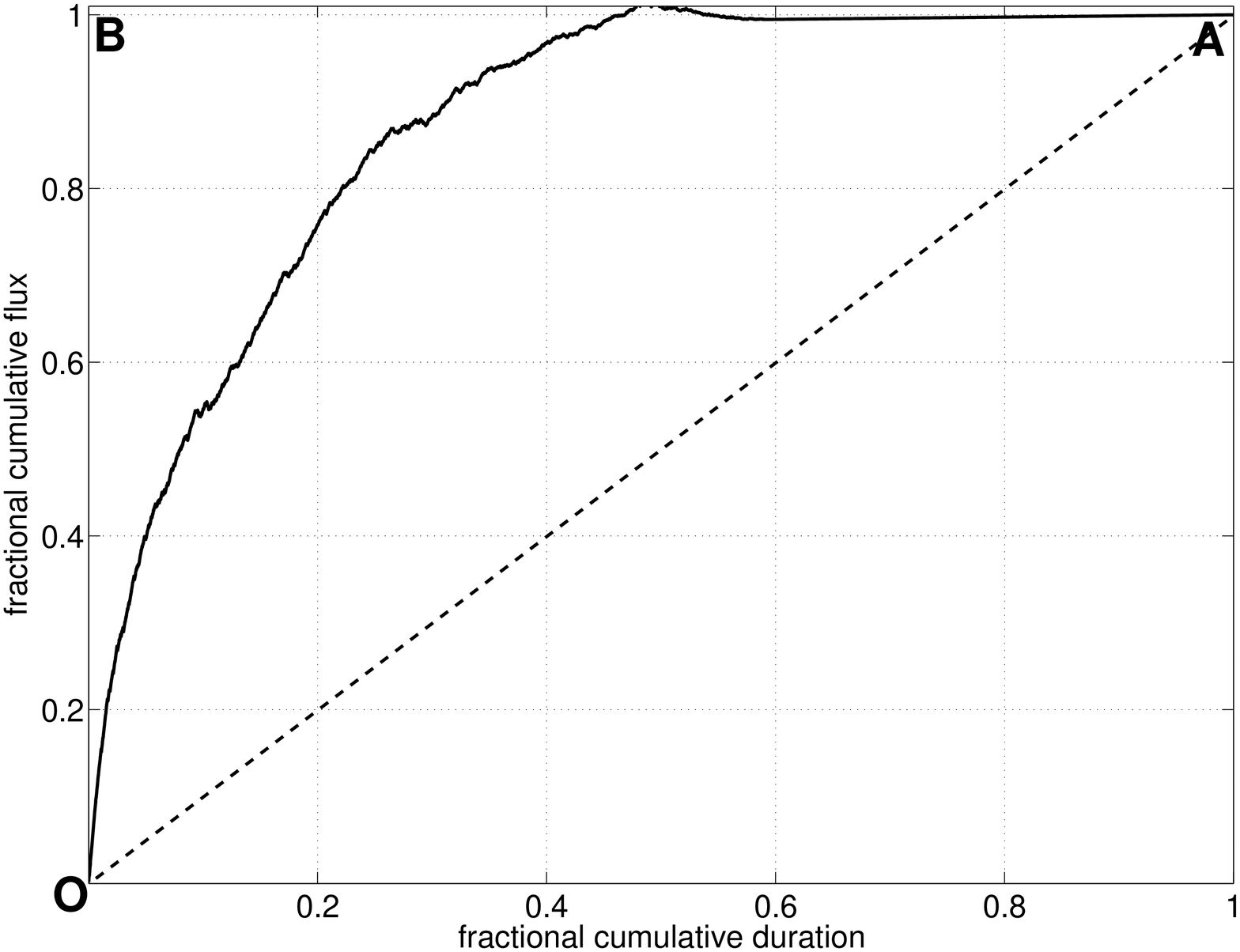}
\caption{Burstiness curve: Variation of cumulative mass flux as a function of cumulative duration for an event detection threshold $k$ of $0.3$. The events were ordered in the descending order of the absolute magnitude, for calculating the cumulative mass flux and the cumulative duration.}
\label{fig:event_fig6}
\end{figure}

We may also choose to arrange the events ordered according to the absolute
magnitude. With such an arrangement, the burstiness curve never exceeds unity  and has small
discontinuities, as shown in
Fig.~\ref{fig:event_fig6}. Following Narasimha \etal (2007), the
burstiness factor $B$ is then defined as the ratio of the area between the
curve and the line AO to the area of the triangle ABO. If $B=0$ for an
event series, then the burstiness curve is represented by the dashed
line AO. This indicates that all events have equal contributions over equal
durations for the entire dataset. On the other hand, if $B=1$, then the
burstiness curve would be the vertical axis BO, which implies
that the events are like a series of delta functions with zero net
duration, i.e. all the contributions occur instantaneously with zero
duration with the amplitude approaching infinity such that the net
contribution is finite. For the present data shown in
Fig.~\ref{fig:event_fig5}, the burstiness factor is $0.75$, which
indicates that there are significant intermittent periods of intense
activity at the R/IR interface that dictate the net mass flux entraining into the
turbulent core.

The visual and quantitative anaylsis presented here so far suggests that the velocity field outside the T/NT interface is organized and are induced by the large scale ordered motions in the turbulent flow. The irrotational fluid is drawn towards the T/NT interface by the large scale coherent motions and not by viscosity-dominated nibbling. Moreover, the induced flow field could either be entraining or detraining depending on the organization and orientation of the vortices near the T/NT interface. The induced flow near the T/NT interface contorts the interface, and the irrotational fluid appears to be entraining into the turbulent core through spatially and temporally localized intense events wherein viscous action imparts vorticity to the irrotational fluid as it crosses the T/NT interface. Thus entrainment is a two step process. In the first step, which is also the rate limiting step, irrotational fluid is driven by the large-scale dynamics into the turbulent core through the gulf or well-like regions in the T/NT interface,  and in the second step vorticity is imparted to fluid parcels through viscous processes as they cross the T/NT interface.

The two-step entrainment process discussed above is consistent with the views expressed by Plourde \etal (2008), Philip \& Marusic (2012), Mistry \etal (2016) and Burridge \etal (2017). Mistry \etal conclude that entrainment is a continuous multiscale process wherein at large scales the entrainment velocity is high to account for the smooth surface area and at smaller scales the entrainment velocity is relatively low, transporting flux occurring across a highly contorted fractal surface. Burridge \etal analyse the data on plumes and highlight the influence of the large scale motions in the entrainment process through conditional statistics. They show that vertical mass transport is decreased or increased in the absence or presence of large scale eddies and conclude that these eddies impart significant vertical velocities to the ambient fluid near the T/NT interface. We relate it to motions induced by vortices in the coherent structures, especially the toroidal vortex at the base of the structure. Moreover, Burridge \etal also show that the horizontal velocities are significantly enhanced in the absence of large scale eddies which in our analysis correspond to the region between the toroidal vortices of successive/neighbouring arrow head shaped coherent structures as in Fig.~\ref{fig:vel-vort}. The azimuthal variation of mass flux near the T/NT interface discussed in the present paper points to the episodic nature of entrainment. Further study will be required to determine whether all events may be associated with the large scale motions close to the T/NT interface and also to describe the shape of these events. It may be worthwhile to track the motions of the passive particles which originate from the irrotational region and study how they acquire vorticity near the T/NT interface. 
 
\section{Concluding remarks}
The objective of the present paper has been to describe the entrainment process in a round turbulent jet, using data provided by a well-resolved DNS study. As a preliminary step towards achieving this objective, detailed analyses of the DNS data have been made in the near-neighbourhood of the turbulent/non-turbulent (T/NT) interface, on the ambient as well as in the turbulent core neighbourhood of the interface. It is found necessary to distinguish a rotational / irrotational (R/IR) interface from the T/NT interface, the region in-between being rotational but not turbulent. Interestingly, it is found that the region between the core and the ambient have a length scale varying from the Kolmogorov to the Taylor length scales, indicating that both scales are relevant. The flow in the ambient neighbourhood of the T/NT interface is found to consist of several well-ordered circulating motions. By the analysis of such a flow field in a relatively simple case,
wherein an approximately two-dimensional blob of vorticity is present, it has been shown that the ordered motion in the ambient flow is due to the induced velocity of vorticity elements in coherent structures in the core-side neighbourhood. When and where the ordered motions are relatively intense in the ambient, the T/NT interface is dented towards the core, which may develop into a deep and twisted well or gulf. It is shown that the entrainment process can be broadly divided into an engulfment phase near the edge of the jet, followed by nibbling across the T/NT interface deeper in the well, supporting a similar conclusion of Burridge \etal (2017) regarding the episodic
nature of the entrainment process. It is convenient to think of this process as
constituting an entrainment ‘event’, intermittent in time and concentrated in space. These events are amenable to the method of analysis used for momentum flux events in the atmospheric boundary layer (Narasimha \etal (2007)). It is found that the burstiness of the signal is of the order of about 75\%, with entrainment about 40\% of the time, detrainment over about 10\% of the time and very little activity over the rest of the time. This picture suggests that for applications where greater entrainment is desired, for example, one may have to devise means by which the intensity and duration of entrainment events can be enhanced.

\providecommand{\noopsort}[1]{}\providecommand{\singleletter}[1]{#1}%

\end{document}